\documentclass[a4paper,10pt]{article}

\usepackage[T1]{fontenc}
\usepackage{lmodern}
\usepackage[utf8]{inputenc}
\usepackage{amsmath,amsfonts,amssymb,mathtools,mathrsfs,dsfont}
\usepackage{xcolor}
\usepackage{hyperref,url} 
\usepackage{setspace}
\usepackage{graphicx, subfig, float}
\usepackage{tikz}
\usetikzlibrary{decorations.pathmorphing}
\tikzset{snake it/.style={decorate, decoration=snake}}
\definecolor{dark-red}{RGB}{175, 0, 0}

\usepackage{comment}
\usepackage[margin=1.0in]{geometry}

\graphicspath{{./Figures/CTCs}{./Figures/Plot-RN}{./Figures/Plot-KN}}

\hypersetup{
  colorlinks=true,
  citecolor=violet,
  linkcolor=blue,
  urlcolor=blue}

\numberwithin{equation}{section}
\allowdisplaybreaks

\newcommand{\myparagraph}[1]{\paragraph{#1}\mbox{}\\ \vspace{-0.35cm} \\}

\newcommand{\beqNN}{\begin{equation*}}
\newcommand{\eeqNN}{\end{equation*}}

\DeclareMathAlphabet{\pazocal}{OMS}{zplm}{m}{n}
\DeclareMathAlphabet{\mathbfcal}{OMS}{cmsy}{b}{n}

\newcommand{\RR}{\mathbb{R}}

\newcommand{\E}{\pazocal{E}}

\DeclareMathSymbol{\Gamma}{\mathalpha}{operators}{0}
\DeclareMathSymbol{\Delta}{\mathalpha}{operators}{1}
\DeclareMathSymbol{\Theta}{\mathalpha}{operators}{2}
\DeclareMathSymbol{\Lambda}{\mathalpha}{operators}{3}
\DeclareMathSymbol{\Xi}{\mathalpha}{operators}{4}
\DeclareMathSymbol{\Pi}{\mathalpha}{operators}{5}
\DeclareMathSymbol{\Sigma}{\mathalpha}{operators}{6}
\DeclareMathSymbol{\Upsilon}{\mathalpha}{operators}{7}
\DeclareMathSymbol{\Phi}{\mathalpha}{operators}{8}
\DeclareMathSymbol{\Psi}{\mathalpha}{operators}{9}
\DeclareMathSymbol{\Omega}{\mathalpha}{operators}{10}

\newcommand{\CORRECTION}[1]{{\color{red}#1}}

\begin{document}

\begin{titlepage}


\vspace{2cm}

\begin{center}
\renewcommand{\thefootnote}{\fnsymbol{footnote}}
{\Huge \bf Kerr-Newman black hole  
\vskip 5mm
in a Melvin-swirling universe}
\vskip 32mm
{\large {Andrea Di Pinto$^{a, b, c}$\footnote{andrea.dipinto@mi.infn.it}},
Silke Klemm$^{a, b}$\footnote{silke.klemm@mi.infn.it} and
Adriano Vigan\`o$^{b}$\footnote{adriano.vigano@mi.infn.it}}\\

\renewcommand{\thefootnote}{\arabic{footnote}}
\setcounter{footnote}{0}
\vskip 10mm
\vspace{0.2 cm}
{\small \textit{$^{a}$Dipartimento di Fisica, Universit\`a degli Studi di Milano \\
Via Celoria 16, I-20133 Milano, Italy}} \\
\vspace{0.2 cm}
{\small \textit{$^{b}$Istituto Nazionale di Fisica Nucleare (INFN), Sezione di Milano \\
Via Celoria 16, I-20133 Milano, Italy}} \\
\vspace{0.2 cm}
{\small \textit{$^{c}$Dipartimento di Scienza e Alta Tecnologia (DISAT), Universit\`a degli Studi dell'Insubria \\
Via Valleggio 11, I-22100 Como, Italy}} \\
\vspace{0.2 cm}
\end{center}
\vspace{5.4 cm}

\begin{center}
{\bf Abstract}
\end{center}
We present a new exact solution of Einstein-Maxwell field equations which represents a rotating black hole with both electric and magnetic charges immersed in a universe which itself is also rotating and magnetized, i.e.~the dyonic Kerr-Newman black hole in a Melvin-swirling universe.

We show that the solution is completely regular and free of any type of singularity;
then we analyze its physical properties, such as the ergoregions and the shape of the event horizons.
Finally we present the extremal near horizon geometry of the metric and study its entropy via the Kerr/CFT correspondence.

\end{titlepage}

\addtocounter{page}{1}

\tableofcontents

\newpage

\section{Introduction}

Black holes are fundamental objects in our universe:
they might represent the keystone to understand quantum gravity, through the observation of gravitational waves produced by black holes merging.
As such, it is of utmost importance to construct exact solutions that describe such objects and to study their physical and mathematical properties.

The most general class of type D black holes is represented by the Pleba\'nski-Demia\'nski spacetime~\cite{Plebanski:1976gy,Griffiths:2005qp}, which describes charged, rotating and accelerating black holes with NUT parameter in the presence of a cosmological constant.
The properties of this metric have been extensively studied and it is known in many coordinate systems~\cite{Ovcharenko:2024yyu,Ovcharenko:2025fxg}.

Recently~\cite{Astorino:2022aam}, a novel class of black hole spacetimes, dubbed swirling black holes, was discovered.
The background spacetime underlying these metrics was first mentioned in~\cite{Gibbons:2013yq}, but the authors did not discuss its properties, nor considered the possibility to embed a black hole.

After the construction of the Schwarzschild and Kerr-swirling solutions~\cite{Astorino:2022aam}, some other extensions have been discovered~\cite{Astorino:2022prj,Astorino:2023elf,Cisterna:2023uqf,Illy:2023iau,Barrientos:2024pkt,Barrientos:2024umq,Astorino:2025zse}, and many physical properties of the swirling universe have been studied~\cite{Capobianco:2023kse,Moreira:2024sjq,Chen:2024jsv,Capobianco:2024jhe,Cao:2024pdb,Gu:2024dna}, proving that these spacetimes are a fertile subject of investigation.

The aim of the present paper is to broaden the swirling family by constructing a quite general solution, which encompasses many interesting sub-cases.

The main tool for embedding black holes in the swirling universe is the Ehlers transformation~\cite{Ehlers}.
As it is well known, the Einstein-Maxwell equations for a stationary and axisymmetric metric can be rewritten in an equivalent form, the celebrated Ernst equations~\cite{Ernst:1967wx,Ernst:1967by}.
The latter make manifest some Lie point symmetries, which allow us to map a solution of the Einstein-Maxwell system into another one.
The Ehlers map is one of these symmetries, and its action in the context of the swirling universe was clarified in~\cite{Astorino:2022aam}\CORRECTION{.}
Here, we apply the Ehlers transformation to the dyonic Kerr-Newman-Melvin metric~\cite{Gibbons:2013yq} to add the swirling rotation, thus obtaining a general solution that encompasses many interesting sub-cases.

The structure of the present paper is as follows.
In Sec.~\ref{sec:generation}, we present the Ernst equations and the Ehlers transformation needed to construct the novel solution.
In Sec.~\ref{sec:DKNMS}, we present the dyonic Kerr-Newman-Melvin-swirling spacetime and investigate its properties.
In particular, we study the singularity structure of the solution, focusing on the tuning of the parameters which allows us to completely remove the conical singularities and the Dirac strings.
Such a choice, as we show in Sec.~\ref{sec:coordinate-sing}, removes the typical curvature singularity hidden behind the horizon, thus leaving us with a completely regular manifold.
We also list, for the various sub-cases, when it is possible to choose the parameters to remove the singularities and when it is not.
Thus, the general solution that we construct offers the possibility to systematize the swirling family and to clarify the role of the parameters.

Then, we consider the frame-dragging and compute the angular velocity of the horizon in Sec.~\ref{sec:frame-drag}, which turns out to be useful to compute the temperature and the entropy in Sec.~\ref{sec:entropy-temp}.
We study the geometry of the solution in Sec.~\ref{sec:hor-ergo} and focus on the behaviour of the ergoregions.
Finally, we write down the near-horizon limit of the full black hole metric in Sec.~\ref{sec:nh}.

The present paper is largely based on the thesis~\cite{DiPinto:2024axv}.

\section{Generation of the solution}
\label{sec:generation}
The Einstein-Maxwell equations for stationary and axisymmetric spacetimes are equivalent to the complex Ernst equations~\cite{Ernst:1967wx,Ernst:1967by}
\begin{subequations}
\label{ernst-eq}
\begin{align}
(\text{Re}\,\E + |\Phi|^2) \, \nabla^2 \E & =
\vec{\nabla} \E \cdot \bigl( \vec{\nabla} \E + 2 \Phi^* \vec{\nabla}\Phi \bigr) \,, \\
(\text{Re}\,\E + |\Phi|^2) \, \nabla^2 \Phi & =
\vec{\nabla} \Phi \cdot \bigl( \vec{\nabla} \E + 2 \Phi^* \vec{\nabla}\Phi \bigr) \,,
\end{align}
\end{subequations}
where $\Vec{\nabla}$ and the other functions are understood in Euclidean space with cylindrical coordinates $(\rho,\varphi,z)$.

Stationarity and axisymmetry hold for the Lewis-Weyl-Papapetrou (LWP) metric, which can be written in two inequivalent forms:
the \emph{electric} one
\begin{equation}
\label{lwp-electric}
{ds}^2 =
f^{-1} \bigl[ e^{2\gamma} \bigl( {d\rho}^2 + {dz}^2 \bigr)
+ \rho^2 {d\phi}^2 \bigr] -f (dt - \omega d\phi)^2 \,,
\end{equation}
and the \emph{magnetic} one
\begin{equation}
\label{lwp-magnetic}
ds^2 = f^{-1} \Bigl[ - \rho^2 dt^2 + e^{2\gamma}\bigl( d\rho^2 + dz^2 \bigr)\Bigr]
+ f \bigl( d\varphi - \omega dt \bigr)^2 \,.
\end{equation}
In the following, we will make use of the magnetic form~\eqref{lwp-magnetic} only.

The Ernst potentials are defined using the components of the magnetic LWP metric,
\begin{equation}
\label{ernst-potentials}
\E = - f - |\Phi|^2 + i h \,, \qquad
\Phi = \tilde{A}_{t} - i A_{\varphi} \,,
\end{equation}
together with the twisted potentials
\begin{align}
\label{Atilde}
\vec{e}_{\varphi} \times  \vec{\nabla} \tilde{A}_{t} & =
- \frac{f}{\rho} \Bigl( \vec{\nabla} A_{t} + \omega \vec{\nabla} A_{\varphi} \Bigr) \,, \\
\label{h}
\vec{e}_{\varphi} \times \vec{\nabla} h & =
- \frac{f^2}{\rho} \vec{\nabla} \omega - 2\, \vec{e}_{\varphi} \times \text{Im} \bigl (\Phi^{*} \vec{\nabla}\Phi \bigr) \,.
\end{align}
The Ernst equations~\eqref{ernst-eq} enjoy many symmetries, i.e.~transformations of the potentials which leave the equations invariant, the non-trivial ones being the Ehlers map and the Harrison map.
These transformations act differently once they are applied to the electric or to the magnetic LWP metric:
The Ehlers map adds NUT charge (electric case) or swirling parameter (magnetic case), while the Harrison map adds dyonic charge (electric case) or electromagnetic Melvin background (magnetic case) to a given spacetime~\cite{Vigano:2022hrg}.
Explicitly, the Ehlers and Harrison transformations are respectively given by
\begin{align}
\E' & = \frac{\E}{1 + i \,\jmath \,\E} \,, \qquad\qquad\qquad \,
\Phi' = \frac{\Phi}{1 + i\,\jmath \,\E} \,, \label{ehlers} \\
\E' & = \frac{\E}{1 - 2\alpha^*\Phi - |\alpha|^2\E} \,, \qquad
\Phi' = \frac{\alpha\E + \Phi}{1 - 2\alpha^*\Phi - |\alpha|^2\E} \,, \label{harrison}
\end{align}
We are interested in the Ehlers transformation~\eqref{ehlers}, since we aim to add the swirling parameter~\cite{Astorino:2022aam}.

Our seed spacetime is the dyonic Kerr-Newman-Melvin metric~\cite{Gibbons:2013yq}, written in the following form
\begin{subequations}
\label{seed}
\begin{align}
ds^2 & = F \biggl[ -\frac{\Delta}{\Sigma}\,dt^2 + \frac{dr^2}{\Delta} + d\theta^2\biggr] + \frac{\Sigma\sin^2\theta}{F}\biggl[ d\varphi - \frac{\Omega}{\Sigma} \,dt \biggr]^2 \,, \\
A & = \frac{A_0}{\Sigma} \,dt + \frac{A_{3}}{F}\biggl[ d\varphi - \frac{\Omega}{\Sigma}\,dt \biggr] \,,
\end{align}
\end{subequations}
whose complete expression is reported in Appendix~\ref{app:seed-dknm}, and that represents a black hole of mass $M$, rotational parameter $a$, electric charge $Q$, magnetic charge $H$, embedded in a magnetic Melvin field $B$.

We can generate the dyonic Kerr-Newman-Melvin swirling spacetime by applying the Ehlers transformation~\eqref{ehlers} to the seed~\eqref{seed}, once it is written in the magnetic Lewis-Weyl-Papapetrou form~\eqref{lwp-magnetic}.
This is easily done by defining the Weyl coordinates
\begin{equation}
\rho = \sqrt{\Delta}\sin\theta \,, \qquad
z = \bigl(r - M \bigl) \cos\theta \,,
\end{equation}
and the functions
\begin{equation}
\label{metric-functions}
\omega = \frac{\Omega}{\Sigma} \,, \qquad
f = \frac{\Sigma \sin^2\theta}{F} \,, \qquad
\gamma = \frac{1}{2}
\log\biggl[ \frac{ \Sigma \sin^2\theta }{ \Delta - \bigl( a^2 + Z^2 - M^2 \bigr)\sin^2\theta }\biggr] \,.
\end{equation}
Therefore, by integrating equations~\eqref{Atilde} and~\eqref{h}, we obtain the twisted potential $\tilde{A}_{t}$ and $h$ for the seed metric~\eqref{seed}, whose explicit result is reported in Appendix~\ref{app:dknm-twisted potentials}, and consequently the Ernst potentials $\E$ and $\Phi$~\eqref{ernst-potentials}, to which we can apply the Ehlers map~\eqref{ehlers}.

It is interesting to note that, in this construction, the Ehlers and Harrison transformations commute.
In fact, it is possible to obtain our result by applying the Harrison map to the swirling Kerr-Newman solution.
This is a highly non-trivial result, since the Ehlers and Harrison maps belong to $\mathrm{SU(2,1)}$, which is not abelian.
This point is discussed in~\cite{Astorino:2023ifg}.

Finally, to obtain the new spacetime, we need to solve for the new Ehlers-transformed solution, by using again the definition of the Ernst potentials~\eqref{ernst-potentials} and by integrating again the equations~\eqref{Atilde} and~\eqref{h} which define the twisted potentials.
The result of the computation is shown in the next section.

\section{Dyonic Kerr-Newman in a Melvin-swirling universe}
\label{sec:DKNMS}
The metric and the Maxwell field resulting from the application of the Ehlers map to the seed~\eqref{seed} are
\begin{subequations}
\label{DKNMS}
\begin{align}
\label{DKNMS-metric}
ds^2 & = F \biggl[ -\frac{\Delta}{\Sigma}\,dt^2 + \frac{dr^2}{\Delta} + d\theta^2\biggr] + \frac{\Sigma\sin^2\theta}{F}\biggl[ d\varphi - \frac{\Omega}{\Sigma} \,dt \biggr]^2 \,, \\
\label{DKNMS-potential}
A & = \frac{A_0}{\Sigma} \,dt + \frac{A_{3}}{F}\biggl[ d\varphi - \frac{\Omega}{\Sigma}\,dt \biggr] \,,
\end{align}
\end{subequations}
where we defined the functions
\begin{align}
F & = R^2 + 2B\,\phi_{(0)} +  \frac{B^2}{2}\,\phi_{(1)} + \frac{B^3}{2}\,\phi_{(2)} + \biggl[\jmath^2 + \frac{B^4}{16} \biggr]\phi_{(3)} + 2 \jmath B \, \phi_{(4)} + \jmath\, F_{(1)} \,, \label{def-F} \\
\Omega &= a\,\lambda + 2B\,\chi_{(0)} +  \frac{B^2}{2}\,\chi_{(1)} + \frac{B^3}{2}\,\chi_{(2)}  + \biggl[\jmath^2 + \frac{B^4}{16} \biggr]\chi_{(3)} + 2 \jmath B \,\chi_{(4)} + \jmath \, \Omega_{(1)} \,, \\
A_0 &= \chi_{(0)} + \frac{B}{2}\,\chi_{(1)} + \frac{3B^2}{4}\,\chi_{(2)} + \frac{B^3}{8}\,\chi_{(3)} + \jmath\,\chi_{(4)} \,, \\
A_3 &= \phi_{(0)} +  \frac{B}{2}\,\phi_{(1)} + \frac{3B^2}{4}\,\phi_{(2)} + \frac{B^3}{8}\,\phi_{(3)} + \jmath\,\phi_{(4)} \,,
\end{align}
and
\begin{align}
Z^2 & = Q^2 + H^2 \,, \qquad\qquad\qquad\qquad\quad
\Xi = \bigl(r^2 + a^2\bigr )\sin^2\theta + Z^2\cos^2\theta \,, \\
\Delta & = r^2 - 2Mr + Z^2 + a^2 \,, \qquad\quad\quad
\lambda = r^2 + a^2 - \Delta = 2Mr - Z^2 \,, \\
\label{def-sigma}
\Sigma & = \bigl(r^2 + a^2\bigr)^2 - \Delta a^2 \sin^2\theta \,, \qquad\;\;
R^2 = r^2 + a^2\cos^2\theta \,.
\end{align}
The above functions are defined as the expansion of the quantities
\begin{subequations}
\begin{align}
\chi_{(0)} & = a H\Delta \cos \theta - Q r \bigl(r^2+a^2\bigr) \,, \\
\chi_{(1)} & = -3 a Z^2 \Bigl( \lambda + \Delta \bigl(1+\cos^2\theta \bigr) \Bigr) \,, \\
\begin{split}
\chi_{(2)} & = Q \biggl[ r^3 \Bigl( \lambda + \Delta \bigl(1+\cos^2\theta \bigr) \Bigr) + a^2 \Bigl( \Delta \cos^2\theta \bigl( 3r - 4 M \bigr) -  r \bigl( Z^2 + \Delta \bigr)   \Bigr) - 2 M a^4 \biggr] \\
&\quad + a H \Delta \cos\theta  \bigl( \, \Xi + 2 R^2 \bigr)  \,,
\end{split}
\\
\begin{split}
\chi_{(3)} & = a \biggl[6 M r^5 - a^2 \Delta \cos^2\theta \Bigl ( \bigl( Z^2 + 4 M^2 - 6 M r \bigr)\cos^2\theta + Z^2 + 12 M^2 - 12 M r - 6 r^2 \Bigr) - 2 a^4 M \bigl( 2 M + r \bigr) \\
& \quad - a^2 Z^2\Delta + 4 a^2 M r \bigl( r^2 - 2 Z^2 + 3 M r \bigr) - \Delta \cos^2\theta \Bigl( 6 r^2 \bigl(\Delta - r^2 \bigr) + \bigl( Z^4 + 2 M r^3 - 3 Z^2 r^2 \bigr) \cos^2\theta \Bigr)\biggr] \,,
\end{split}
\\
\begin{split}
\chi_{(4)} & = - a Q \Delta \cos\theta \bigl( \, \Xi + 2 R^2 \bigr) \\
&\quad + H \biggl[ r^3 \bigl( \Delta \cos^2\theta + r^2 \bigr) - a^2 \Bigl(r \bigl( Z^2 - r^2 \bigr) + 4 M \Delta \cos^2\theta + r \Delta \bigl( 1 - 3 \cos^2\theta \bigr) \Bigr) - 2 a^4 M\biggr] \,,
\end{split}
\end{align}
\end{subequations}
and
\begin{subequations}
\begin{align}
\phi_{(0)} & = a Q r \sin^2\theta - H \bigl( r^2 + a^2 \bigr) \cos\theta \,, \\
\phi_{(1)} & = \Sigma \sin^2 \theta + 3 Z^2 \bigl( r ^2\cos^2\theta + a^2 \bigr) \,, \\
\begin{split}
\phi_{(2)} & = a Q \biggl[\bigl( 1 + \cos^2\theta \bigr) \Bigl( r^3 +  \bigl (2 M + r \bigr) a^2 \Bigr) + r \cos^2\theta \Bigl(2 Z^2 - \Delta \bigl(3 - \cos^2\theta \bigr) \Bigr)\biggr] \\
&\quad + H \cos\theta \Bigl[ 2 a^2 \lambda \sin^2\theta - \bigl(r^2 +a^2) \, \Xi \Bigr] \,,
\end{split}
\\
\begin{split}
\phi_{(3)} & = Z^2 \biggl[2 a^4 \bigl( 1 + \cos^2\theta \bigr)^2 +  r^2 \cos^2\theta \bigl( \, \Xi + R^2 \sin^2\theta \bigr) \\
&\quad + a^2 \cos^2\theta \Bigl( 2\, \Xi + 3 Z^2 + r^2 \bigl( 5 + 6 \sin^2\theta + 3 \cos^4\theta \bigr) - 8 \Delta \Bigr) \biggr] \\
&\quad + a^2 \Bigl( \lambda^2 \cos^2\theta \bigl( 3 - \cos^2\theta \bigr)^2  + r^3 \sin^6\theta ( 4 M - r )\Bigr) + 2 a^4 \Bigl( 2 M^2 \bigl( 1 + \cos^2\theta \bigr)^2 - \Delta \sin^6\theta \Bigr) \\
&\quad + a^6 \sin^6\theta + \bigl( r^2 + a^2 \bigr)^3 \sin^4\theta \,,
\end{split}
\\
\phi_{(4)} & = a H \biggl[2 M \Bigl(a^2 +  \cos^2\theta \bigl(2 r^2 + a^2 \bigr) \Bigr) + r \sin^2\theta \bigl( \lambda + \Delta \sin^2\theta \bigr)\biggr] + Q \cos\theta \Bigl[ \bigl( r^2+ a^2 \bigr)\, \Xi - 2 \lambda a^2 \sin^2\theta \Bigr] \,,
\end{align}
\end{subequations}
and finally
\begin{subequations}
\begin{align}
\Omega_{(1)} & = -4 \Delta \cos \theta  \biggl[r^3+a^2 \Bigl(\bigl(r-M\bigr) \cos^2\theta - M \Bigr)\biggr] \,, \\
F_{(1)} & = -4 a \cos\theta \Bigl[M \bigl( 1 + \cos^2\theta \bigr) \bigl( r^2 + a^2 \bigr)+\lambda\, r \sin^2\theta \Bigr] \,.
\end{align}
\end{subequations}
The spacetime~\eqref{DKNMS} represents an electromagnetically charged and rotating black hole, embedded in a swirling and Melvin background\footnote{We remind the reader that the Melvin background represents a spacetime filled with a magnetic field, and it was firstly found and studied in~\cite{Bonnor:1954,Melvin:1963qx}.}.
As such, it is characterized by six parameters:
the mass $M$, the charges $Q$ and $H$, the angular momentum $a$, the external magnetic field $B$ and the swirling parameter $\jmath$.

For the sake of completeness, the integration of the equations~\eqref{Atilde} and~\eqref{h} after the Ehlers transformation~\eqref{ehlers} provides us with the following twisted potentials
\begin{align}
\tilde{A}_{t} & = \frac{1}{F}\biggl[ \tilde{\chi}_{(0)} + B \tilde{\chi}_{(1)} + \frac{B^2}{4} \, \tilde{\chi}_{(2)} + \jmath \, \tilde{\chi}_{(3)} + \frac{\jmath B}{2}\, \tilde{\chi}_{(4)} \biggr] \,, \\
h & = -\frac{1}{F}\biggl[ 2 \tilde{\chi}_{(1)} + B\, \tilde{\chi}_{(2)} + \jmath \, \tilde{\chi}_{(4)} \biggr] \,,
\end{align}
with $F$ as in~\eqref{def-F}, and where we have defined the ancillary functions
\begin{subequations}
\begin{align}
\tilde{\chi}_{(0)} & = - Q \cos\theta \bigl( r^2 + a^2 \bigr) - a H r \sin^2\theta \,, \\
\tilde{\chi}_{(1)} & = - a \cos\theta \biggl[ M \Bigl(\bigl( 3 r^2 + a^2 \bigr) - \bigl( r^2 - a^2\bigr)\cos^2\theta \Bigr) - Z^2 r \sin^2\theta \biggr] \,, \\
\tilde{\chi}_{(2)} & = a H \biggl[ 2 M \Bigl( a^2 + \cos^2\theta \bigl( 2 r^2 + a^2 \bigr) \Bigr) + r \sin^2\theta \bigl( \lambda + \Delta \sin^2\theta \bigr)\biggr] + Q \cos\theta \Bigl[\, \Xi  \bigl( r^2 + a^2 \bigr) - 2 a^2 \lambda  \sin^2\theta \Bigr] \,, \\
\tilde{\chi}_{(3)} & = a Q \biggl[ 2 M \Bigl( a^2 + \cos^2\theta \bigl( 2 r^2 + a^2 \bigr) \Bigr) + r \sin^2\theta  \bigl( \lambda + \Delta \sin^2\theta \bigr)\biggr] - H \cos\theta \Bigl[\, \Xi  \bigl( r^2 + a^2 \bigr) - 2 a^2 \lambda  \sin^2\theta \Bigr] \,, \\
\begin{split}
\tilde{\chi}_{(4)} & = a^2 \cos^2\theta \biggl[ \Delta^2 \bigl( 2-\cos^2\theta \bigr)^2  - 4 M r \Bigl(Z^2 + 2 \cos^2\theta \bigl( M r - Z^2 \bigr) \Bigr) + 2 r^2 \Bigl( 10 M^2 - \Delta \sin^2\theta - Z^2 \Bigr) - r^4 \biggr] \\
& \quad + 4 M a^2 r^3+ 2 a^4 \biggl[2 M^2 \bigl( 1 + \cos^2\theta \bigr)^2 - \Delta  \bigl( 2 - \cos^2\theta \bigr) \cos^2\theta + 2 M r \sin^2\theta \biggr] + a^6 \cos^2\theta + r^2 \Xi ^2 \,.
\end{split}
\end{align}
\end{subequations}
\subsection{Singularities}

In this section, we will analyze the properties of the novel solution~\eqref{DKNMS}.
We will encompass all the singularities that are usually found in spacetimes, i.e.~conical singularities, Dirac and Misner strings, and curvature singularities.
Moreover, we will check the presence of closed timelike curves (CTCs).

In particular, we will show that the new spacetime~\eqref{DKNMS} is completely regular outside the event horizon according to an appropriate choice of the parameters:
a constraint on the parameters will allow us to remove both the conical singularities and the Dirac string (which naturally appear in the presence of a magnetic charge), while the Misner string is not present because there is no NUT parameter.
In addition, this choice of parameters will also make the new spacetime free of curvature singularities, while CTCs will still be present, but hidden behind the event horizon.

\subsubsection{Coordinate singularities and horizons}
\label{sec:coordinate-sing}
It is straightforward to notice that the metric~\eqref{DKNMS} is in the same form as the seed~\eqref{seed}.
This implies that the new solution is also singular whenever $F=0$, $\theta=0$, $\Delta=0$, or $\Sigma=0$.
By analyzing the Kretschmann scalar, the only possible curvature singularity is for $F=0$, which will be discussed in detail in Sec.~\ref{sec:curvature}, while the others are all coordinate singularities\footnote{To be more precise, $\theta=0$ can also be a curvature singularity in three very peculiar sub-cases, which will be analyzed in Sec.~\ref{sec:curv-along-axis}.
However, this singularity also requires $F=0$.}.
In particular, the functions corresponding to these coordinate singularities are not modified by the swirling or Melvin parameters.
 
Therefore, the event horizons are found in the same locations of the corresponding black hole in Minkowski background
\begin{equation}
\label{horizon-DKNMS}
\Delta = 0 \quad \Rightarrow \quad r_{\pm} = M\pm\sqrt{M^2-a^2-Z^2} \,,
\end{equation}
which also implies that the extremality condition is $M_{ext}^2 = a^2 + Z^2$.

\subsubsection{Conical singularities}
\label{sec:conical-sing}
Axial conical singularities are a type of spacetime pathology due to a defect of the azimuthal angle, which results in a non-regular symmetry axis.
Furthermore, these singularities always appear in a pair, one for each half of the symmetry axis.

To compute the angular defects, we consider the ratio between the perimeter of a small circle around the symmetry axis and its radius~\cite{Astorino:2022aam}, for both halves of the symmetry axis, $\theta=0$ and $\theta=\pi$:
\begin{equation}
\delta_0 = \lim_{\theta \to 0} \frac{1}{\theta} \int_{0}^{2 \pi} \sqrt{\frac{g_{\phi\phi}}{g_{\theta\theta}}}d\phi \,, \quad
\delta_\pi =  \lim_{\theta \to \pi} \frac{1}{\pi-\theta} \int_{0}^{2 \pi} \sqrt{\frac{g_{\phi\phi}}{g_{\theta\theta}}}d\phi \,.
\end{equation}
Therefore, the condition for a spacetime to not possess conical singularities is
\begin{equation}
\label{conical-condition}
\delta_0 = \delta_\pi = 2\pi \,.
\end{equation}
It is always possible to fix at least one conical defect by choosing a new periodicity $\frac{2 \pi}{\delta \phi}$ for the azimuthal angle $\phi$ or, equivalently, by rescaling this angle as $\phi\mapsto\frac{2 \pi}{\delta \phi}\phi$ while maintaining the usual periodicity of $2\pi$.
Indeed, with this method, it is always possible to set at least one between $\delta_0$ and $\delta_\pi$ equal to $2 \pi$.
After this procedure, the other conical singularity, if still present, can be fixed by constraining the parameters of the solution;
however, such a constraint might be unphysical and thus unacceptable.


Furthermore, the presence of a non-removable conical singularity implies that a cosmic string or a strut has to be postulated to compensate for the effect induced by the spin-spin interaction between the black hole and the background, which would otherwise tend to add an acceleration to the black hole.

For the new metric~\eqref{DKNMS} we find
\begin{subequations}
\label{con-sing-1}
\begin{align}
\delta_0 & = \frac{32 \pi}{D_{(0)} + 16 \bigl(1 - 4 \jmath a M \bigr)^2 + 32 B\bigl( \jmath Q Z^2 - H\bigr) - 8 B^3 H Z^2} \,, \\
\delta_\pi & = \frac{32 \pi}{D_{(0)} + 16 \bigl(1 + 4 \jmath a M \bigr)^2 - 32 B\bigl(  \jmath Q Z^2 - H\bigr) + 8 B^3 H Z^2} \,,
\end{align}
\end{subequations}
where
\begin{equation}
D_{(0)} \coloneqq 24 B^2 Z^2 + 32 B^3 a M Q + B^4 \bigl(Z^4 + 16 a^2 M^2 \bigr) +  128 \jmath B a M H+ 16 \jmath ^2 Z^4 \,.
\end{equation}
Such conditions can be solved by choosing an appropriate value for the swirling parameter, to give
\begin{subequations}
\label{free-con-sing}
\begin{align}
\jmath & = - \frac{ B H \bigl( 4 + B^2 Z^2 \bigr)}{16 a M - 4 B Q Z^2} \,, \\
\delta\phi & = \frac{32 \pi \bigl( B Q Z^2 - 4 a M \bigr)^2\Bigl( 16 a^2 M^2 - 8 a B M Q Z^2 + B^2 Z^6\Bigr)^{-1}}{16 + 8 B^2 \bigl( H^2 + 3 Q^2 \bigr) + 32 a B^3 M Q + B^4 \bigl(16 a^2 M^2 + Z^4 \bigr)} \,.
\end{align}
\end{subequations}
Thus, upon the above choice, the solution~\eqref{DKNMS} is free of conical singularities.
In particular, the condition on $\jmath$ is physical, since the swirling parameter can take any sign.

A detailed discussion of the various sub-cases, comprehensive of the choices on the physical parameters, can be found in the Appendix~\ref{app:con-sing}.
If we look at them, there are some for which the conical singularities are non-removable, others for which they are never present, and some particular cases for which these singularities are removable, as summarized in Table~\ref{table:Conical-singularities}.
{\begin{table}[H]
\begin{center}
\onehalfspacing
\small
\begin{tabular}{ |p{4.65cm}||p{1.9cm}|p{1.9cm}|p{1.9cm}|p{2.5cm}|  }
\hline
\multicolumn{5}{|c|}{Conical Singularities} \\
\hline
Spacetime                                                  & Minkowski   & Melvin          & Swirling        & Melvin-swirling \\
\hline
Background                                                & No               & No                & No                & No              \\
Schwarzschild                                            & No               & No                & No                & No               \\
Electric\,\,\,\,\,\,Reissner-Nordstr\"{o}m      & No               & Removable   & Removable  & Yes               \\
Magnetic\,\,Reissner-Nordstr\"{o}m           & No               & Yes               & Removable   & Yes               \\
Dyonic\,\,\,\,\,\,\,\,Reissner-Nordstr\"{o}m   & No               & Yes               & Removable   & Removable    \\
Kerr                                                             & No              & Removable   & Yes               & Yes                 \\
Electric\,\,\,\,\,\,Kerr-Newman                      & No              & Removable   & Yes               & Removable     \\
Magnetic\,\,Kerr-Newman                           & No              & Yes                & Yes              & Removable      \\
Dyonic\,\,\,\,\,\,\,\,Kerr-Newman                   & No              & Yes                & Yes             & Removable      \\
\hline
\end{tabular}
\caption{\small Summary for the sub-cases whether the conical singularities are never present (No), non-removable (Yes), or removable by a certain choice of the parameters and/or a rescaling of the azimuthal angle (Removable).}
\label{table:Conical-singularities}
\end{center}
\end{table}}
If both the Melvin and swirling parameters are present, the conical singularities are removable only if both of the charges are non-null, $Q,H\neq0$, or if the rotation parameter is non-zero, $a\neq0$.
Additionally, for these Melvin-swirling sub-cases, the conical singularities are never present only if the charges and the rotation parameters are all null, $a=Q=H=0$.
Moreover, for the swirling sub-cases, $B=0$, the conical singularities are never present or removable only if the rotation parameter is equal to zero, $a=0$.
Similarly, for the Melvin sub-cases, $\jmath=0$, these singularities are removable or never present only if the magnetic charge is null, $H=0$.

\subsubsection{Dirac strings}
\label{sec:dirac-strings}
Dirac strings arise in the presence of magnetic charges~\cite{Dirac:1931kp,Dirac:1948um}.
Indeed, the presence of the magnetic charge cannot be described by a smooth magnetic potential and its discontinuity generates a line singularity, which is the Dirac string.
These strings can be seen as one-dimensional curves in space that connect two Dirac monopoles with opposite magnetic charges, or similarly from one magnetic monopole out to infinity.

The condition for a Maxwell potential $A_{\mu}$ to \emph{not} present Dirac strings is 
\begin{equation}
\label{dirac-condition}
\lim_{\theta \to 0} A_{\phi} = \lim_{\theta \to \pi} A_{\phi} = 0 \,.
\end{equation}
If one wants to remove both the conical singularities and the Dirac strings, it is more convenient to simultaneously consider the rescaling of the azimuthal coordinate $\phi\mapsto\frac{2 \pi}{\delta \phi}\phi$ with the gauge transformation as $A_{\phi} \mapsto \frac{2 \pi}{\delta \phi} A_{\phi} - \delta A_{\phi}$.

In our case, we find
\begin{subequations}
\begin{align}
\label{Dirac-sing-1}
\lim_{\theta \to 0} A_{\phi} & =\frac{\delta_0}{16 \pi} \Bigl( A_{(0)} + A_{(1)}\Bigr) \,, \\
\lim_{\theta \to \pi} A_{\phi}  & = \frac{\delta_\pi}{16 \pi} \Bigl( A_{(0)} - A_{(1)}\Bigr) \,,
\end{align}
\end{subequations}
where $\delta_{0}$ and $\delta_{1}$ are those which define the condition for conical singularities~\eqref{con-sing-1}, and
\begin{subequations}
\begin{align}
A_{(0)} & \coloneqq 32 \jmath a H M + 12 B Z^2 + 24 B^2 a M Q + B^3 \bigl(16 a^2 M^2 + Z^4\bigr)  \,,\\
A_{(1)} & \coloneqq - 8 H + 8 \jmath Q Z^2 - 6 B^2 H Z^2 \,.
\end{align}
\end{subequations}
An appropriate choice to remove both the conical singularities and the Dirac strings is given by
\begin{subequations}
\label{reg-DKNMS}
\begin{align}
\jmath & = \frac{ H \bigl(4 + 3 B^2 Z^2\bigr)}{4 Q Z^2} \,, \\
a & = \frac{ Q B^3 Z^4}{2 M \bigl(4 + 3 B^2 Z^2\bigr)} \,, \label{reg-DKNMS-a}\\
\delta \phi & = \frac{32 \pi Q^2 \bigl(4 + 3 B^2 Z^2\bigr)^2 \Bigl[ 16 + 8 B^2 \bigl( 3 H^2 + Q^2 \bigr) + B^4 Z^2 \bigl( 9 H^2 + Q^2 ) \Bigr]^{-1}}{Z^2  \Bigl(16 + 8 B^2 \bigl( 3 H^2 + 5 Q^2 \bigr) + B^4 Z^2 \bigl( 9 H^2 + 25 Q^2 \bigr) + 4 B^6 Q^2 Z^4 \Bigr)} \,, \\
\delta A_{\phi} & = \frac{B Z^2 \Bigl[ 192 + 16 B^2 \bigl(23 H^2 + 19 Q^2 \bigr) + 12 B^4 Z^2 \bigl(19 H^2 + 15 Q^2 \bigr) + 45 B^6 Z^6 + 4 B^8 Q^2 Z^6 \Bigr]}{8 \bigl(4 + 3 B^2 Z^2 \bigr)^2} \,.
\end{align}
\end{subequations}
We can notice that the expressions in Eqs.~\eqref{reg-DKNMS} are physically well-posed only for
\begin{equation}
\label{physical-req}
a \,Q\, B > 0 \,.
\end{equation}
Indeed, the mass only appears in the constraint on the angular momentum~\eqref{reg-DKNMS-a}, from which we can guarantee that the mass is positive, $M>0$, only when the product of the parameters $a$, $Q$ and $B$ is positive as well.
Equivalently, if the mass $M$ is taken as a free positive parameter, the requirement~\eqref{physical-req} directly determines the direction of the black hole rotation through the constraint provided by Eq.~\eqref{reg-DKNMS-a}.

The sub-cases are studied in the Appendix~\ref{app:dirac} and summarised in Table~\ref{table:Dirac-strings}.
{\begin{table}[H]
\begin{center}
\onehalfspacing
\small
\begin{tabular}{ |p{4.65cm}||p{1.9cm}|p{1.9cm}|p{1.9cm}|p{2.5cm}|  }
\hline
\multicolumn{5}{|c|}{Dirac Strings} \\
\hline
Spacetime                                                  & Minkowski    & Melvin                  & Swirling                        & Melvin-swirling \\
\hline
Background                                                & No                & No                         & No                               & No                           \\
Schwarzschild                                            & No                & No                         & No                               & No                            \\
Electric\,\,\,\,\,\,Reissner-Nordstr\"{o}m      & No                & No                         & Yes                              & Removable$^{*}$     \\
Magnetic\,\,Reissner-Nordstr\"{o}m           & Yes               & Yes                        & Yes                              & Removable$^{*}$     \\
Dyonic\,\,\,\,\,\,\,\,Reissner-Nordstr\"{o}m   & Yes               & Removable            & Removable$^{**}$      & Removable$^{***}$    \\
Kerr                                                             & No                & No                         & No                               & Yes                            \\
Electric\,\,\,\,\,\,Kerr-Newman                      & No                & No                         & Yes                              & Removable$^{***}$    \\
Magnetic\,\,Kerr-Newman                           & Yes               & Removable$^{*}$  & Removable                  & Removable$^{***}$    \\
Dyonic\,\,\,\,\,\,\,\,Kerr-Newman                   & Yes               & Removable$^{*}$  & Removable$^{*}$        & Removable$^{**}$      \\
\hline
\end{tabular}
\caption{\small Table summarizing for all the sub-cases whether the Dirac strings are never present (No), non-removable (Yes), or removable by a certain choice of the parameters and/or a gauge transformation.\\
$*$ In these cases, there are additional constraints in addition to the relations between the parameters necessary to remove the Dirac strings.
For example, for the magnetic Reissner–Nordstr\"{o}m black hole in a Melvin-swirling universe it is also required that $|BH|>2$.\\
$**$ In these cases, the Dirac Strings are removable with constraints on the parameters that are compatible with those necessary to remove the conical singularities.\\
$***$ In these cases, the conditions to remove the Dirac Strings are not compatible with the respective conditions needed in order to remove the conical singularities.}
\label{table:Dirac-strings}
\end{center}
\end{table}}
Collecting the results of Table~\ref{table:Conical-singularities} and Table~\ref{table:Dirac-strings}, the only two possible sub-cases for which it is possible to remove both the Dirac strings and the conical singularities\footnote{With this statement we are considering only the spacetimes which initially have both of these singularities, otherwise there are other sub-cases that can result in a spacetime without any of these singularities.} are the new dyonic Kerr-Newman black hole in a Melvin-swirling universe~\eqref{DKNMS}, and its non-rotating and non-Melvin sub-case, the dyonic Reissner-Nordstr\"om black hole in a swirling universe.

For the other sub-cases, we have that if both the Melvin and swirling parameters are present, the Dirac strings are always removable or never present, except if the black hole is rotating but not charged, which is quite a surprising result since the Dirac strings usually arise in black holes that do carry a magnetic charge.
On the other hand, if only the swirling parameter is present, it is possible to have Dirac strings if any type of charge is present, which however is an expected result since, in a certain sense, the addition of the swirling rotation generates a magnetic charge from an electric one, and vice-versa\footnote{In Appendix~\ref{Dyonic-Reissner-Nordstrom-Melvin-swirling} is reported the dyonic Reissner-Nordstr\"om black hole in a Melvin-swirling universe with a clarification on this topic in Appendix~\ref{app:nheg}.}.

Finally, for the Melvin sub-cases, we have that the addition of the uniform magnetic field to the Minkowski background has the effect of making the Dirac strings always removable except for the non-rotating magnetic black hole.

\subsubsection{Closed timelike curves}
\label{sec:CTCs}
Closed timelike curves (CTCs) are timelike curves that loop back on themselves, thus violating causality.
For an axisymmetric spacetime with Killing vector $m=\partial_{\phi}$, whose orbits are closed, the existence of CTCs is thus determined by the region for which $m$ becomes timelike, i.e.
\begin{equation}
m^2 = g_{\mu \nu} m^{\mu}m^{\nu} = g_{\phi \phi} < 0 \,.
\end{equation}
In the case of solution~\eqref{DKNMS} we have that 
\begin{equation}
g_{\phi \phi} = \frac{\Sigma \sin^2\theta}{F} \,,
\end{equation}
with functions $F$ and $\Sigma$ respectively defined by Eq.~\eqref{def-F} and Eq.~\eqref{def-sigma}.
However, using the argument explained in Appendix~\ref{app:ernst-property}, we have that the regions in which CTCs are allowed do not depend on either the Melvin parameter $B$ or the swirling parameter $\jmath$, and are the same as those of the Kerr-Newman black hole, which are determined only by the condition
\begin{equation}
\Sigma = \bigl(r^2 + a^2\bigr)^2 - \bigl(r^2 - 2Mr + Q^2 + H^2 + a^2\bigr)\, a^2 \sin^2\theta < 0 \,.
\end{equation}
Hence, CTCs are allowed whenever the rotation parameter $a$ and at least one of the charges $Q$ or $H$ are non-zero, i.e.~for the Kerr-Newman black hole\footnote{To be more precise, if the radial coordinate $r$ is allowed to be negative, $r<0$, the same argument also applies for the rotating uncharged Kerr black hole.},
and in the same regions independently of a possible embedding of this black hole in a Melvin, swirling, or Melvin-swirling universe.
Therefore, CTCs are allowed only in the finite ``doughnut'' region near the origin $r=0$ inside the (inner) event horizon.

\begin{figure}[H]
\captionsetup[subfigure]{labelformat=empty}
\centering
\hspace{-0.75cm} 
\subfloat[\hspace{1cm} CTCs Cross-section $y=0$]{{\hspace{0.5cm}\includegraphics[width=0.3\textwidth]{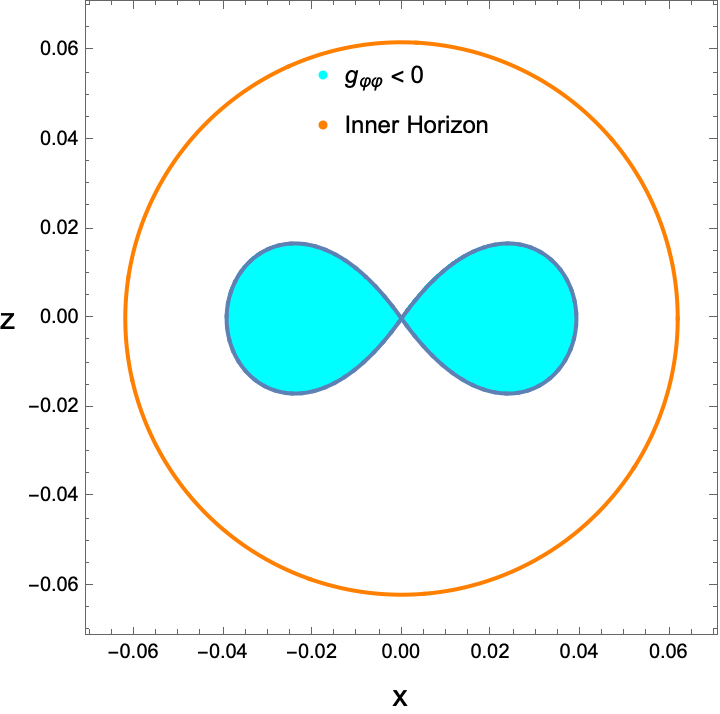}}}
\subfloat[\hspace{1cm} CTCs]{{\hspace{0.5cm}\includegraphics[width=0.3\textwidth]{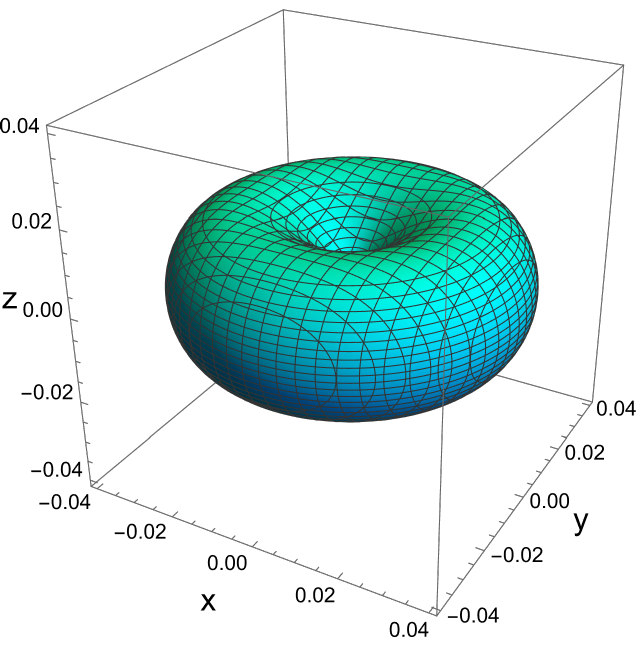}}}
\subfloat[\hspace{1cm} CTCs Cross-section $z=0$]{{\hspace{0.5cm}\includegraphics[width=0.3\textwidth]{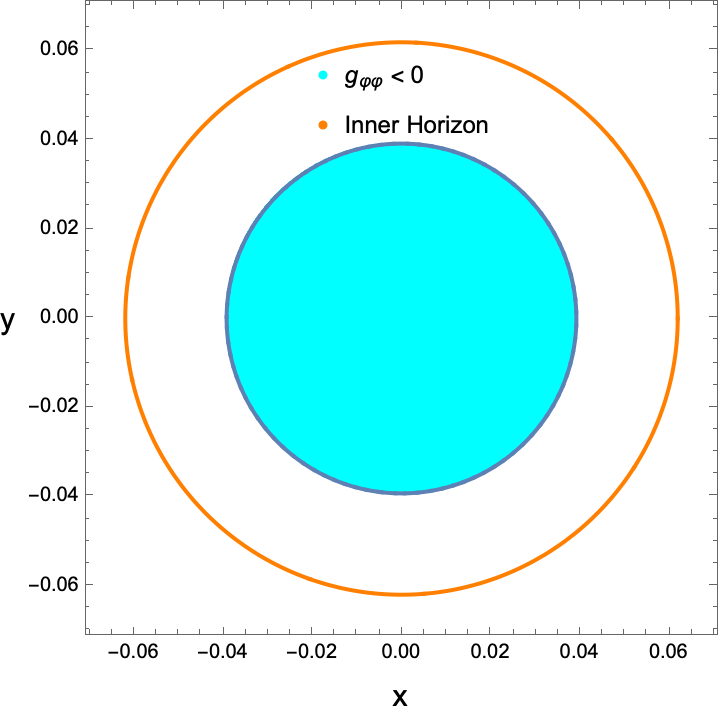}}}
\caption{\small Plots depicting the ``doughnut'' region in which CTCs are allowed, for the dyonic Kerr-Newman black hole with parameters $M=1$, $Q=1/5$, $H=1/5$, $a=1/5$ and then $\pazocal{I}=0.12$, independently of the choice of the swirling and Melvin parameters $\jmath$ and $B$.
The left and right panels show the cross-sections of this region and of the corresponding inner horizon, respectively for $y=0$ and $z=0$, in order to prove that this region is located inside the inner horizon.}
\label{CTC-KN}
\end{figure}

\begin{figure}[H]
\captionsetup[subfigure]{labelformat=empty}
\centering
\hspace{-0.75cm}
\subfloat[\hspace{1cm} CTCs Cross-section $y=0$]{{\hspace{0.5cm}\includegraphics[width=0.3\textwidth]{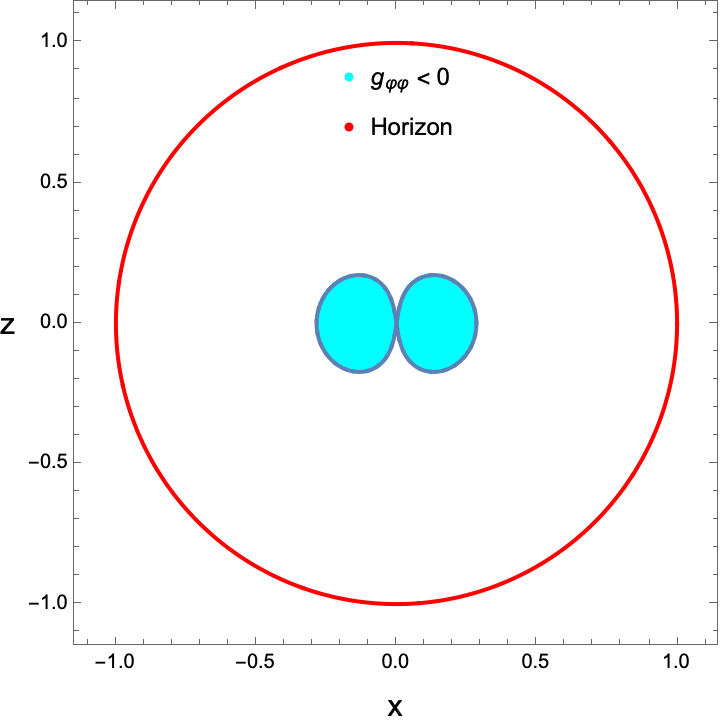}}}
\subfloat[\hspace{0.5cm} CTCs]{{\hspace{0.5cm}\includegraphics[width=0.3\textwidth]{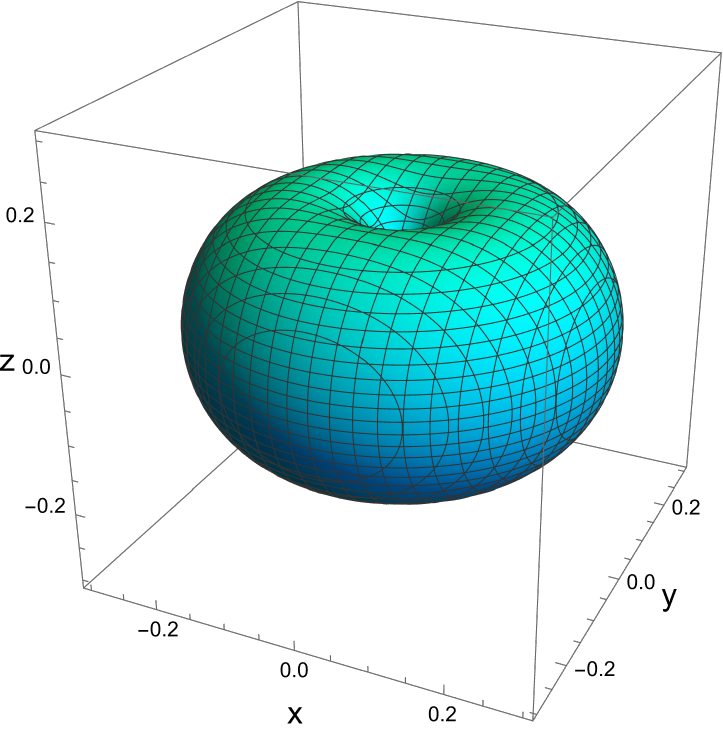}}}
\subfloat[\hspace{1cm} CTCs Cross-section $z=0$]{{\hspace{0.5cm}\includegraphics[width=0.3\textwidth]{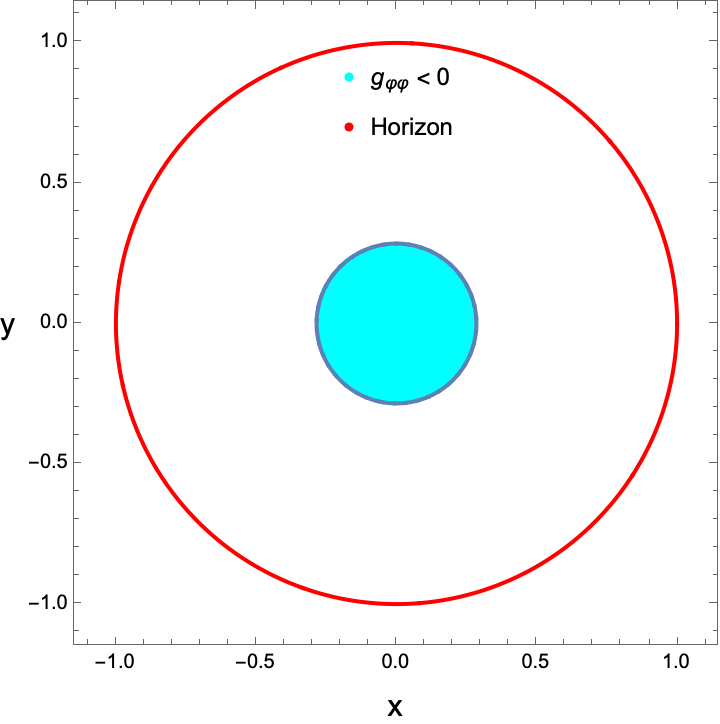}}}
\caption{\small Plots depicting the ``doughnut'' region in which CTCs are allowed, for the extremal dyonic Kerr-Newman black hole with parameters $M=1$, $Q=7/10$, $H=7/10$, $a=\sqrt{2}/10$ and $\pazocal{I}=1$, independently of the choice of the swirling and Melvin parameters $\jmath$ and $B$.
The left and right panels show the cross-sections of this region and of the corresponding event horizon, respectively for $y=0$ and $z=0$, in order to prove that this region is located inside the event horizon.}
\label{CTC-KN-2}
\end{figure}

In order to visualize this region, it is useful to define the rectangular coordinates $(x,y,z)$:
\begin{equation}
\label{rectangular-coordinates}
r = \sqrt{x^2 + y^2 + z^2} \,, \quad
\cos\theta = \frac{z^2}{x^2 + y^2 + z^2} \,.
\end{equation}
Additionally, we define the ``extremality index'' $\pazocal{I}$, which represents how close the black hole is to extremality, in the sense that $\pazocal{I}=1$ if the black hole is extremal:
\begin{equation}
\label{extremality-index}
\pazocal{I}=\frac{Q^2+H^2+a^2}{M^2} \,.
\end{equation}
With these definitions, we can perform a numerical analysis of the function $g_{\phi \phi}$, which is pictured in Fig.~\ref{CTC-KN} for the non-extremal case, and in Fig.~\ref{CTC-KN-2} for the extremal one.


However, as we can see from these plots, the use of the rectangular coordinates~\eqref{rectangular-coordinates} leads to distorted results since, for example, the horizons are always spheres.
Nevertheless, these coordinates are still useful for qualitative analysis.
Moreover, as already pointed out, with these coordinates the symmetry axis is entirely mapped to the $z-$axis.

\subsubsection{Misner strings}
\label{sec:misner}
Misner strings~\cite{Misner:1963fr,Astefanesei:2004kn}, which are the gravitational analog of the Dirac strings, appear in rotating spacetimes when the non-diagonal function $g_{t\phi}/g_{tt}$ is not regular on the symmetry axis~\cite{Alekseev:2019kcf}, taking different values on the two hemispheres $\theta=0$ and $\theta=\pi$.
Furthermore, the presence of a Misner string also leads to CTCs and causality issues.

Hence, the condition which guarantees the absence of the singularity is given by
\begin{equation}
\lim_{\theta \to 0} \frac{g_{t\phi}}{g_{tt}} = \lim_{\theta \to \pi} \frac{g_{t\phi}}{g_{tt}} \,.
\end{equation}
This kind of defect is usually associated with the presence of the NUT parameter, and since the latter is absent in our class of spacetimes, we find that, as expected, the solution~\eqref{DKNMS} and none of its sub-cases are characterized by Misner strings.

\subsection{Curvature}
\label{sec:curvature}
Finally, we study the curvature by analyzing the Kretschmann scalar
$K=R_{\mu \nu \rho \sigma}R^{\mu \nu \rho \sigma}$.
For large radial distances, $r\to\infty$, the Kretschmann scalar of the dyonic Kerr-Newman metric~ decays as
\begin{equation}
K\big\rvert_{\jmath = B = 0} \overset{r\to \infty}{\approx} \frac{48 M^2}{r^6} \,.
\end{equation}
Conversely, for the Melvin-swirling solutions, we have that this scalar invariant decays faster if at least one between the Melvin or swirling parameters is non-zero
\begin{equation}
K \overset{r\to \infty}{\approx} \frac{49152}{\bigl[B^4 + 16 \jmath^2\bigr]^2\sin^{12}\theta\, r^{12}} \,.
\end{equation}
Therefore, if these parameters are present, the spacetime is locally asymptotically flat.
Moreover, these spacetimes have also an asymptotic constant curvature on the symmetry axis.

For the most general case~\eqref{DKNMS}, we have
\begin{subequations}
\label{kretch-inf-poles}
\begin{align}
K\big\rvert_{\theta = 0} \,\xrightarrow{r\to \infty}\,  &  \biggl[\frac{\delta_0}{2\pi}\biggl]^4\Bigl[ K_{(0)} + K_{(1)}\Bigr] \,, \\
K\big\rvert_{\theta = \pi} \,\xrightarrow{r\to \infty}\,&  \biggl[\frac{\delta_\pi}{2\pi}\biggl]^4\Bigl[ K_{(0)} - K_{(1)}\Bigr] \,,
\end{align}
\end{subequations}
where $\delta_{0}$ and $\delta_{1}$ are the same which define the condition for conical singularities~\eqref{con-sing-1}, and
\begin{subequations}
\begin{align}
\begin{split}
K_{(0)} & \coloneqq  2 B^4\Bigr[10 + B^2 \bigl(15 H^2 - Q^2\bigr) - 128 \jmath H Q\Bigr] - 8 \jmath^2 \Bigl[24 + 4 B^2\bigl(H^2 - 15 Q^2\bigr) \Bigr] \\
& \quad  - 24 \Bigl[B^4 + 16 \jmath^2\Bigr] \bigl(B a M \bigr)\bigl( 4 \jmath H + B^2 Q\bigr) + \frac{1}{4}\Bigl[B^4 + 16 \jmath^2\Bigr]^2 \bigl(5 Z^4 - 48 a^2 M^2\bigr) \,,
\end{split}\\
\begin{split}
K_{(1)} & \coloneqq  - 8 B \bigl(5 B^4 H - 32 \jmath B^2 Q - 48 \jmath^2 H \bigl) \\
& \quad + 2 \Bigl[B^4 + 16 \jmath^2\Bigr] \Bigl[48 \jmath a M + 5 B Z^2\big(4\jmath Q - B^2 H\bigr) \Bigr] \,.
\end{split}
\end{align}
\end{subequations}
More precisely, as we can see from Eqs.~\eqref{kretch-inf-poles}, the two hemispheres $\theta=0$ and $\theta=\pi$ do not asymptotically converge to the same value of constant curvature, except if we remove the conical singularities and possibly impose some additional constraints.

Using the conditions necessary to remove the conical singularities~\eqref{free-con-sing}, we obtain, by definition, $\delta_0=\delta_\pi=2\pi$.
Moreover, we obtain that $K_{(1)} = 0$ if
\begin{equation}
\jmath = \frac{B^2 H}{4 Q} \,, \qquad
a = -\frac{Q}{B M}\,.
\end{equation}
However, these constraints are not compatible with those needed to remove the Dirac strings~\eqref{reg-DKNMS}.

\subsubsection{Curvature singularities at the ``center'' of the black hole}
\label{sec:curv-center}
{\begin{table}[H]
\begin{center}
\onehalfspacing
\small
\begin{tabular}{ |p{4.65cm}||p{1.9cm}|p{1.9cm}|p{1.9cm}|p{2.5cm}|  }
\hline
\multicolumn{5}{|c|}{Curvature Singularity for $r = \cos\theta = 0$} \\
\hline
Spacetime                                                  & Minkowski    & Melvin                         & Swirling       & Melvin-swirling \\
\hline
Background                                                & No                & No                               & No               & No                                  \\
Schwarzschild                                            & Yes               & Yes                              & Yes              & Yes                                 \\
Electric\,\,\,\,\,\,Reissner-Nordstr\"{o}m      & Yes               & Yes                              & Yes              & Yes                                  \\
Magnetic\,\,Reissner-Nordstr\"{o}m           & Yes               & Yes                              & Yes              & Yes                                  \\
Dyonic\,\,\,\,\,\,\,\,Reissner-Nordstr\"{o}m   & Yes               & Yes                              & Yes              & Yes                                  \\
Kerr                                                             & Yes               & No                               & No               & No                                   \\
Electric\,\,\,\,\,\,Kerr-Newman                      & Yes               & Condition$^{\star}$     & No               & No                                    \\
Magnetic\,\,Kerr-Newman                           & Yes               & No                               & No               & No                                     \\
Dyonic\,\,\,\,\,\,\,\,Kerr-Newman                   & Yes               & No                               & No               & Condition$^{\star \star}$   \\
\hline
\end{tabular}
\caption{Summary of the cases in which the ``center'' of the black hole $r=\cos\theta=0$ represents a curvature singularity. \\
$\star$ In this case, it is possible to have a curvature singularity for $r=\cos\theta=0$ if the following relation between the parameters holds:
$a=-\frac{2Q}{BM}$. \\
$\star\star$ In this case, there can be a curvature singularity for $r=\cos\theta=0$ only if the parameters satisfy both of the following constraints:
$a=-\frac{2Q}{BM}$, $\jmath=\frac{HB^2}{4Q}$.
However, these conditions are not compatible with those needed in order to remove the other pathologies.}
\label{table:Curvature:singularities}\end{center}
\end{table}}

As summarized in Table~\ref{table:Curvature:singularities}, the ``center'' of the black hole $R=0$, i.e.~$r=\cos\theta=0$, does not always represent a curvature singularity, in contrast to the common notion that these objects always present such a singularity at their center.
In general, we have that all the rotating black holes embedded in a Melvin or swirling universe do not possess a curvature singularity at their center, which was already known for the black holes embedded in a Melvin universe~\cite{Ghezelbash:2021lcf}, but still not verified for the swirling or Melvin-swirling universe.

To be more precise, the Kerr-Newman black hole~\eqref{DKNMS} might in principle be singular for $R=0$, when the parameters satisfy both of the following constraints
\begin{equation}
\label{curvature-sing-DKNMS}
a = -\frac{2 Q}{B M} \,, \qquad
\jmath = \frac{H B^2 }{4 Q} \,.
\end{equation}
However, we can see that the conditions necessary to remove the other pathologies~\eqref{reg-DKNMS}, or just simply those that remove the conical singularities~\eqref{free-con-sing}, are not compatible with the constraints needed for $R=0$ to be a curvature singularity~\eqref{curvature-sing-DKNMS}.
This results in a completely non-singular spacetime when~\eqref{reg-DKNMS} are satisfied.

In the following, we list the behavior of the Kretschmann scalar in the limit $R\to0$, in order to discuss the order of divergence, or non-divergence, of the curvature singularity at that point.
We begin by recalling the results for black holes without swirling and Melvin parameters.
\paragraph{Schwarzschild ($a=Q=H=0$) and Kerr ($Q=H=0$):}
\begin{equation}
K\big\rvert_{\jmath = B = 0} \overset{R \to 0}{\approx} \frac{48 M^2}{r^6} \,.
\end{equation}
\paragraph{Reissner-Nordstr\"{o}m ($a=0$) and Kerr-Newman:}
\begin{equation}
K\big\rvert_{\jmath = B = 0} \overset{R \to 0}{\approx} \frac{56 Z^4}{r^8} \,.
\end{equation}
On the other hand, for black holes in a Melvin-swirling universe we obtain
\paragraph{Schwarzschild in a Melvin-swirling universe ($a=Q=H=0$):}
\begin{equation}
K \overset{R \to 0}{\approx} \frac{48 M^2}{r^6} \,.
\end{equation}
Thus, this result is not modified by the embedding in the Melvin-swirling universe.
\paragraph{Reissner-Nordstr\"{o}m in a Melvin-swirling universe ($a=0$):}
\begin{equation}
K \overset{R \to 0}{\approx} \frac{56 Z^4}{r^8} \,.
\end{equation}
Hence, if the black hole is not rotating, the presence of the swirling or Melvin parameters does not modify the behavior of the Kretschmann scalar for $R\to0$.
\paragraph{Kerr in a Melvin-swirling universe ($Q=H=0$):}
\begin{equation}
K \overset{R \to 0}{\propto} \frac{1}{a^{8}\Bigl[ \bigl(B^4+16\jmath^2)\, a^2 M^2 \Bigr]^6} \,.
\end{equation}
\paragraph{Kerr-Newman in a Melvin-swirling universe:}
\begin{equation}
K \overset{R \to 0}{\propto} \frac{1}{a^{8} \Bigl[4 B^2 Z^2 + 4 a M B\bigl(4 \jmath H + B^2 Q\bigr) + \bigl(B^4+16\jmath^2) \,a^2 M^2 \Bigr]^6} \,.
\end{equation}
As we can see, if the black hole is rotating, the spin-spin interaction between the black hole $(a M)^2$ and the background $\bigl(B^4+16\jmath^2)$ removes the curvature singularity for $R=0$, with additional corrections if the black hole also possesses an electromagnetic charge.
\myparagraph{Singular dyonic Kerr-Newman in a Melvin-swirling universe:}
However, as we have already pointed out, for the most general solution we found~\eqref{DKNMS}, it is still possible to introduce a curvature singularity for $R=0$ by constraining the parameters as in Eq.~\eqref{curvature-sing-DKNMS}, which indeed yields
\begin{equation}
K\big\rvert_{\eqref{curvature-sing-DKNMS}} \overset{R \to 0}{\propto} \frac{1}{r^8} \,.
\end{equation}
Moreover, the same result is also obtained for the electric Kerr-Newman black hole in a Melvin universe, using the same conditions~\eqref{curvature-sing-DKNMS} with $H=0$, which in fact results in $\jmath=0$.

\subsubsection{Curvature singularities along the symmetry axis}
\label{sec:curv-along-axis}
There are some peculiar sub-cases where the origin $R=0$ is not the unique place for curvature singularities.
Indeed, there are three cases for which $F=0$ represents a curvature singularity on an entire half of the symmetry axis for some special values of the parameters, as sketched in Fig.~\ref{fig:line-singularities}.
However, these constraints on the parameters are in general not compatible with those necessary to remove the other possible pathologies.
{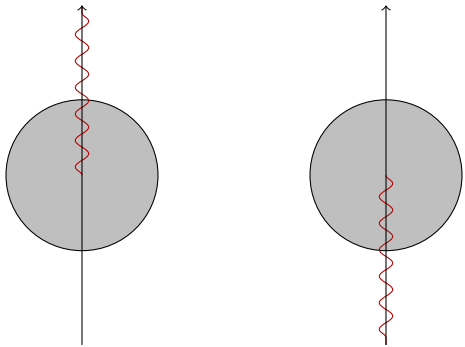
\begin{figure}[H]
\begin{center}
\begin{tikzpicture}
\filldraw[color = black , fill = black!25] (-2,0) circle (1cm);
\path [draw = dark-red, snake it] (-2,0) -- (-2,2.25);
\path [draw = black, ->] (-2,-2.25) -- (-2, 2.25);
\filldraw[color = black , fill = black!25] (2,0) circle (1cm);
\path [draw = dark-red, snake it] (2,0) -- (2,-2.25);
\path [draw = black, ->] (2,-2.25) -- (2, 2.25);
\end{tikzpicture}
\caption{The left figure depicts a curvature singularity along the symmetry axis in the entire North hemisphere $\theta=0$.
Similarly, the right figure depicts a curvature singularity in the entire South hemisphere $\theta=\pi$ of the symmetry axis.}
\label{fig:line-singularities}
\end{center} 
\end{figure} 
Given the peculiarity of these pathologies, we report here the possible sub-cases, with their respective constraints, for which these singularities are present\footnote{The upper sign represents the curvature singularity for $\theta=0$ while the lower sign is for $\theta=\pi$.}.
\paragraph{Dyonic Kerr-Newman in a Melvin-swirling universe:}
\begin{subequations}
\begin{align}
B & = \pm \frac{2 H}{Z^2} \,, \\
a & = \mp \frac{Q Z^2}{4 H M} \,, \\
\jmath &  = - \frac{HQ}{Z^4} \,.
\end{align}
\end{subequations}
These values of the parameters are not compatible with the constraints necessary to remove the conical singularities~\eqref{free-con-sing}.
Moreover, these conditions are in general not compatible with the removal of any other pathology.
\paragraph{Kerr in a swirling universe ($B=Q=H=0$):}
\begin{equation}
\jmath = \pm \frac{1}{4aM} \,.
\end{equation}
These values of $\jmath$ are the same which remove one conical singularity, the one in the same hemisphere of this curvature singularity, as can be verified from Eqs.~\eqref{kerr-melvin-swirling-conical} with $B=0$.
\paragraph{Magnetic Reissner-Nordstr\"{o}m in a Melvin universe ($\jmath=a=Q=0$):}
\begin{equation}
B = \pm \frac{2}{H} \,.
\end{equation}
These values of $B$ are the same which remove the conical singularity and the Dirac string located in the same hemisphere of the curvature singularity added along the symmetry axis.

\subsection{Frame-dragging}
\label{sec:frame-drag}
The frame-dragging of the whole spacetime~\eqref{DKNMS} is given by
\begin{equation}
\omega = - \frac{g_{t\phi}}{g_{\phi \phi}} = \frac{ \Omega}{\Sigma} \,.
\end{equation}
In particular, for large radial distances $r\to\infty$, we find
\begin{align}
\begin{split}
\omega & \overset{r\to \infty}{\approx} \Bigl[ - 4 \jmath \cos\theta + 2 \jmath B H \bigl( 1 + \cos^2\theta \bigr) + \frac{Q B^3}{2}  \bigl( 1 + \cos^2\theta \bigr) \\
& \quad + \frac{a M}{8} \bigl( B^4 + 16 \jmath^2 \bigr) \bigl( 3 + 6\cos^2\theta - \cos^4\theta \bigr)  \Bigr] \, r  \,,
\end{split}
\end{align}
which thus receives contributions from the swirling parameter $\jmath$, from a coupling between the Melvin-swirling background and the magnetic charge of the black hole ($\jmath BH$), from the coupling between the electric charge and the Melvin background $(Q B^3)$, and from the spin-spin interaction between the angular momentum of the black hole ($aM$) and the background ($B^4+16\jmath^2$).

Similarly, we can also obtain the approximation for large radial distances $r\to\infty$ of the angular velocity on the two halves of the symmetry axis
\begin{subequations}
\begin{align}
\omega \big\rvert_{\theta = 0} \overset{r\to \infty}{\approx} \Bigl[ 4 \jmath \bigl( B H - 1\bigr) + Q B^3 + a M \bigl( B^4 + 16 \jmath^2 \bigr) \Bigr] \, r \,, \\ 
\omega \big\rvert_{\theta = \pi} \overset{r\to \infty}{\approx} \Bigl[ 4 \jmath \bigl( B H + 1\bigr) + Q B^3 + a M \bigl( B^4 + 16 \jmath^2 \bigr)\Bigr] \, r \,,
\end{align}
\end{subequations}
and on the equatorial plane
\begin{equation}
\omega  \big\rvert_{\theta = \frac{\pi}{2}} \overset{r\to \infty}{\approx} \Bigl[ 2 \jmath H B  + \frac{Q B^3}{2} + \frac{3 a M}{8} \bigl( B^4 + 16 \jmath^2 \bigr) \Bigr] \, r \,,
\end{equation}
which, as expected, are not constant.

Since the value of the gravitational frame-dragging grows unbounded, and thus can easily exceed 1, it would be natural to conclude that there exist superluminal observers that violate causality.
However, as we proved in Sec.~\ref{sec:CTCs}, there is no possible occurrence of CTCs outside the event horizon.
Therefore, the apparent paradox of the superluminal observers can be attributed to a poor choice of coordinates, in the sense that a set of coordinates adapted to timelike observers does not experience an unbounded growth of the frame-dragging, as it happens, for example, for the Alcubierre spacetime~\cite{Alcubierre:1994tu}.

It is necessary to constrain some parameters and perform a rescaling of the azimuthal coordinates $\phi\mapsto\frac{2\pi}{\delta\phi}\phi$, in order to remove the various pathologies.
For these cases, the correct frame-dragging to consider is therefore
\begin{equation}
\tilde{\omega}  = \frac{\delta \phi}{2 \pi} \, \omega\big\rvert_{\text{“regular”}} \,.
\end{equation}

\subsubsection{Angular velocity of the event horizon}
\label{sec:angular-horizon}
It is interesting to compute the angular velocity of the horizons $r_{\pm}$~\eqref{horizon-DKNMS}, which is thus given by
\begin{equation}
\label{angular-dknms}
\omega_{\pm} \coloneqq \omega\big\rvert_{r=r_{\pm}} = \frac{2 \Gamma_{\pm} }{\lambda_{\pm}} \,,
\end{equation}
where
\begin{subequations}
\label{angular-def}
\begin{align}
\lambda_{\pm} & = r_{\pm}^2 + a^2 \,, \\
\Gamma_{\pm} &= \frac{a}{2} - BQr_{\pm} -  \frac{3 B^2}{4}  a Z^2 + \frac{B^3}{4} Q\gamma_{\pm} \,+ a M \biggr( \jmath^2 + \frac{B^4}{16} \biggr)\Bigl(\gamma_{\pm} + 2 r_{\pm} \lambda_{\pm}\Bigr) + \jmath B H \gamma_{\pm} \,, \\
\gamma_{\pm} & = r_{\pm}^3 - \frac{a^2}{\lambda_{\pm}}\Bigl(2 a^2 M + Z^2 r_{\pm} \Bigr) \,,
\end{align}
\end{subequations}
which, as expected, is constant.

\subsection{Entropy and temperature}
\label{sec:entropy-temp}
It is known~\cite{Astorino:2016hls,Astorino:2022prj} that both the Melvin parameter $B$ and the swirling parameter $\jmath$ do not modify the area and the surface gravity of the black holes.
Nevertheless, we can prove that this also holds for the new solution if both parameters are present.

Solution~\eqref{DKNMS} possesses the event horizons $r_{\pm}$~\eqref{horizon-DKNMS}, which are Killing horizons for the Killing vectors
\begin{equation}
\xi^{\mu}_{\pm} = \bigl(1, 0 , 0, \omega_{\pm} \bigr) \,,
\end{equation}
where $\omega_{\pm}$ is the angular velocity of the horizons~\eqref{angular-dknms}.
Therefore, the surface gravity $\kappa^2=-\frac{1}{2} (\nabla_\mu\xi^\nu)^2$ of these horizons is given by
\begin{equation}
\kappa_{\pm} = \frac{r_{\pm}-r_{\mp}}{ 2 \bigl( r_{\pm}^2 + a^2 \bigr)} \,,
\end{equation}
which is the same surface gravity of the Kerr-Newman black hole, and thus results in the same Hawking temperature
\begin{equation}
\label{temp-hawk}
T = \frac{\kappa_{+}}{2 \pi} \,.
\end{equation}
Moreover, the very same phenomenon occurs for the area and the entropy of the black hole
\begin{equation}
\label{entropy-DKNMS}
\pazocal{S} = \frac{\pazocal{A}_{+}}{4}
= \pi \bigl( r_{\pm}^2 + a^{2} \bigr) \,,
\end{equation}
which are indeed the same area and entropy of the Kerr-Newman black hole.

However, as discussed in Sec.~\ref{sec:frame-drag}, these quantities might depend on the Melvin and swirling parameters after the removal of the various pathologies.
Thus, the regular quantities are obtained upon substitution of constraints~\eqref{reg-DKNMS}.

\section{Ergoregions and geometry of the horizons}
\label{sec:hor-ergo}
We will now analyze the ergoregions, i.e.~the regions for which $g_{tt} > 0$, which are the regions of spacetime where the frame dragging makes it impossible for an observer to remain static; and the event horizons.
Moreover, we will also provide various plots, in order to perform a qualitative analysis of these quantities.
More specifically, we will start by analyzing the dyonic Kerr-Newman black hole in a Melvin-swirling universe~\eqref{DKNMS}, with the singularities removed through Eqs.~\eqref{reg-DKNMS}.

For the ergoregions, we will perform a numerical analysis of the function $g_{tt}$, utilizing the rectangular coordinates~\eqref{rectangular-coordinates}.
On the other hand, for the outer event horizon we will use a proper embedding in the Euclidean three-dimensional space $\mathbb{E}^3$, as in~\cite{Gnecchi:2013mja}.

In particular, the diagrams will consist of (at least) one three-dimensional plot for the event horizon and one for the ergoregions, plus additional plots representing the (same) two-dimensional cross-section of the ergoregions for $y=0$, in order to compare the position of the ergoregions with that of the horizons.
Since these plots will include both extremal and non-extremal black holes, we will take advantage of the extremality index~\eqref{extremality-index}.

Before showing the plots, it is interesting to notice that the $g_{tt}$ component of the solution~\eqref{DKNMS} can be written in the following form:
\begin{equation}
\label{gtt}
g_{tt} =  g_{(1)}\bigl(1 - \cos^2\theta\bigr) + g_{(2)}\Delta \,,
\end{equation}
where $g_{(1)}$ and $g_{(1)}$ are some differentiable functions, and where we recall that $\Delta=0$ defines the locations of the horizons~\eqref{horizon-DKNMS}.
From this result, we can conclude that the poles of the horizons ($\Delta=0$, $\cos^2\theta=1$) can never be part of the ergoregion, since in these points $g_{tt} = 0$.
Moreover, it can be proved that on the symmetry axis ($\cos^2\theta=1$) it holds $g_{(2)}<0$.
Therefore, we have that the symmetry axis is never part of the ergoregions outside the outer horizon and also inside the inner horizon, since in these regions $\Delta>0$.
Analogously, the portion of the symmetry axis between the outer and inner horizons is always part of the ergoregions, since between the horizons we have $\Delta<0$.

Furthermore, it holds that $g_{(1)}=0$ if $a=B=0$.
Thus, we can conclude that $g_{tt}=0$ for the entirety of the horizons ($\Delta=0$) for all the non-rotating black holes in a swirling universe sub-cases, independently from the specific value of the parameters.
Due to this last consideration, we will also present plots for the dyonic Reissner-Nordstr\"{o}m black hole in a swirling and Melvin-swirling universe in Appendix~\ref{app:ergo-RN}, using the same values of the parameters, except for the magnetic field $B$, which is fixed by the constraints needed to remove the conical singularities for the Melvin-swirling black hole~\eqref{conical-DRNMS}, while it is obviously zero for the swirling sub-case.
\begin{figure}[H]
\captionsetup[subfigure]{labelformat=empty}
\centering
\hspace{2cm} \subfloat[\mbox{\hspace{-0.25cm} Event Horizon}]{{\includegraphics[height=7.5cm]{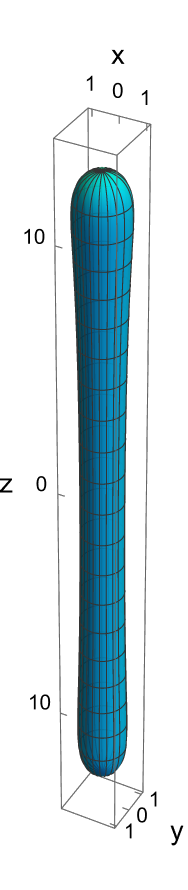}}}
\hspace{2.6cm} \subfloat[\hspace{0.3cm} Ergoregions]{{\hspace{0.5cm}\includegraphics[height=7cm]{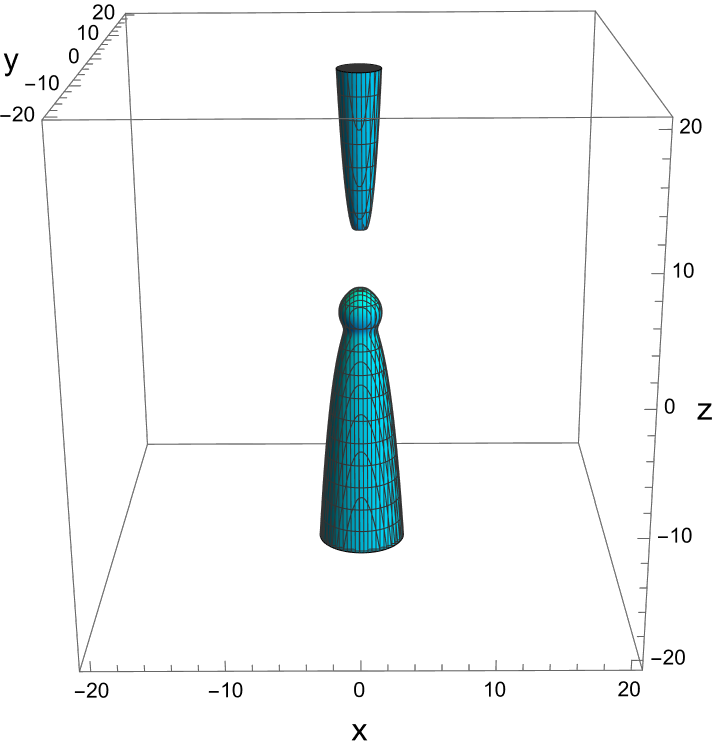}}}\\
\null \, \\
\hspace{-1.5cm} \subfloat[\hspace{1cm} Ergoregions Cross-section $y=0$]{{\hspace{0.5cm}\includegraphics[height=6.5cm]{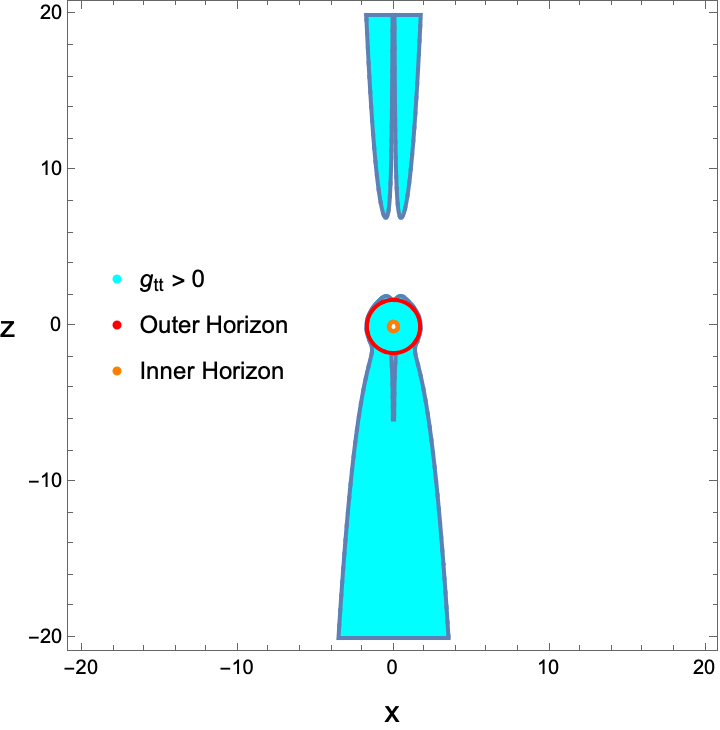}}}
\hspace{1cm} \subfloat[ \hspace{0.5cm}Ergoregions Cross-section $y=0$]{{\includegraphics[height=6.5cm]{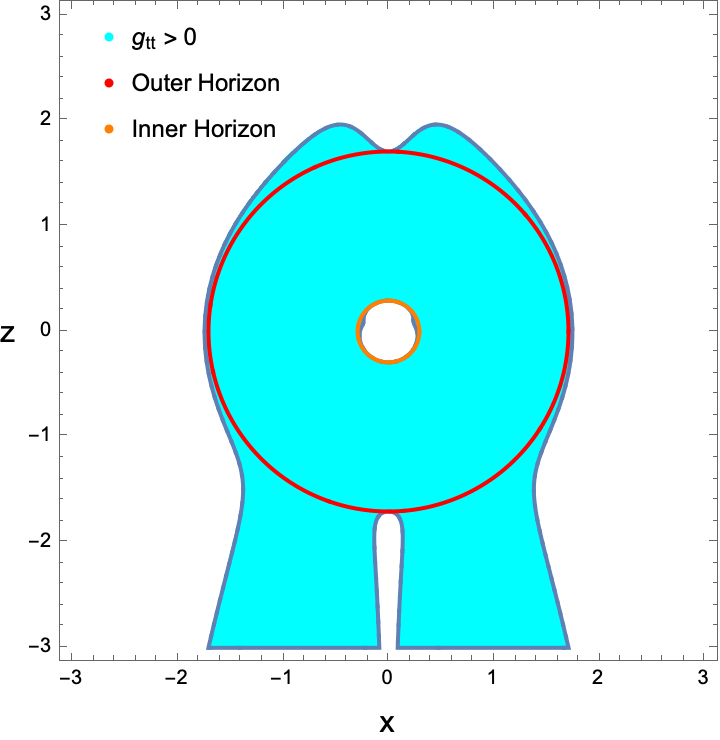}}}
\caption{\small Ergoregions and event horizon of the dyonic Kerr-Newman black hole in a Melvin-swirling universe, for which both conical singularities and Dirac strings have been removed, with parameters:
$M=1$, $Q=1/2$, $H=1/2$, $B=1$ and $a\simeq0.01$, $\jmath=2.75$, $\pazocal{I}\simeq0.5$.}
\label{Plot-DKNMS1}
\end{figure}
As we can see in Fig.~\ref{Plot-DKNMS1}, the presence of the Melvin-swirling background has the effect of making the horizon more prolate, and also of extending through infinity the ergoregions for all the vicinity of the symmetry axis, which is also a common feature of black holes embedded in either just a Melvin or a swirling universe.
\begin{figure}[H]
\captionsetup[subfigure]{labelformat=empty}
\centering
\hspace{0.5cm} \subfloat[Event Horizon]{{\includegraphics[height=6cm]{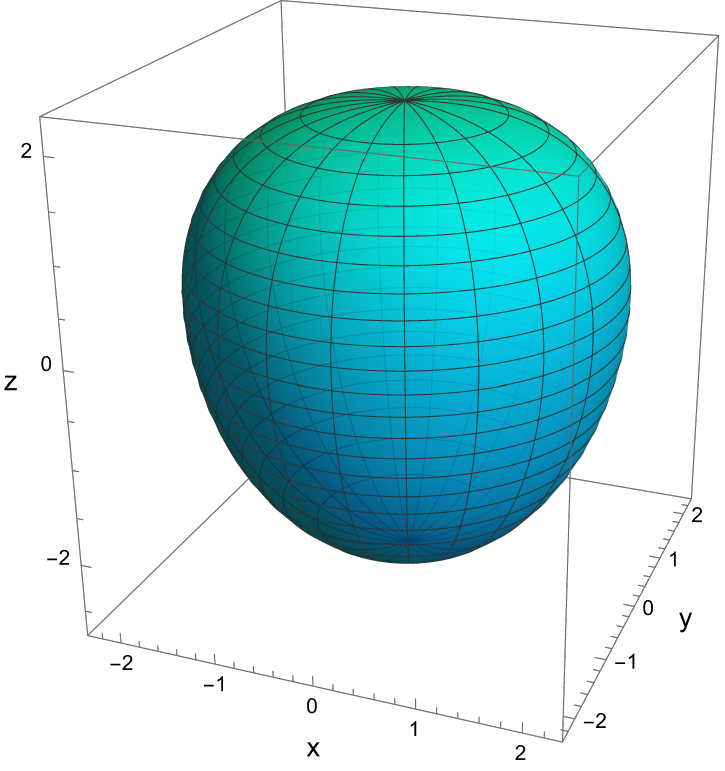}}}
\hspace{0.3cm} \subfloat[\hspace{0.2cm} Ergoregions]{{\hspace{0.5cm}\includegraphics[height=6cm]{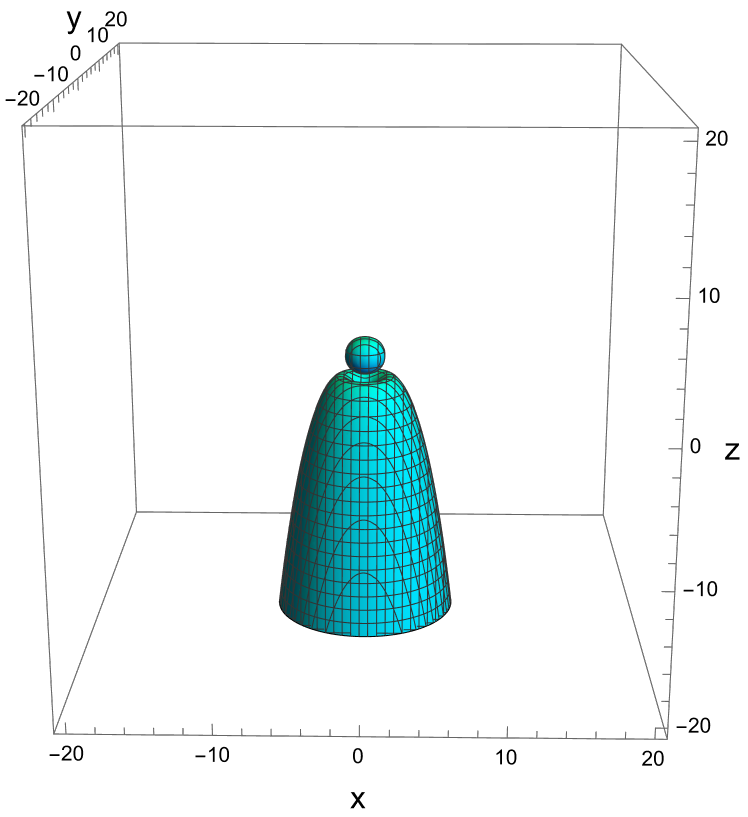}}} \\
\subfloat[\hspace{1cm} Ergoregions Cross-section $y=0$]{{\hspace{0.5cm}\includegraphics[height=6cm]{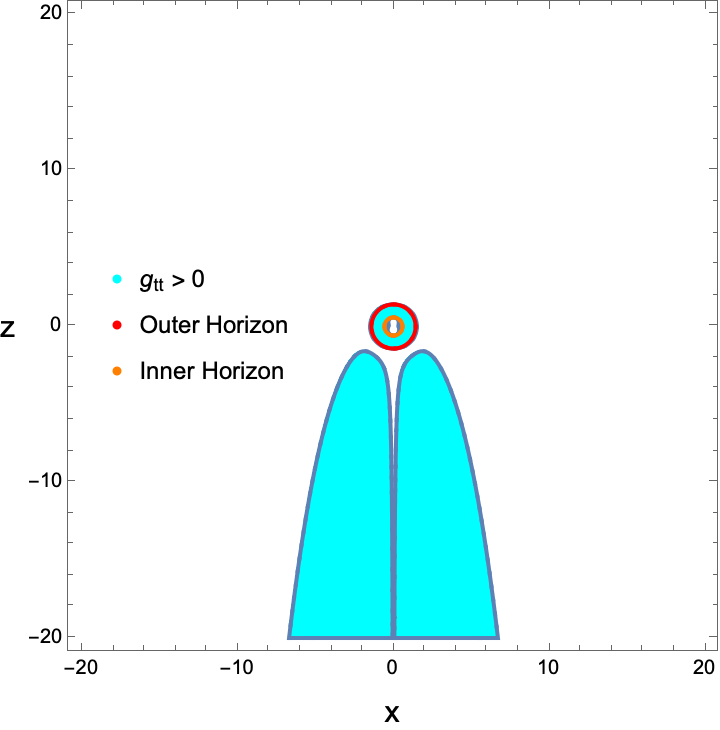}}}
\subfloat[\hspace{1cm} Ergoregions Cross-section $y=0$]{{\hspace{0.5cm}\includegraphics[height=6cm]{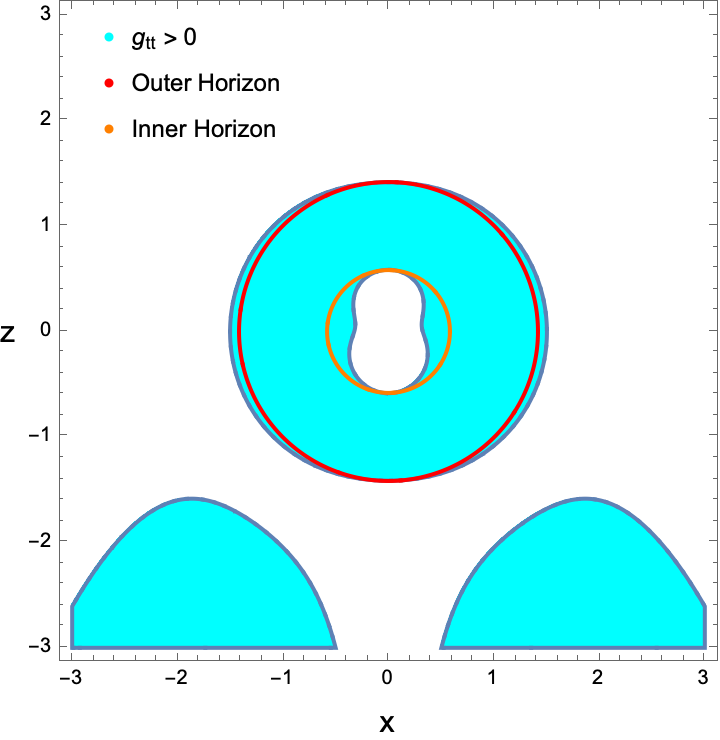}}} \\
\subfloat[\hspace{1cm} Ergoregions Cross-section $y=0$]{{\hspace{0.5cm}\includegraphics[height=6cm]{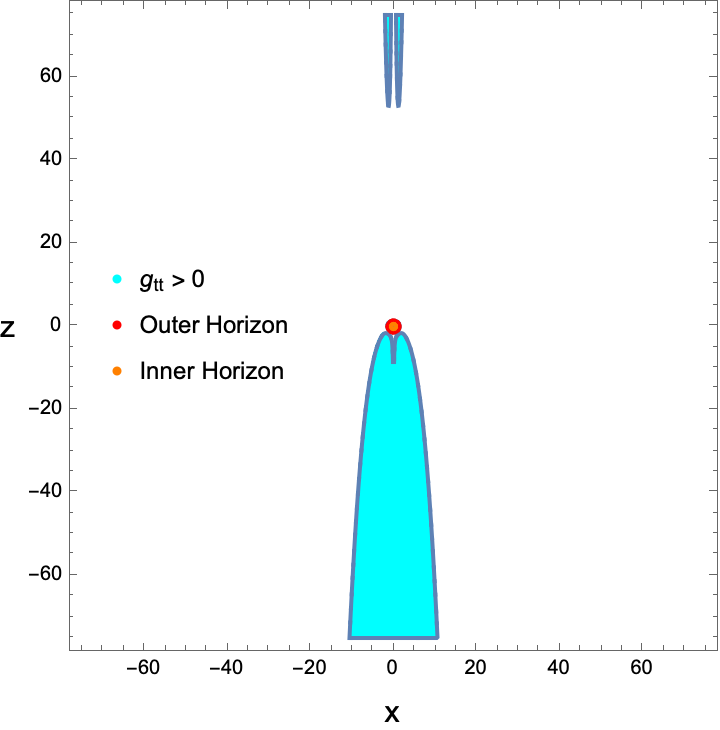}}}
\hspace{0.3cm} \subfloat[\hspace{0.2cm} Ergoregions]{{\hspace{0.5cm}\includegraphics[height=6cm]{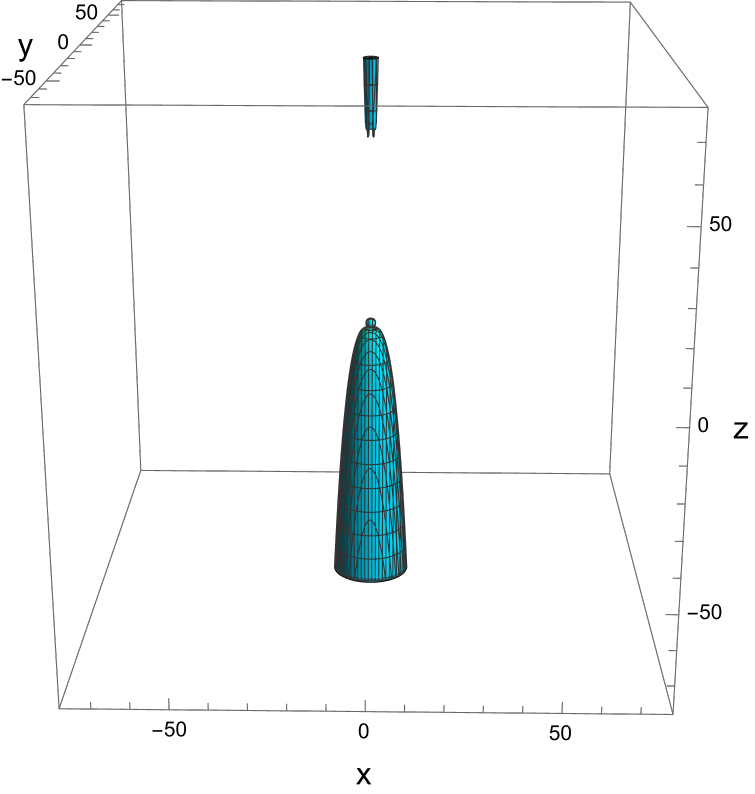}}}
\caption{\small Ergoregions and the event horizon of the dyonic Kerr-Newman black hole in a Melvin-swirling universe, for which both conical singularities and Dirac strings have been removed, with parameters:
$M=1$, $Q=9/10$, $H=1/8$, $B=1$ and $a\simeq0.047$, $\jmath\simeq0.272$, $\pazocal{I}\simeq0.828$.} \label{Plot-DKNMS2}
\end{figure}
\clearpage
\begin{figure}[H]
\captionsetup[subfigure]{labelformat=empty}
\centering
\hspace{0.75cm} \subfloat[Event Horizon]{{\includegraphics[height=7.5cm]{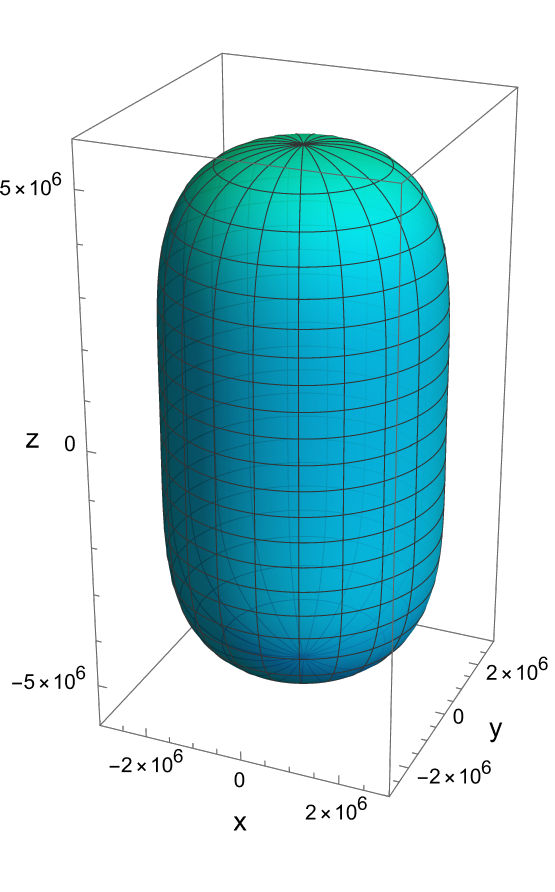}}}
\hspace{1.25cm} \subfloat[\hspace{0.5cm} Ergoregions]{{\hspace{0.5cm}\includegraphics[height=7cm]{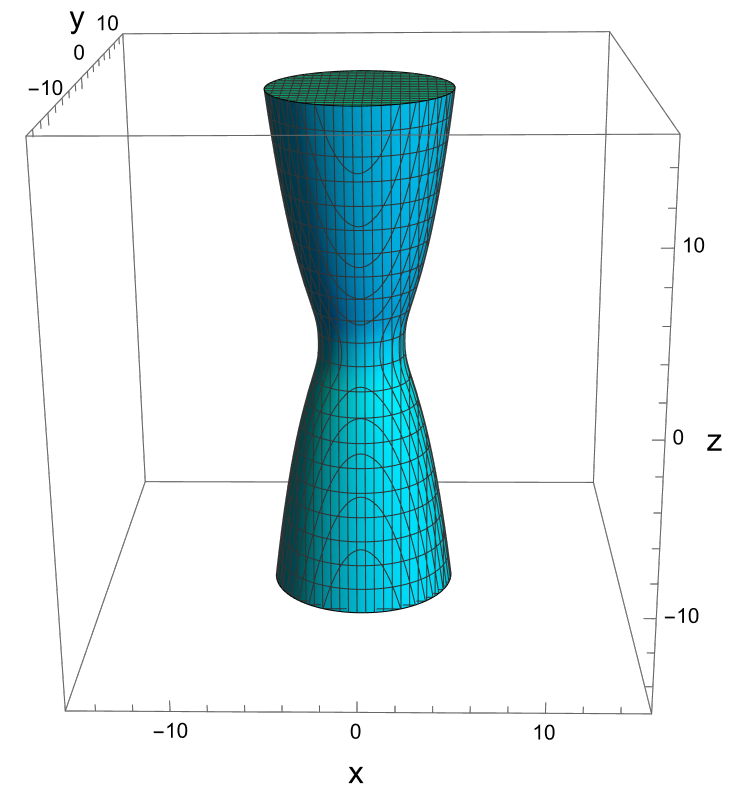}}} \\
\null \,
\hspace{-1.5cm} \subfloat[\hspace{1cm} Ergoregions Cross-section $y=0$]{{\hspace{0.5cm}\includegraphics[height=6.5cm]{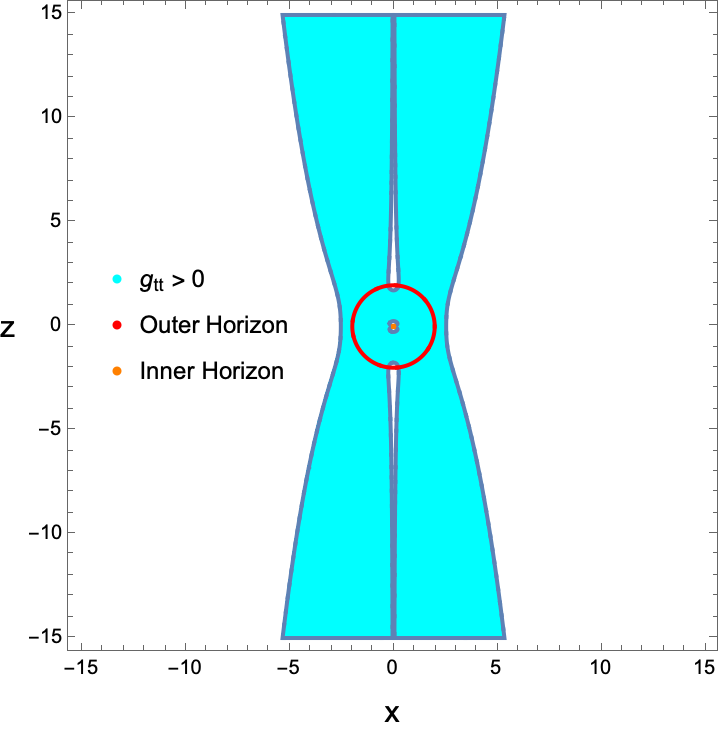}}}
\hspace{1cm} \subfloat[\hspace{0.5cm} Ergoregions Cross-section $y=0$]{{\includegraphics[height=6.5cm]{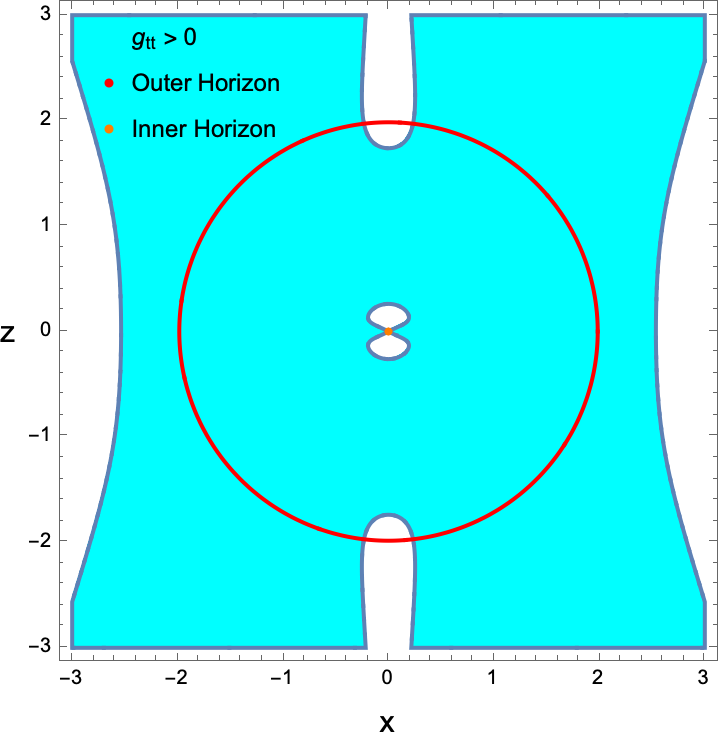}}}
\caption{\small Ergoregions and the event horizon of the dyonic Kerr-Newman black hole in a Melvin-swirling universe, for which both conical singularities and Dirac strings have been removed, with parameters:
$M=1$, $Q=1/8$, $H=1/8$, $B=10^3$ and $a\simeq0.65$, $\jmath\simeq7.5\times10^5$, $\pazocal{I}\simeq0.455$.}
\label{Plot-DKNMS3}
\end{figure}
In the case provided in Fig.~\ref{Plot-DKNMS2}, we have that the horizon is not symmetric with respect to the equatorial plane, and that the ergoregion in the upper hemisphere is pushed away from the black hole, which are both effects due to coupling between the charges of the black hole and the Melvin background.
On the other hand, in Fig.~\ref{Plot-DKNMS3}, we can see that if the rotation of the black hole is not negligible, the shape of the horizon becomes less oblate, despite a possible enormous value of the Melvin or swirling parameters.
\begin{figure}[H]
\captionsetup[subfigure]{labelformat=empty}
\centering
\hspace{0.125cm} \subfloat[Event Horizon]{{\includegraphics[height=5.8cm]{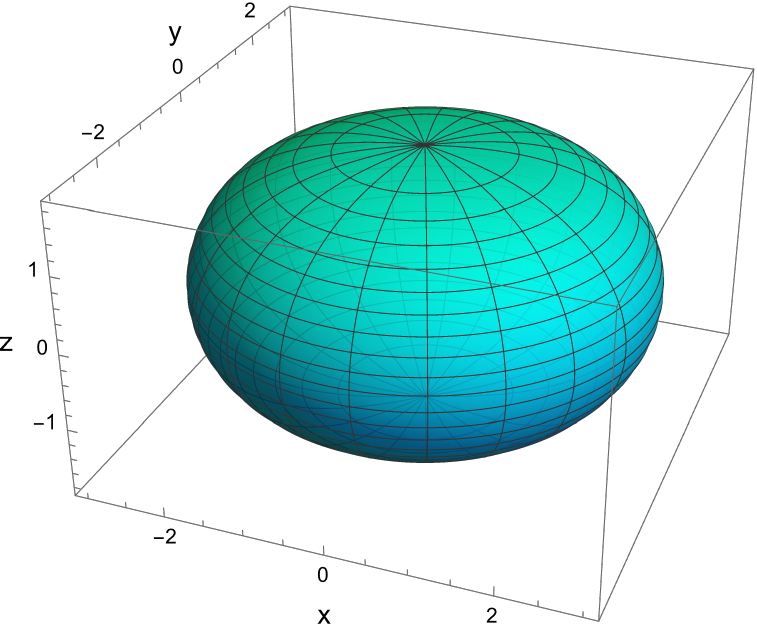}}}
\hspace{0.125cm} \subfloat[\hspace{0.5cm} Ergoregions]{{\hspace{0.5cm}\includegraphics[height=7cm]{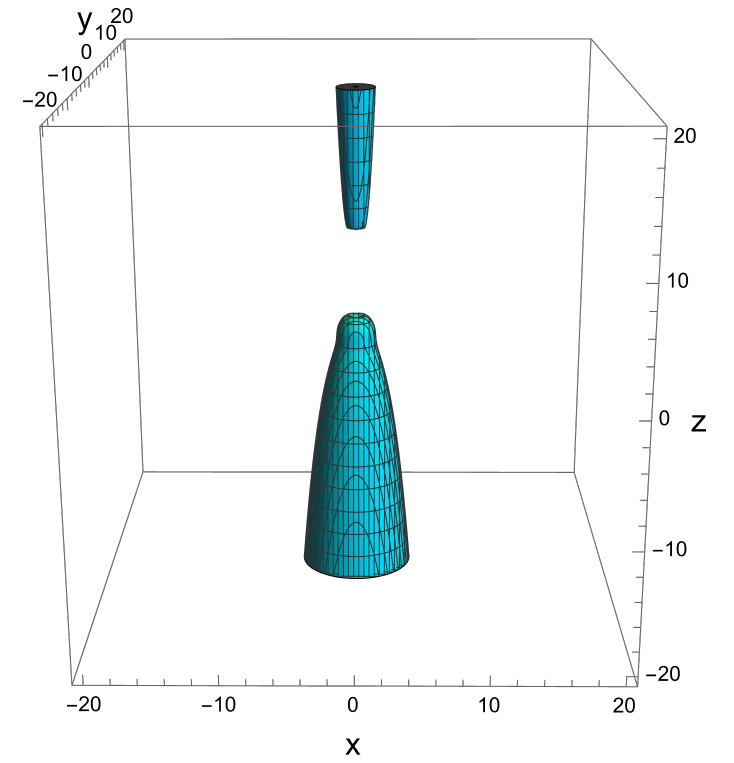}}}\\
\null \, \\
\hspace{-1cm} \subfloat[\hspace{1cm} Ergoregions Cross-section $y=0$]{{\hspace{0.5cm}\includegraphics[height=6.5cm]{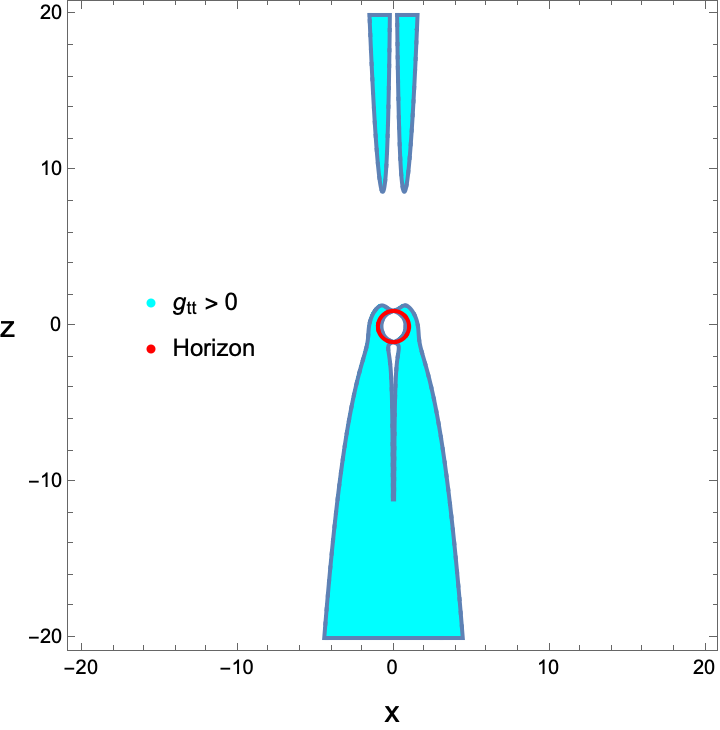}}}
\hspace{0.5cm} \subfloat[\hspace{1cm} Ergoregions Cross-section $y=0$]{{\hspace{0.5cm}\includegraphics[height=6.5cm]{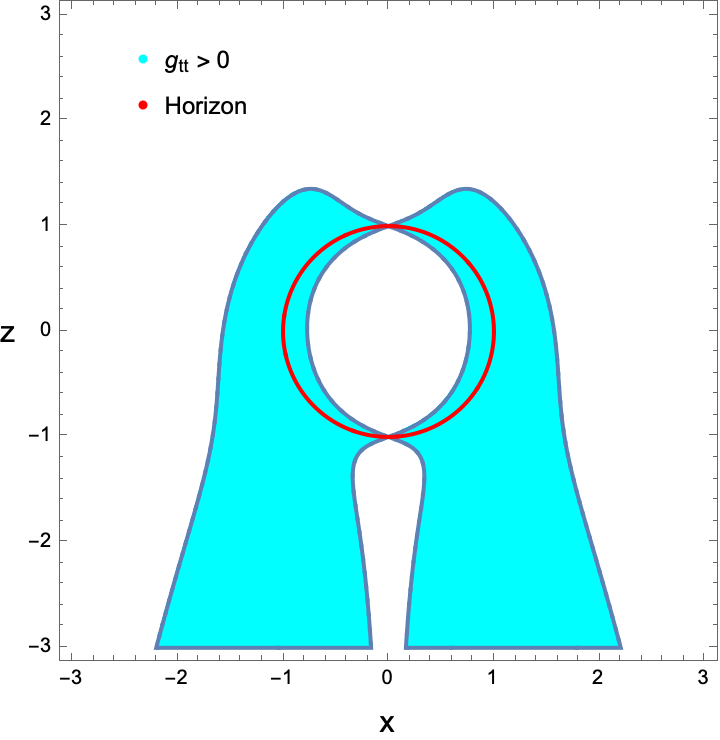}}}
\caption{Ergoregions and the event horizon of the extremal dyonic Kerr-Newman black hole in a Melvin-swirling universe, for which both conical singularities and Dirac strings have been removed, with parameters:
$M=1$, $Q=0.706$, $H=0.706$, $B=1$ and $a\simeq0.05$, $\jmath\simeq1.75$, $\pazocal{I}=1$.}
\label{Plot-DKNMS-EXT}
\end{figure}
As we can see in Figure~\ref{Plot-DKNMS-EXT}, the shape of the horizon in the extremal case tends to the shape of the black hole in Minkowski spacetime.
However, as we will prove later, the shape of the horizon is not exactly the same as that of the non-Melvin-swirling black hole, at least not for all the sub-cases.

In order to better understand the results of these plots, we can analyze the geometrical properties of the horizon itself.
In particular, we will determine whether in the extremal case the shape of the event horizon returns to that of the same black hole not embedded in the Melvin-swirling universe.
As we have already seen~\eqref{horizon-DKNMS}, the extremality condition for~\eqref{DKNMS} is given by
\begin{align}
M^2 & = Z^2 + a^2 \qquad
\Longrightarrow \qquad
r_{\pm} = M \,.
\end{align}
Therefore, we have that the area, the length of the equator, and the length of a meridian of the event horizon in the extremal case are, respectively
\begin{subequations}
\begin{align}
\pazocal{A} & =  \int_{0}^{2 \pi}\!d\phi \int_{0}^{\pi} \!d\theta\, \sqrt{g_{\theta \theta} g_{\phi \phi}} \,\bigg\rvert_{r=M} \,, \\
\pazocal{C}_\text{equator} & =  \int_{0}^{2 \pi}\!d\phi\, \sqrt{g_{\phi \phi}} \,\bigg\rvert_{r=M,\,\, \theta=\frac{\pi}{2}} \,, \qquad
\pazocal{C}_\text{meridian} =  \int_{0}^{\pi} \!d\theta\,\sqrt{g_{\theta \theta}} \,\bigg\rvert_{r=M} \,,
\end{align}
\end{subequations}
where it is understood that they are evaluated with regular parameters~\eqref{reg-DKNMS}.
We define the following ratios
\begin{equation}
\pazocal{R}_{1} \coloneqq \frac{\pazocal{A}}{2\,\pazocal{C}_{equator}} \,, \qquad
\pazocal{R}_{2}  \coloneqq \frac{2\, \pazocal{C}_{meridian}}{\pazocal{C}_{equator}} \,.
\end{equation}
In particular, $\pazocal{R}_{2}$ indicates the degree of ``sphericity'' of the event horizon.
Indeed, it is straightforward to verify that $\pazocal{R}_{2}=1$ for a perfect sphere, it increases as the horizon becomes more prolate, and it decreases as it becomes more oblate.
Similarly, $\pazocal{R}_{1}$ is exactly the physical radius when the horizon is a perfect sphere, and in general we have that the volume of the black hole is proportional to $\pazocal{R}_{1}$.
Moreover, $\pazocal{R}_{1}$ is also independent from the rescaling of the azimuthal coordinate needed to remove the conical singularities.

For the extremal dyonic Kerr-Newman black hole in Melvin-swirling background~\eqref{DKNMS}
\begin{subequations}
\begin{align}
\pazocal{A}^\text{(ext)} & = 4 \pi \bigl( M^2 + a^{2} \bigr) \,, \quad
\pazocal{C}_\text{equator}^\text{(ext)} = \frac{ 2 \pi \bigl( M^2 + a^2 \bigr) }{\sqrt{ |\pazocal{C}_\text{den} |}} \,, \\
\pazocal{R}_{1} & = \sqrt{| \pazocal{C}_\text{den} |} \,, \qquad\quad
\pazocal{R}_{2} \neq 1 \,,
\end{align}
\end{subequations}
where
\begin{align}
\begin{split}
\pazocal{C}_\text{den} & = \frac{M}{16} \bigl( M^2 + 3 a^2\bigr)\biggl[ 8B^2 M + 8 B^3 a Q + \Bigl(B^4 + 16\jmath^2\Bigr) \bigl( M^2 + 3 a^2\bigr) M + 32 \jmath B a H \biggl] \\
&\quad + M^2 + 2 B a M Q + B^2 a^2 Z^2 \,.
\end{split}
\end{align}
Therefore, the volume of the black hole is proportional to the swirling and Melvin parameters, and not only to the physical quantities characterizing the black hole.

From a numerical evaluation, we have that $\pazocal{R}_{2}$ results in a different value if we set the swirling and Melvin parameters to zero.
Hence, despite seeming visually similar, the shape of the event horizon in the extremal case is not exactly equal to that of the corresponding extremal dyonic Kerr-Newman black hole.

\section{Near-horizon extremal geometry and Kerr/CFT correspondence}
\label{sec:nh}
It is well known~\cite{Kunduri:2007vf,Kunduri:2013gce} that the geometry near the event horizon in the extremal case has some enhanced symmetry properties. 
For this reason, we will now study the geometry of the spacetime~\eqref{DKNMS} near the outer event horizon of the extremal black hole, defined as the limit $\varepsilon\rightarrow0$ of the metric after the
coordinate transformation~\cite{Bardeen:1999px}
\begin{equation}
r \mapsto M + \varepsilon \, r_{0} \, \hat{r} \,, \quad
t \mapsto  \frac{r_{0}}{\varepsilon} \hat{t} \,, \quad
\phi \mapsto \hat{\phi} + \frac{\Omega_{H}}{r_{0}^4} \hat{t} \,,
\end{equation}
together with the gauge transformation
\begin{equation}
A_{t} \mapsto A_{t} - \frac{A_{0 H}}{r_{0}^4} \,,\\
\end{equation}
where
\begin{subequations}
\begin{align}
M^2 & = Z^2 + a^2 \,, \\
r_{0}^2 & = M^2 + a^2 \,, \\
A_{0 H} & = - r_{0}^2 \biggl[ Q M + \frac{3 B}{2} a Z^2 - \frac{3 B^2}{4} M Q Z^2
- \frac{B^3}{4}\Bigl( 3 Z^4 + 7 a^2 Z^2 + 4 a^4 \Bigr)a - \jmath H M Z^2 \biggr] \,,
\\
\Omega_{H} & = r_{0}^2 \Biggl[ a - 2 B Q M + 2 \jmath B M H Z^2 - \frac{3 B^2}{2} a Z^2
+ \frac{B^3}{2} M Q Z^2 + \biggr(2 \jmath^2 + \frac{B^4}{8} \biggr) \bigl( 3 Z^4 + 7 a^2 Z^2 + 4 a^4 \bigr)a \Biggr] \,.
\end{align}
\end{subequations}
Now we can derive the near-horizon extremal geometry of the solution~\eqref{DKNMS}, which is
\begin{subequations}
\label{NHEDKNMS}
\begin{align}
d\hat{s}^2 & = \Gamma(\theta) \biggl[ -\hat{r}^2 \,d\hat{t}^2 + \frac{d\hat{r}^2}{\hat{r}^2} + \alpha^2(\theta) d\theta^2 + \gamma^2(\theta) \Bigl( d\hat{\phi} + \tilde{\kappa} \, \hat{r}\,d\hat{t} \Bigr)^2 \biggr] \,, \\
\hat{A} & = \frac{ \hat{A}_0 }{ r_{0}^2 } \,  \hat{r} \, d\hat{t} + \frac{2\pi}{\delta\phi} \frac{  \hat{A}_{3} }{ \Gamma(\theta) }\Bigl( d\hat{\phi} + \tilde{\kappa}  \, \hat{r} \,d\hat{t} \Bigr) \,,
\end{align}
\end{subequations}
with
\begin{subequations}
\begin{align}
\Gamma(\theta) & = \Gamma_{(0)} + 2B\,\hat{\phi}_{(0)} +  \frac{B^2}{2}\,\hat{\phi}_{(1)} + \frac{B^3}{2}\,\hat{\phi}_{(2)} + \biggr( \jmath^2 + \frac{B^4}{16} \biggr)\hat{\phi}_{(3)} + 2 \jmath B \, \hat{\phi}_{(4)} + \jmath\, \Gamma_{(1)} \,, \\
\tilde{\kappa} & = - \frac{\delta\phi}{2\pi} \frac{\kappa}{r_0^2} \,, \qquad
\gamma(\theta) = \frac{2\pi}{\delta\phi} \frac{r_{0}^{2}\sin\theta}{\Gamma(\theta)} \,, \qquad
\alpha(\theta) = 1 \,, \\
\kappa   &  = \kappa_{(0)}    + 2B\,\hat{\chi}_{(0)} +  \frac{B^2}{2}\,\hat{\chi}_{(1)} + \frac{B^3}{2}\,\hat{\chi}_{(2)}  + \biggr( \jmath^2 + \frac{B^4}{16} \biggr)\hat{\chi}_{(3)} + 2 \jmath B \,\hat{\chi}_{(4)} \,, \\
\hat{A}_0 &= \hat{\chi}_{(0)} + \frac{B}{2}\,\hat{\chi}_{(1)} + \frac{3B^2}{4}\,\hat{\chi}_{(2)} + \frac{B^3}{8}\,\hat{\chi}_{(3)} + \jmath\,\hat{\chi}_{(4)} \,, \\
\hat{A}_3 &= \hat{\phi}_{(0)} +  \frac{B}{2}\,\hat{\phi}_{(1)} + \frac{3B^2}{4}\,\hat{\phi}_{(2)} + \frac{B^3}{8}\,\hat{\phi}_{(3)} + \jmath\,\hat{\phi}_{(4)} \,,
\end{align}
\end{subequations}
where the auxiliary functions have been defined as
\begin{subequations}
\begin{align}
\hat{\chi}_{(0)} = \,& Q Z^2 \,, \\
\hat{\chi}_{(1)} = \,& 6 a M Z^2 \,, \\
\hat{\chi}_{(2)} = \, & Q \bigl(8 a^2 M^2 + Z^4\bigr) \,, \\
\hat{\chi}_{(3)} = \,& 2 a M \bigl( M^2 + 3 a^2 \bigr) \bigl( 3 M^2 + a^2 \bigr) \,, \\
\hat{\chi}_{(4)} = \,&  H \bigl(8 a^2 M^2 + Z^4\bigr) \,, \\
\hat{\phi}_{(0)} = \,& a Q M \sin^2\theta - H \bigl( M^2 + a^2 \bigr) \cos\theta \,,\\
\hat{\phi}_{(1)} = \,& 4 a^4 \sin^2\theta + a^2 Z^2 \bigl(7 -  \cos^2\theta \bigr) +Z^4 \bigl(1 +  2\cos^2\theta \bigr) \,,\\
\hat{\phi}_{(2)} = \,& a Q M \Bigl[ M^2 + 3 a^2 + \bigl( 3 M^2 + a^2 \bigr)\cos^2\theta \Bigr] - H Z^2 \bigl( M^2 + a^2 \bigr) \cos\theta \,, \\
\hat{\phi}_{(3)} = \, & Z^6 + a^2 \bigl( 3 M^2 + a^2 \bigr)^2 \bigl(1 + \cos^2\theta \bigr) \,,\\
\hat{\phi}_{(4)} = \,& a H M \Bigl[ M^2 + 3 a^2 + \bigl( 3 M^2 + a^2 \bigr)\cos^2\theta \Bigr] + Q Z^2 \bigl( M^2 + a^2 \bigr) \cos\theta\,, \\
\kappa_{(0)} = \, & - 2 a M \,, \\
\Gamma_{(0)} = \, & M^2 + a^2 \cos^2\theta \,, \\
\Gamma_{(1)} = \, & - 8 a M \bigl( M^2 + a^2 \bigr) \cos\theta \,.
\end{align}
\end{subequations}
We have reinserted the factor $2\pi/\delta\phi$, since the regularity of the solution (guaranteed by~\eqref{reg-DKNMS}) will be crucial in the following.
Remarkably, the near-horizon geometry~\eqref{NHEDKNMS} is in the same form as the near-horizon geometry of the standard Kerr-Newman black hole.
Therefore, it possesses a $SL(2,\RR) \times U(1)$ isometry, which is a warped and twisted product of $AdS_{2} \times S^{2}$.
The near-horizon extremal geometry for the sub-cases are in Appendix~\ref{app:nheg}.

We can make use of the metric~\eqref{NHEDKNMS} to study the Kerr/CFT correspondence~\cite{Guica:2008mu,Compere:2012jk}:
such a correspondence provides a connection between the near-horizon geometry and a conformal field theory defined on the boundary of the near-horizon metric.
The technique proved to be useful in the context of Melvin~\cite{Astorino:2015naa} (even in the presence of NUT charge~\cite{Siahaan:2021uqo}) and swirling~\cite{Astorino:2022prj} backgrounds;
now, we have the chance to check the validity of the correspondence in a Melvin-swirling spacetime and to provide a further example of the goodness of the conjecture.

According to the Kerr/CFT correspondence~\cite{Guica:2008mu}, the asymptotic region of the near-horizon geometry~\eqref{NHEDKNMS} gives information about the microscopic entropy of the extremal black hole.
The boundary of the near-horizon fields enjoy a conformal symmetry, which is encoded in the following central charge of the left Virasoro algebra
\begin{equation}
c_L = 3 \tilde{\kappa} \int_0^\pi d\theta \; \Gamma(\theta) \, \alpha(\theta) \, \gamma(\theta)
= -6 \kappa
\end{equation}
Thanks to the correspondence, we can use the Cardy formula
\begin{equation}
\pazocal{S}_\text{CFT} = \frac{\pi^2}{3} c_L T_L \,,
\end{equation}
where $T_L$ is the left temperature.
The Hawking temperature~\eqref{temp-hawk} is null in the extremal case, thus we make use of the Frolov-Thorne temperature~\cite{Frolov:1989jh} to take into account the rotational degrees of freedom.
The Frolov-Thorne temperature is defined as
\begin{equation}
\label{temp-frolov}
T_L = \lim_{r_+\to r_e} \frac{T_H}{\omega_+^\text{ext} - \omega_+} \,,
\end{equation}
where $\omega_+$ is the angular velocity of the horizon~\eqref{angular-dknms} evaluated with the constraints~\eqref{reg-DKNMS}, i.e.
\begin{equation}
\omega_+ = \frac{\delta\phi}{2\pi} \frac{2\Gamma_+}{\lambda_+} \,,
\end{equation}
and $\Gamma_+$ and $\lambda_+$ are defined in~\eqref{angular-def}.
$\omega_+^\text{ext}$ represents the extreme limit of the angular velocity $\omega_+$.

In order to obtain~\eqref{temp-frolov}, we first compute $\omega_+^\text{ext}$ by taking the limit $r_+\to r_e$, plug the result into~\eqref{temp-frolov} and then take again the limit to find the Frolov-Thorne temperature
\begin{equation}
T_L = -\frac{M^2+a^2}{2\pi\kappa} \,.
\end{equation}
One can check that the resulting microscopic entropy satisfies
\begin{equation}
\pazocal{S}_\text{CFT} = \pi \bigl(M^2 + a^2\bigr)
= \pazocal{S}_\text{BH} \,,
\end{equation}
and thus it equals the Bekenstein-Hawking entropy as computed in~\eqref{entropy-DKNMS}.
We stress that the computation works only if the constraints~\eqref{reg-DKNMS} hold\footnote{Metric~\eqref{NHEDKNMS} in the presence of conical singularities may be diffeomorphic to the near-horizon metric of the accelerating Kerr-Newman black hole, according to the conjecture stated in~\cite{Astorino:2025zse}.
If this is the case, such a diffeomorphism might be used to compute the mass and the angular momentum of our solution.
We leave this point, however, for future improvements.}:
this means that the the above quantities must be evaluated for the regular solution\footnote{The various quantities are not written explicitly in terms of the regularised parameters for sake of brevity.}.

The Kerr/CFT correspondence is confirmed to be a consistent technique to extract physical information from the near-horizon geometry and provides a link with its dual conformal field theory.

\section{Conclusions}

We presented a novel family of solutions representing rotating black holes embedded in a swirling and Melvin background.
Such a family was constructed by means of the Ernst technique, in particular by adding the swirling background to the Kerr-Newman-Melvin spacetime.
A large part of our study was devoted to the investigation of the singularity structure of the new solution:
we found that, upon an appropriate choice of the parameters, the metric is completely free of string singularities and even of curvature singularities, thus leading to a spacetime smooth everywhere.
We also took advantage of the analysis to classify all the various sub-cases according to their regularity.

Some physical (angular velocity, entropy, temperature) and geometric (ergoregions) properties of the spacetime have been investigated:
we produced many diagrams to clarify the deformations the horizons and the ergoregions undergo in the presence of both Melvin and swirling backgrounds.
Finally, we wrote down the near-horizon metric for the extremal black hole, and used it to check the Kerr/CFT correspondence.

There are some possible extensions of our work that are worth to be investigated:
it is possible to write down the most general solution in a Melvin-swirling background, by adding the NUT and acceleration parameters to the black hole, and by including the electric field in the Melvin background.
These can be accomplished by means of the Ehlers and Harrison transformations (to add NUT and electric field, respectively), while the acceleration parameter should be included in the seed solution.
This operation is, in principle, easy to be done, but we expect the resulting metric to be very involved.
Moreover, the swirling black hole with cosmological constant is still to be found:
in this case, there are no suitable Ernst transformations, thus such a metric should be found by means of an educated ansatz.

Finally, one may wonder if a supersymmetric extension is possible.
This topic was investigated in~\cite{DiPinto:2024axv}, where it was found that whenever either the swirling parameter or the Melvin parameter $B$ is present, a supersymmetric extension of any sub-case of the dyonic
Kerr-Newman black hole in a Melvin-swirling universe is never possible, regardless of the
value of the black hole parameters under consideration.
It is not clear, however, if the addition of other parameters, such as the cosmological constant, the NUT parameter and the acceleration, may modify this result.

\section*{Aknowledgements}

AV thanks Marco Astorino for useful discussions.
This work was partly supported by INFN.

\clearpage

\appendix
\section{Dyonic Kerr-Newman in a Melvin universe}
\label{app:seed-dknm}
The explicit expression of the dyonic Kerr-Newman Melvin spacetime~\eqref{seed} is given by:
\begin{subequations}
\label{DKNM}
\begin{align}
\label{DKNM-metric}
ds^2 & = F \biggl[ -\frac{\Delta}{\Sigma}\,dt^2 + \frac{dr^2}{\Delta} + d\theta^2\biggr] + \frac{\Sigma\sin^2\theta}{F}\biggl[ d\varphi - \frac{\Omega}{\Sigma} \,dt \biggr]^2 \,, \\
\label{DKNM-potential}
A & = \frac{A_0}{\Sigma} \,dt + \frac{A_{3}}{F}\biggl[ d\varphi - \frac{\Omega}{\Sigma}\,dt \biggr] \,,
\end{align}
\end{subequations}
where
\begin{align}
Z^2 & = Q^2 + H^2 \,, \qquad\qquad\qquad\qquad\quad
\Xi = \bigl(r^2 + a^2\bigr )\sin^2\theta + Z^2\cos^2\theta \,, \\
\Delta & = r^2 - 2Mr + Z^2 + a^2 \,, \qquad\quad\quad
\lambda = r^2 + a^2 - \Delta = 2Mr - Z^2 \,, \\
\Sigma & = \bigl(r^2 + a^2\bigr)^2 - \Delta a^2 \sin^2\theta \,, \qquad\;\;
R^2 = r^2 + a^2\cos^2\theta \,.
\end{align}
and
\begin{subequations}
\begin{align}
F & = R^2 + 2B\,\phi_{(0)} +  \frac{B^2}{2}\,\phi_{(1)} + \frac{B^3}{2}\,\phi_{(2)} + \frac{B^4}{16}\phi_{(3)} \,, \label{def-F-dknm} \\
\Omega &= a\,\lambda + 2B\,\chi_{(0)} +  \frac{B^2}{2}\,\chi_{(1)} + \frac{B^3}{2}\,\chi_{(2)} + \frac{B^4}{16}\chi_{(3)} \,, \\
A_0 &= \chi_{(0)} + \frac{B}{2}\,\chi_{(1)} + \frac{3B^2}{4}\,\chi_{(2)} + \frac{B^3}{8}\,\chi_{(3)} \,, \\
A_3 &= \phi_{(0)} +  \frac{B}{2}\,\phi_{(1)} + \frac{3B^2}{4}\,\phi_{(2)} + \frac{B^3}{8}\,\phi_{(3)} \,.
\end{align}
\end{subequations}
The above functions are defined as the expansion of the following quantities:
\begin{subequations}
\begin{align}
\chi_{(0)} & = a H\Delta \cos \theta - Q r \bigl(r^2+a^2\bigr) \,, \\
\chi_{(1)} & = -3 a Z^2 \Bigl( \lambda + \Delta \bigl(1+\cos^2\theta \bigr) \Bigr) \,, \\
\begin{split}
\chi_{(2)} & = Q \biggl[ r^3 \Bigl( \lambda + \Delta \bigl(1+\cos^2\theta \bigr) \Bigr) + a^2 \Bigl( \Delta \cos^2\theta \bigl( 3r - 4 M \bigr) -  r \bigl( Z^2 + \Delta \bigr)   \Bigr) - 2 M a^4 \biggr] \\
&\quad + a H \Delta \cos\theta  \bigl( \, \Xi + 2 R^2 \bigr)  \,,
\end{split}
\\
\begin{split}
\chi_{(3)} & = a \biggl[6 M r^5 - a^2 \Delta \cos^2\theta \Bigl ( \bigl( Z^2 + 4 M^2 - 6 M r \bigr)\cos^2\theta + Z^2 + 12 M^2 - 12 M r - 6 r^2 \Bigr) - 2 a^4 M \bigl( 2 M + r \bigr) \\
& \quad - a^2 Z^2\Delta + 4 a^2 M r \bigl( r^2 - 2 Z^2 + 3 M r \bigr) - \Delta \cos^2\theta \Bigl( 6 r^2 \bigl(\Delta - r^2 \bigr) + \bigl( Z^4 + 2 M r^3 - 3 Z^2 r^2 \bigr) \cos^2\theta \Bigr)\biggr] \,,
\end{split}
\\
\phi_{(0)} & = a Q r \sin^2\theta - H \bigl( r^2 + a^2 \bigr) \cos\theta \,, \\
\phi_{(1)} & = \Sigma \sin^2 \theta + 3 Z^2 \bigl( r ^2\cos^2\theta + a^2 \bigr) \,, \\
\begin{split}
\phi_{(2)} & = a Q \biggl[\bigl( 1 + \cos^2\theta \bigr) \Bigl( r^3 +  \bigl (2 M + r \bigr) a^2 \Bigr) + r \cos^2\theta \Bigl(2 Z^2 - \Delta \bigl(3 - \cos^2\theta \bigr) \Bigr)\biggr] \\
&\quad + H \cos\theta \Bigl[ 2 a^2 \lambda \sin^2\theta - \bigl(r^2 +a^2) \, \Xi \Bigr] \,,
\end{split}
\\
\begin{split}
\phi_{(3)} & = Z^2 \biggl[2 a^4 \bigl( 1 + \cos^2\theta \bigr)^2 +  r^2 \cos^2\theta \bigl( \, \Xi + R^2 \sin^2\theta \bigr) \\
&\quad + a^2 \cos^2\theta \Bigl( 2\, \Xi + 3 Z^2 + r^2 \bigl( 5 + 6 \sin^2\theta + 3 \cos^4\theta \bigr) - 8 \Delta \Bigr) \biggr] \\
&\quad + a^2 \Bigl( \lambda^2 \cos^2\theta \bigl( 3 - \cos^2\theta \bigr)^2  + r^3 \sin^6\theta ( 4 M - r )\Bigr) + 2 a^4 \Bigl( 2 M^2 \bigl( 1 + \cos^2\theta \bigr)^2 - \Delta \sin^6\theta \Bigr) \\
&\quad + a^6 \sin^6\theta + \bigl( r^2 + a^2 \bigr)^3 \sin^4\theta \,.
\end{split}
\end{align}
\end{subequations}
\subsection{Seed twisted potentials}
The twisted potential~\eqref{Atilde} and~\eqref{h} corresponding to the seed metric~\eqref{DKNM} are
\label{app:dknm-twisted potentials}
\begin{align}
\tilde{A}_{t} & = \frac{1}{F}\biggl[ \tilde{\chi}_{(0)} + B \tilde{\chi}_{(1)} + \frac{B^2}{4} \, \tilde{\chi}_{(2)} \biggr] \,, \\
h & = -\frac{1}{F}\biggl[ 2 \tilde{\chi}_{(1)} + B\, \tilde{\chi}_{(2)} \biggr] \,,
\end{align}
with $F$ as in~\eqref{def-F-dknm}, and where we have defined the auxiliary functions
\begin{subequations}
\begin{align}
\tilde{\chi}_{(0)} & = - Q \cos\theta \bigl( r^2 + a^2 \bigr) - a H r \sin^2\theta \,, \\
\tilde{\chi}_{(1)} & = - a \cos\theta \biggl[ M \Bigl(\bigl( 3 r^2 + a^2 \bigr) - \bigl( r^2 - a^2\bigr)\cos^2\theta \Bigr) - Z^2 r \sin^2\theta \biggr] \,, \\
\tilde{\chi}_{(2)} & = a H \biggl[ 2 M \Bigl( a^2 + \cos^2\theta \bigl( 2 r^2 + a^2 \bigr) \Bigr) + r \sin^2\theta \bigl( \lambda + \Delta \sin^2\theta \bigr)\biggr] + Q \cos\theta \Bigl[\, \Xi  \bigl( r^2 + a^2 \bigr) - 2 a^2 \lambda  \sin^2\theta \Bigr] \,.
\end{align}
\end{subequations}

\section{New black holes from sub-cases of the new solution}
\label{app:subcases}
In this Appendix, we report some of the new solutions that we obtained as sub-cases of the dyonic Kerr-Newman black hole in a Melvin-swirling universe~\eqref{DKNMS}.

\subsection{Kerr in a Melvin-swirling Universe}
\label{Kerr-Melvin-swirling}

The {\bfseries Kerr black hole in a Melvin-swirling universe} corresponds to the uncharged sub-case of the new solution~\eqref{DKNMS}, thus representing a black hole with rotation parameter $a$, mass $M$, embedded in a rotating universe with swirling parameter $\jmath$ and permeated by a uniform magnetic field $B$.
This sub-case is obtained by setting $H=Q=0$ in Eq.~\eqref{DKNMS}, which yields
\begin{subequations}
\label{KMS}
\begin{align}
ds^2 & = F \biggl[ -\frac{\Delta}{\Sigma}\,dt^2 + \frac{dr^2}{\Delta} + d\theta^2\biggr] + \frac{\Sigma\sin^2\theta}{F}\biggl[ d\phi - \frac{\Omega}{\Sigma} \,dt \biggr]^2 \,, \label{KMS-metric} \\
A & = \frac{A_0}{\Sigma} \,dt + \frac{A_{3}}{F}\biggl[ d\phi - \frac{\Omega}{\Sigma}\,dt \biggr] \,, \label{KMS-potential}
\end{align}
\end{subequations}
where the functions have been defined as
\begin{subequations}
\begin{align}
A_0 &=  \frac{B^3}{8}\,\chi_{(3)} \,, \\
A_3 &=\frac{B}{2}\,\varphi_{(1)}+ \frac{B^3}{8}\,\varphi_{(3)}\,, \\
\Omega &= a\,\lambda\, +\biggr[ \jmath^2 + \frac{B^4}{16} \biggr]\chi_{(3)}+ \jmath \, \Omega_{(1)} \,, \\
F & = R^2 +  \frac{B^2}{2}\,\varphi_{(1)}+ \biggr[ \jmath^2 + \frac{B^4}{16} \biggr]\varphi_{(3)} + \jmath\, F_{(1)} \,, \\
\Sigma & = \bigl(r^2 + a^2\bigr)^2 - \Delta a^2 \sin^2\theta \,, \\
\Delta & = r^2 - 2Mr + a^2 \,,
\end{align}
\end{subequations}
\clearpage
\noindent and
\vspace{-0.3cm}
\begin{subequations}
\begin{align}
\lambda & = r^2 + a^2 - \Delta = 2Mr \,, \\
R^2 & = r^2 + a^2\cos^2\theta \,, \\
\begin{split}
\chi_{(3)} & = -a \biggl[  a^2 \Delta \cos^2\theta \Bigl ( \bigl( 4 M^2 - 6 M r \bigr)\cos^2\theta + 12 M^2 - 12 M r - 6 r^2 \Bigr) \\
&\quad + 2 a^4 M \bigl( 2 M + r \bigr) - 4 a^2 M r \bigl( r^2+ 3 M r \bigr) - 6 M r^5 \\
& \quad + \Delta \cos^2\theta \Bigl( 6 r^2 \bigl(\Delta - r^2 \bigr) + 2 M r^3\cos^2\theta \Bigr)\biggr] \,,
\end{split}
\\
\varphi_{(1)} & = \Sigma \sin^2 \theta \,, \\
\begin{split}
\varphi_{(3)} & = a^6 \sin^6\theta + a^2 \Bigl[ \lambda^2 \cos^2\theta \bigl( 3 - \cos^2\theta \bigr)^2  + r^3 \sin^6\theta ( 4 M - r )\Bigr] \\
&\quad +2 a^4 \Bigl[ 2 M^2 \bigl( 1 + \cos^2\theta \bigr)^2 - \Delta \sin^6\theta \Bigr]  + \bigl( r^2 + a^2 \bigr)^3 \sin^4\theta \,,
\end{split}
\\
\Omega_{(1)} & = -4 \Delta \cos \theta  \biggl[r^3+a^2 \Bigl(\bigl(r-M\bigr) \cos^2\theta - M \Bigr)\biggr] \,, \\
F_{(1)} & = -4 a \cos\theta \Bigl[M \bigl( 1 + \cos^2\theta \bigr) \bigl( r^2 + a^2 \bigr)+\lambda\, r \sin^2\theta \Bigr] \,.
\end{align}
\end{subequations}

\subsection{Dyonic Reissner-Nordstr\"om in a Melvin-swirling Universe}
\label{Dyonic-Reissner-Nordstrom-Melvin-swirling}

Similarly, the {\bfseries dyonic Reissner-Nordstr\"om black hole in a Melvin-swirling universe} is the non-rotating sub-case of the new solution~\eqref{DKNMS}, therefore corresponding to a black hole with mass $M$, electric charge $Q$ and magnetic charge $H$, embedded in a universe with swirling parameter $\jmath$ and permeated by a uniform magnetic field $B$.
Thus, setting $a=0$ in Eq.~\eqref{DKNMS} results in
\vspace{-0.2cm}
\begin{subequations}
\label{DRNMS}
\begin{align}
ds^2 & = F \biggl[ -\frac{\Delta}{r^2}\,dt^2 + \frac{r^2dr^2}{\Delta} + r^2 d\theta^2\biggr] + \frac{r^2\sin^2\theta}{F}\biggl[ d\phi - \frac{\Omega}{r^2} \,dt \biggr]^2 \,, \label{DRNMS-metric} \\
A & = \frac{A_0}{r^2} \,dt + \frac{A_{3}}{F}\biggl[ d\phi - \frac{\Omega}{r^2}\,dt \biggr] \,, \label{DRNMS-potential}
\end{align}
\end{subequations}
\vspace{-0.32cm}
where
\vspace{-0.2cm}
\begin{subequations}
\begin{align}
A_0 &= \chi_{(0)} + \frac{3B^2}{4}\,\chi_{(2)} + \jmath\,\chi_{(4)} \,, \\
A_3 &= \varphi_{(0)} +  \frac{B}{2}\,\varphi_{(1)} + \frac{3B^2}{4}\,\varphi_{(2)} + \frac{B^3}{8}\,\varphi_{(3)} + \jmath\,\varphi_{(4)} \,, \\
\Omega &= 2B\,\chi_{(0)} + \frac{B^3}{2}\,\chi_{(2)}  + 2 \jmath B \,\chi_{(4)} + \jmath \, \Omega_{(1)} \,, \\
F & = 1 + 2B\,\varphi_{(0)} +  \frac{B^2}{2}\,\varphi_{(1)} + \frac{B^3}{2}\,\varphi_{(2)} + \biggr[ \jmath^2 + \frac{B^4}{16} \biggr]\varphi_{(3)} + 2 \jmath B \, \varphi_{(4)} \,, \\
\Delta & = r^2 - 2Mr + Z^2 \,, \\
Z^2 & = Q^2 + H^2 \,,
\end{align}
\end{subequations}
and
\begin{subequations}
\begin{align}
\Xi & = r^2\sin^2\theta + Z^2\cos^2\theta \,, \\
\lambda & = r^2 - \Delta = 2Mr - Z^2 \,, \\
\chi_{(0)} & = - Q r\,, \\
\chi_{(2)} & = Q r \Bigl[ \lambda + \Delta \bigl(1+\cos^2\theta \bigr) \Bigr] \,,\\
\chi_{(4)} & = H r \bigl( \Delta \cos^2\theta + r^2 \bigr) \,, \\
\varphi_{(0)} & =- H \cos\theta \,, \\
\varphi_{(1)} & = r^2 \sin^2 \theta + 3 Z^2 \cos^2\theta \,, \\
\varphi_{(2)} & =  - H \Xi \, \cos\theta  \,,
\\
\varphi_{(3)} & = Z^2 \cos^2\theta \bigl( \, \Xi + r^2 \sin^2\theta \bigr) \,,
\\
\varphi_{(4)} & = Q\Xi\, \cos\theta \,, \\
\Omega_{(1)} & = -4 r \Delta \cos \theta \,.
\end{align}
\end{subequations}

\subsection{Electric Kerr-Newman in a Melvin-swirling Universe}
\label{Electric-Kerr-Newman-Melvin-swirling}

Moreover, the {\bfseries electric Kerr-Newman in a Melvin-swirling universe} is the sub-case of~\eqref{DKNMS} without a magnetic charge $H=0$, which is then given by 
\begin{subequations}
\label{EKNMS}
\begin{align}
ds^2 & = F \biggl[ -\frac{\Delta}{\Sigma}\,dt^2 + \frac{dr^2}{\Delta} + d\theta^2\biggr] + \frac{\Sigma\sin^2\theta}{F}\biggl[ d\phi - \frac{\Omega}{\Sigma} \,dt \biggr]^2 \,, \label{EKNMS-metric} \\
A & = \frac{A_0}{\Sigma} \,dt + \frac{A_{3}}{F}\biggl[ d\phi - \frac{\Omega}{\Sigma}\,dt \biggr] \,, \label{EKNMS-potential}
\end{align}
\end{subequations}
where
\begin{subequations}
\begin{align}
A_0 &= \chi_{(0)} + \frac{B}{2}\,\chi_{(1)} + \frac{3B^2}{4}\,\chi_{(2)} + \frac{B^3}{8}\,\chi_{(3)} + \jmath\,\chi_{(4)} \,, \\
A_3 &= \varphi_{(0)} +  \frac{B}{2}\,\varphi_{(1)} + \frac{3B^2}{4}\,\varphi_{(2)} + \frac{B^3}{8}\,\varphi_{(3)} + \jmath\,\varphi_{(4)} \,, \\
\Omega &= a\,\lambda + 2B\,\chi_{(0)} +  \frac{B^2}{2}\,\chi_{(1)} + \frac{B^3}{2}\,\chi_{(2)}  + \biggr[ \jmath^2 + \frac{B^4}{16} \biggr]\chi_{(3)} + 2 \jmath B \,\chi_{(4)} + \jmath \, \Omega_{(1)} \,, \\
F & = R^2 + 2B\,\varphi_{(0)} +  \frac{B^2}{2}\,\varphi_{(1)} + \frac{B^3}{2}\,\varphi_{(2)} + \biggr[ \jmath^2 + \frac{B^4}{16} \biggr]\varphi_{(3)} + 2 \jmath B \, \varphi_{(4)} + \jmath\, F_{(1)} \,, \\
\Sigma & = \bigl(r^2 + a^2\bigr)^2 - \Delta a^2 \sin^2\theta \,, \\
\Delta & = r^2 - 2Mr + Q^2 + a^2 \,,
\end{align}
\end{subequations}
and
\begin{subequations}
\begin{align}
\Xi & = \bigl(r^2 + a^2\bigr )\sin^2\theta + Q^2\cos^2\theta \,, \\
\lambda & = r^2 + a^2 - \Delta = 2Mr - Q^2 \,, \\
R^2 & = r^2 + a^2\cos^2\theta \,, \\
\chi_{(0)} & = - Q r \bigl(r^2+a^2\bigr) \,, \\
\chi_{(1)} & = -3 a Q^2 \Bigl[ \lambda + \Delta \bigl(1+\cos^2\theta \bigr)  \Bigr] \,, \\
\begin{split}
\chi_{(2)} & = Q \biggl[ a^2 \Bigl( \Delta \cos^2\theta \bigl( 3r - 4 M \bigr) -  r \bigl( Q^2 + \Delta \bigr) -2 M a^2  \Bigr)  \\
& \quad  + r^3 \Bigl(  \lambda + \Delta \bigl(1+\cos^2\theta \bigr)  \Bigr)\biggr] \,,
\end{split}
\\
\begin{split}
\chi_{(3)} & = -a \Biggl[  a^2 \Delta \cos^2\theta \Bigl ( \bigl( Q^2 + 4 M^2 - 6 M r \bigr)\cos^2\theta + Q^2 + 12 M^2 - 12 M r - 6 r^2 \Bigr) \\
&\quad + 2 a^4 M \bigl( 2 M + r \bigr) + a^2 Q^2\Delta - 4 a^2 M r \bigl( r^2 - 2 Q^2 + 3 M r \bigr) - 6 M r^5 \\
& \quad + \Delta \cos^2\theta \Bigl( 6 r^2 \bigl(\Delta - r^2 \bigr) + \bigl( Q^4 + 2 M r^3 - 3 Q^2 r^2 \bigr) \cos^2\theta \Bigr)\Biggr] \,,
\end{split}
\\
\chi_{(4)} & = - a Q \Delta \cos\theta \bigl( \, \Xi + 2 R^2 \bigr) \,,
\\
\varphi_{(0)} & = a Q r \sin^2\theta \,, \\
\varphi_{(1)} & = \Sigma \sin^2 \theta + 3 Q^2 \bigl( r ^2\cos^2\theta + a^2 \bigr) \,, \\
\varphi_{(2)} & = a Q \biggl[\bigl( 1 + \cos^2\theta \bigr) \Bigl( r^3 +  \bigl (2 M + r \bigr) a^2 \Bigr) + r \cos^2\theta \Bigl(2 Q^2 - \Delta \bigl(3 - \cos^2\theta \bigr) \Bigr)\biggr] \,,
\\
\begin{split}
\varphi_{(3)} & = Q^2 \bigg[2 a^4 \bigl( 1 + \cos^2\theta \bigr)^2 +  r^2 \cos^2\theta \bigl( \, \Xi + R^2 \sin^2\theta \bigr) \\
&\quad + a^2 \cos^2\theta \Bigl( 2\, \Xi + 3 Q^2 + r^2 \bigl( 5 + 6 \sin^2\theta + 3 \cos^4\theta \bigr) - 8 \Delta \Bigr) \biggr] \\
&\quad + a^6 \sin^6\theta + a^2 \Bigl[ \lambda^2 \cos^2\theta \bigl( 3 - \cos^2\theta \bigr)^2  + r^3 \sin^6\theta ( 4 M - r )\Bigr] \\
&\quad +2 a^4 \Bigl[ 2 M^2 \bigl( 1 + \cos^2\theta \bigr)^2 - \Delta \sin^6\theta \Bigr]  + \bigl( r^2 + a^2 \bigr)^3 \sin^4\theta \,,
\end{split}
\\
\varphi_{(4)} & =  Q \cos\theta \Bigl[ \bigl( r^2+ a^2 \bigr)\, \Xi - 2 \lambda a^2 \sin^2\theta \Bigr] \,,
 \\
\Omega_{(1)} & = -4 \Delta \cos \theta  \biggl[r^3+a^2 \Bigl(\bigl(r-M\bigr) \cos^2\theta - M \Bigr)\biggr] \,, \\
F_{(1)} & = -4 a \cos\theta \Bigl[M \bigl( 1 + \cos^2\theta \bigr) \bigl( r^2 + a^2 \bigr)+\lambda\, r \sin^2\theta \Bigr] \,.
\end{align}
\end{subequations}

\subsection{Magnetic Kerr-Newman in a Melvin-swirling Universe}
\label{Magnetic-Kerr-Newman-Melvin-swirling}

Finally, setting the electric charge to zero, $Q=0$, in the new solution~\eqref{DKNMS}, we obtain the {\bfseries magnetic Kerr-Newman black hole in a Melvin-swirling universe}:
\begin{subequations}
\label{MKNMS}
\begin{align}
ds^2 & = F \biggl[ -\frac{\Delta}{\Sigma}\,dt^2 + \frac{dr^2}{\Delta} + d\theta^2\biggr] + \frac{\Sigma\sin^2\theta}{F}\biggl[ d\phi - \frac{\Omega}{\Sigma} \,dt \biggr]^2 \,, \label{MKNMS-metric} \\
A & = \frac{A_0}{\Sigma} \,dt + \frac{A_{3}}{F}\biggl[ d\phi - \frac{\Omega}{\Sigma}\,dt \biggr] \,, \label{MKNMS-potential}
\end{align}
\end{subequations}
where
\begin{subequations}
\begin{align}
A_0 &= \chi_{(0)} + \frac{B}{2}\,\chi_{(1)} + \frac{3B^2}{4}\,\chi_{(2)} + \frac{B^3}{8}\,\chi_{(3)} + \jmath\,\chi_{(4)} \,, \\
A_3 &= \varphi_{(0)} +  \frac{B}{2}\,\varphi_{(1)} + \frac{3B^2}{4}\,\varphi_{(2)} + \frac{B^3}{8}\,\varphi_{(3)} + \jmath\,\varphi_{(4)} \,, \\
\Omega &= a\,\lambda + 2B\,\chi_{(0)} +  \frac{B^2}{2}\,\chi_{(1)} + \frac{B^3}{2}\,\chi_{(2)}  + \biggr[ \jmath^2 + \frac{B^4}{16} \biggr]\chi_{(3)} + 2 \jmath B \,\chi_{(4)} + \jmath \, \Omega_{(1)} \,, \\
F & = R^2 + 2B\,\varphi_{(0)} +  \frac{B^2}{2}\,\varphi_{(1)} + \frac{B^3}{2}\,\varphi_{(2)} + \biggr[ \jmath^2 + \frac{B^4}{16} \biggr]\varphi_{(3)} + 2 \jmath B \, \varphi_{(4)} + \jmath\, F_{(1)} \,, \\
\Sigma & = \bigl(r^2 + a^2\bigr)^2 - \Delta a^2 \sin^2\theta \,, \\
\Delta & = r^2 - 2Mr + H^2 + a^2 \,, \\
\end{align}
\end{subequations}
and
\begin{subequations}
\begin{align}
\Xi & = \bigl(r^2 + a^2\bigr )\sin^2\theta + H^2\cos^2\theta \,, \\
\lambda & = r^2 + a^2 - \Delta = 2Mr - H^2 \,, \\
R^2 & = r^2 + a^2\cos^2\theta \,, \\
\chi_{(0)} & = a H\Delta \cos \theta \,, \\
\chi_{(1)} & = -3 a H^2 \Bigl[ \lambda + \Delta \bigl(1+\cos^2\theta \bigr) \Bigr] \,, \\
\chi_{(2)} & = a H \Delta \cos\theta  \bigl( \, \Xi + 2 R^2 \bigr)  \,, \\
\begin{split}
\chi_{(3)} & = -a \Biggl[  a^2 \Delta \cos^2\theta \Bigl ( \bigl( H^2 + 4 M^2 - 6 M r \bigr)\cos^2\theta + H^2 + 12 M^2 - 12 M r - 6 r^2 \Bigr) \\
&\quad + 2 a^4 M \bigl( 2 M + r \bigr) + a^2 H^2\Delta - 4 a^2 M r \bigl( r^2 - 2 H^2 + 3 M r \bigr) - 6 M r^5 \\
& \quad + \Delta \cos^2\theta \Bigl( 6 r^2 \bigl(\Delta - r^2 \bigr) + \bigl( H^4 + 2 M r^3 - 3 H^2 r^2 \bigr) \cos^2\theta \Bigr)\Biggr] \,,
\end{split}
\\
\begin{split}
\chi_{(4)} & = -H \biggl[ a^2 \Bigl(r \bigl( H^2 - r^2 \bigr) + 4 M \Delta \cos^2\theta + r \Delta \bigl( 1 - 3 \cos^2\theta \bigr) \Bigr) \\
& \quad + 2 a^4 M - r^3 \bigl( \Delta \cos^2\theta + r^2 \bigr) \biggr]  \,,
\end{split}
\\
\varphi_{(0)} & =- H \bigl( r^2 + a^2 \bigr) \cos\theta \,, \\
\varphi_{(1)} & = \Sigma \sin^2 \theta + 3 H^2 \bigl( r ^2\cos^2\theta + a^2 \bigr) \,, \\
\varphi_{(2)} & =H \cos\theta \Bigl[ 2 a^2 \lambda \sin^2\theta - \bigl(r^2 +a^2) \, \Xi \Bigr]  \,,
\\
\begin{split}
\varphi_{(3)} & = H^2 \bigg[2 a^4 \bigl( 1 + \cos^2\theta \bigr)^2 +  r^2 \cos^2\theta \bigl( \, \Xi + R^2 \sin^2\theta \bigr) \\
&\quad + a^2 \cos^2\theta \Bigl( 2\, \Xi + 3 H^2 + r^2 \bigl( 5 + 6 \sin^2\theta + 3 \cos^4\theta \bigr) - 8 \Delta \Bigr) \biggr] \\
&\quad + a^6 \sin^6\theta + a^2 \Bigl[ \lambda^2 \cos^2\theta \bigl( 3 - \cos^2\theta \bigr)^2  + r^3 \sin^6\theta ( 4 M - r )\Bigr] \\
&\quad +2 a^4 \Bigl[ 2 M^2 \bigl( 1 + \cos^2\theta \bigr)^2 - \Delta \sin^6\theta \Bigr]  + \bigl( r^2 + a^2 \bigr)^3 \sin^4\theta \,,
\end{split}
\\
\varphi_{(4)} & = a H \biggl[2 M \Bigl(a^2 +  \cos^2\theta \bigl(2 r^2 + a^2 \bigr) \Bigr) + r \sin^2\theta \bigl( \lambda + \Delta \sin^2\theta \bigr)\biggr] \,, \\
\Omega_{(1)} & = -4 \Delta \cos \theta  \biggl[r^3+a^2 \Bigl(\bigl(r-M\bigr) \cos^2\theta - M \Bigr)\biggr] \,, \\
F_{(1)} & = -4 a \cos\theta \Bigl[M \bigl( 1 + \cos^2\theta \bigr) \bigl( r^2 + a^2 \bigr)+\lambda\, r \sin^2\theta \Bigr] \,.
\end{align}
\end{subequations}

\section{Conical singularities for the sub-cases}
\label{app:con-sing}
\subsection{Sub-cases with non-removable conical singularities}
We list here some of the sub-cases with non-removable conical singularities.
\paragraph{Kerr in a Melvin-swirling universe ($Q=H=0$):}
\label{kerr-melvin-swirling-conical}
\begin{subequations}
\begin{align}
\delta_0 & = \frac{32\pi}{16  - 8 a \jmath M + a^2\bigl(B^4 + 16 \jmath^2\bigr)M^2} \,, \\
\delta_\pi & = \frac{32\pi}{16 + 8 a \jmath M + a^2\bigl(B^4 + 16 \jmath^2\bigr)M^2} \,.
\end{align}
\end{subequations}
We can see that, for this sub-case, the non-removability of the conical singularities exactly arises from the spin-spin interaction between the black hole $(aM)$ and the background $(\jmath)$.
For its non-Melvin sub-case $B=0$, the respective condition was already discussed in~\cite{Astorino:2022aam}.
Nevertheless, we can say that the addition of the uniform magnetic field $B$ is not sufficient to remove the conical singularities of the rotating black hole in a Melvin-swirling universe.
\paragraph{Magnetic Reissner-Nordstr\"{o}m in a Melvin-swirling universe ($a=0$):}
\begin{subequations}
\begin{align}
\delta_0 & = \frac{32\pi}{16 \jmath^2 H^4  + \bigl(2 - BH\bigr)^4} \,, \\
\delta_\pi & = \frac{32\pi}{16 \jmath^2 H^4 + \bigl(2 + BH\bigr)^4} \,.
\end{align}
\end{subequations}
In a similar manner, the non-removable conical singularities are due to the coupling between the magnetic Melvin background $B$ and the black hole magnetic charge $H$.

\subsection{Sub-cases with removable conical singularities}
Analogously, we present here all the sub-cases with the respective conditions under which the conical singularities are removable, also including, for the sake of completeness, the already-known sub-cases for which the swirling parameter is absent, $\jmath = 0$.
\paragraph{Magnetic Kerr-Newman in a Melvin-swirling universe ($Q=0$):}
\begin{subequations}
\label{conical-MKNMS}
\begin{align}
\jmath & = - \frac{ B H \bigl( 4 + B^2 H^2 \bigr)}{16 a M} \,,  \\
\delta\phi & = \frac{512 \pi a^2 M^2}{\Bigl[ 16 a^2 M^2 + B^2 H^6\Bigr] \Bigl[16 a^2 M^2 B^4 + \bigl(4 + B^2 H^2 \bigr)^2\Bigr]} \,.
\end{align}
\end{subequations}
\paragraph{Electric Kerr-Newman in a Melvin-swirling universe ($H=0$):}
\begin{subequations}
\label{conical-EKNMS}
\begin{align}
a & = \frac{ B Q^3}{4 M} \,, \\
\delta\phi & = \frac{32 \pi}{\Bigl[ 1 + B^2 Q^2 \Bigr] \Bigl[16 + 8 B^2 Q^2 + \bigl( B^4 + 16 \jmath^2\bigr) Q^4 \Bigr]} \,.
\end{align}
\end{subequations}
\paragraph{Dyonic Reissner-Nordstr\"{o}m in a Melvin-swirling universe ($a=0$):}
\begin{subequations}
\label{conical-DRNMS}
\begin{align}
\jmath & = \frac{ H \bigl( 4 + B^2 Z^2 \bigr)}{4 Q Z^2} \,, \\
\delta\phi & = \frac{32\pi Q^2 }{Z^2 \Bigl[16 + 8 B^2 \bigl(H^2 +3 Q^2 \bigr) + B^4 Z^4 \Bigr]} \,.
\end{align}
\end{subequations}
\paragraph{Reissner-Nordstr\"{o}m in a swirling universe ($B=a=0$):}
\begin{equation}
\label{conical-DRNS}
\delta \phi = \frac{2 \pi}{1 + \jmath^2 Z^4} \,,
\end{equation}
where it is not specified if the black hole is dyonic, magnetic, or electric because this condition is the same for all three sub-cases.
\paragraph{Electric Kerr-Newman in a Melvin universe ($\,\jmath=H=0$):}
\begin{equation}
\label{conical-EKNM}
\delta\phi = \frac{32 \pi}{16 + 24 B^2 	Q^2 + 32 a B^3 M Q + B^4\bigl(Q^4 + 16 a^2 M^2\bigr)} \,.
\end{equation}
\paragraph{Kerr in a Melvin universe ($\jmath=Q=H=0$):}
\begin{equation}
\label{conical-KM}
\delta\phi = \frac{2 \pi}{1 + a^2 B^4 M^2} \,.
\end{equation}
\paragraph{Electric Reissner-Nordstr\"{o}m in a Melvin universe ($\jmath=a=H=0$):}
\begin{equation}
\label{conical-ERNM}
\delta\phi = \frac{32 \pi}{16 + 24 B^2 	Q^2 + B^4 Q^4} \,.
\end{equation}

\section{Dirac strings for the sub-cases}
\label{app:dirac}
\subsection{Sub-cases with non-removable Dirac strings} 
As summarized in Table~\ref{table:Dirac-strings}, not in every sub-case the constraints necessary to remove the conical singularities are compatible with those to remove the Dirac string.
For this reason, in this part, we list all the sub-cases with the reasons why this is not possible, corresponding to the difference in the value of the $\phi-$component of the electromagnetic potential on the two halves of the symmetry axis.
\paragraph{Magnetic Kerr-Newman in a Melvin-swirling universe ($Q=0$):}
\begin{equation}
\frac{2 \pi}{\delta \phi}\Bigl[A_{\phi}\big\rvert_{\theta=0} - A_{\phi}\big\rvert_{\theta=\pi}\Bigr]\bigg\rvert_{\eqref{conical-MKNMS}} = - \frac{H \bigl(4 + 3 B^2 H^2\bigr)}{2} \,.
\end{equation}
\paragraph{Electric Kerr-Newman in a Melvin-swirling universe ($H=0$):}
\begin{equation}
\frac{2 \pi}{\delta \phi}\Bigl[A_{\phi}\big\rvert_{\theta=0} - A_{\phi}\big\rvert_{\theta=\pi}\Bigr]\bigg\rvert_{\eqref{conical-EKNMS}} = 2 \jmath Q^3 \,.
\end{equation}
\paragraph{Dyonic Reissner-Nordstr\"{o}m in a Melvin-swirling universe ($a=0$):}
\begin{equation}
\frac{2 \pi}{\delta \phi}\Bigl[A_{\phi}\big\rvert_{\theta=0} - A_{\phi}\big\rvert_{\theta=\pi}\Bigr]\bigg\rvert_{\eqref{conical-DRNMS}} = - B^2 H Z^2 \,. \label{not-rem-dirac-DRNMS}
\end{equation}
\paragraph{Magnetic and electric Reissner-Nordstr\"{o}m in a swirling universe ($B=a=0$):}
Using for convenience the result of the dyonic case, we have that
\begin{equation}
\label{dirac-DRNS}
\frac{2 \pi}{\delta \phi}\Bigl[A_{\phi}\big\rvert_{\theta=0} - A_{\phi}\big\rvert_{\theta=\pi}\Bigr]\bigg\rvert_{\eqref{conical-DRNS}} = 2 \bigl(\jmath Q Z^2 - H\bigr) \,,
\end{equation}
which is clearly non-zero if only one charge is present.
Moreover, it also strengthens the interpretation that a charged black hole embedded in a swirling universe acquires a magnetic charge proportional to the electric one, and vice versa.
\paragraph{Kerr in a Melvin-swirling universe ($Q=H=0$):}
It is also interesting to study the only Melvin-swirling sub-case for which both singularities are not removable.
Indeed, as we have already briefly commented, the Kerr black hole in a Melvin-swirling universe also presents non-removable Dirac strings despite not being a charged black hole. 
This Dirac string can be interpreted as a singularity that arises from the interaction between the swirling parameter $\jmath$ and a charge of magnitude $p=a\,MB$, as can be seen from the following result
\begin{equation}
A_{\phi}\big\rvert_{\theta=0} - A_{\phi}\big\rvert_{\theta=\pi} =\frac{32 \jmath \bigl( a M B \bigr)^3}{1+2a^2\big(B^4 - 16 \jmath^2\bigr)M^2 + a^4 \bigl(B^4 +16 \jmath^2\bigr)^2M^4} \,.
\end{equation}

\subsection{Sub-cases with removable Dirac strings} 
As already pointed out, there are only two possible sub-cases for which the conical singularities and the Dirac strings are both removable.
Thus, in the following, we list the two spacetimes with the respective values of the rescaling of the azimuthal coordinate $\phi\mapsto\frac{2 \pi}{\delta \phi}\phi$ and gauge transformation $A_{\phi}\mapsto\frac{2 \pi}{\delta \phi}A_{\phi}-\delta A_{\phi}$, in addition to the constraints on the parameters, needed in order to remove both of these pathologies.
\paragraph{Dyonic Reissner-Nordstr\"{o}m in a swirling universe ($B=a=0$):}
Since, for this sub-case, the condition necessary to remove the conical singularities is just a rescaling of the azimuthal coordinate~\eqref{conical-DRNS}, we have that solving Eq.~\eqref{dirac-DRNS} exactly corresponds to removing the Dirac strings, independently from the fact that the conical singularities are still present or not.
Therefore, the constraints necessary to remove both the Dirac strings and the conical singularities are given by
\begin{subequations}
\label{reg-DRNS}
\begin{align}
\jmath = \frac{H}{Q Z^2} \,, \label{dirac-DRNS-cond}\\
\delta \phi = \frac{2 \pi Q^2}{Z^2} \,,
\end{align}
\end{subequations}
where the condition on the parameters~\eqref{dirac-DRNS-cond} is the only one needed in order to remove the Dirac strings.

\section{Curvature for the sub-cases}

\myparagraph{Reissner-Nordstr\"om in a swirling universe ($B=a=0$):}
For the non-rotating charged black hole in a swirling universe, there are no additional conditions other than those necessary to remove the coordinate singularities~\eqref{conical-DRNMS}, and we obtain
\begin{equation}
K_{(0)} = - 64 \jmath^2 \bigl( 3 - 5\jmath^2 Z^4\bigr) \,.
\end{equation}
In particular, in the dyonic case, when both charges $Q$ and $H$ are present, it is also possible to remove the Dirac strings~\eqref{reg-DRNS}, which yields
\begin{equation}
K_{(0)} = -\frac{64 H^2 \bigl(3 Q^2 - 5 H^2\bigr)}{Q^4 Z^4} \,,
\end{equation}
that could be set to zero with the following additional constraint
\begin{equation}
\label{cond-asympt-rns}
3 Q^2 = 5 H^2 \,,
\end{equation}
corresponding to
\begin{equation}
\jmath = \pm \frac{\sqrt{15}}{8 Q^2}\,.
\end{equation}
\myparagraph{Schwarzschild black holes ($a=Q=H=0$):}
For the Schwarzschild black holes we have that
\begin{equation}
K_{(0)} = 4 \bigl(5 B^4 - 48 \jmath^2\bigl) \,.
\end{equation}
Therefore, we can obtain $K_{(0)}=0$ by imposing
\begin{equation}
\label{cond-asympt-back}
5 B^4 = 48 \jmath^2 \,.
\end{equation}
\myparagraph{Backgrounds ($M=a=Q=H=0$):}
For the massless backgrounds, $M=0$, we obtain the same results as for the Schwarzschild black holes.
However, in addition, the curvature on the symmetry axis is \emph{exactly} constant for the backgrounds, and not just an asymptotic result for large radial distances $r \to \infty$.
\begin{equation}
K\big\rvert_{\theta = 0, \pi} \equiv 4 \bigl(5 B^4 - 48 \jmath^2\bigl) \,.
\end{equation}
\myparagraph{Other Melvin sub-cases ($\,\jmath=0$):}
Similarly, for the other three Melvin sub-cases which can result in a spacetime free of any singularity, at least outside of the event horizon, i.e.~the electric Kerr-Newman ($H=0$)~\eqref{conical-EKNM}, Kerr ($Q=H=0$)~\eqref{conical-KM} and electric Reissner-Nordstr\"om ($a=H=0$)~\eqref{conical-ERNM}, we obtain the following asymptotic constant curvature $K_{(0)}$, without any additional constraint:
\begin{equation}
K_{(0)} = 20 B^4 - 2 B^6 Q^2 - 24 B^7 a M Q + \frac{1}{4} \bigl(5 Q^4 - 12 a^2 M^2\bigr) \,,
\end{equation}
which can also be set to zero in all three sub-cases.

\section{Ergoregions for dyonic Reissner-Nordstr\"om in a (Melvin-)\hspace{0.00000001pt}swirling universe}
\label{app:ergo-RN}
\myparagraph{Dyonic Reissner-Nordstr\"om in a (Melvin-)swirling universe ($a=0$):}
\null
\begin{figure}[H]
\captionsetup[subfigure]{labelformat=empty}
\centering
\subfloat[\hspace{-0.25cm} \mbox{Event Horizon}]{{\includegraphics[height=0.4\textwidth]{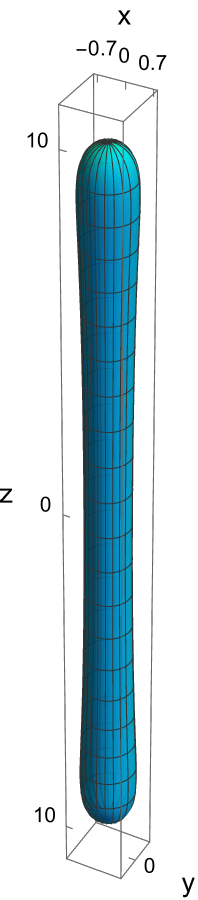}}}
\subfloat[\hspace{0.4cm} Ergoregions]{{\hspace{0.5cm}\includegraphics[width=0.3\textwidth]{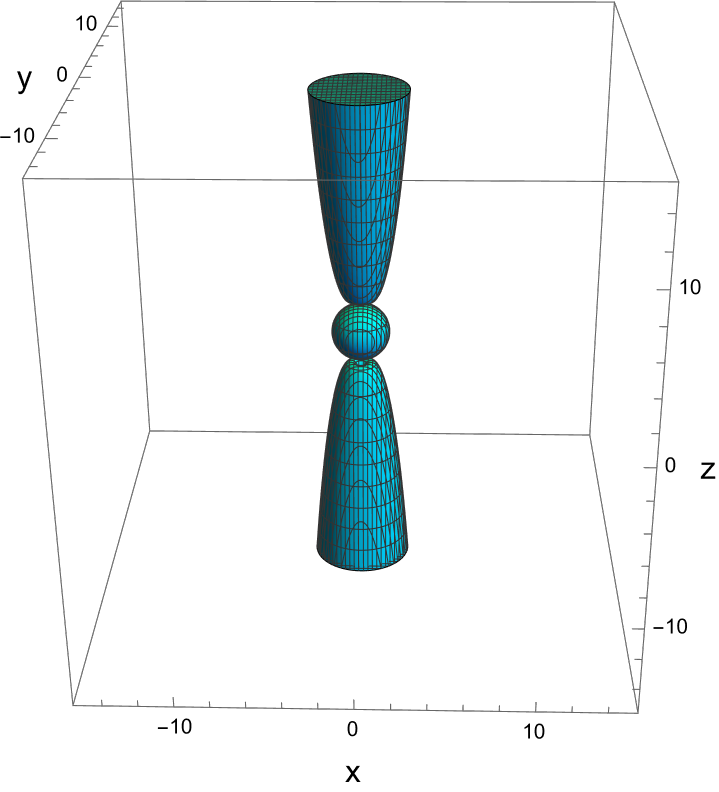}}}
\subfloat[\hspace{0.6cm} Ergoregions Cross-section $y=0$]{{\hspace{0.5cm}\includegraphics[width=0.3\textwidth]{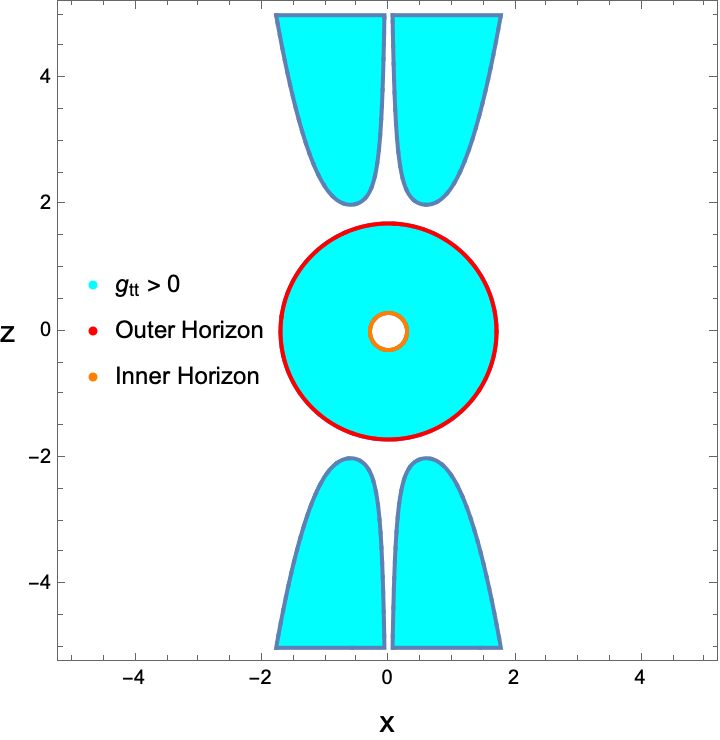}}}
\caption{\small Ergoregions and the event horizon of the dyonic Reissner-Nordstr\"om hole in a swirling universe with parameters:
$B=0$, $M=1$, $Q=1/2$, $H=1/2$, $\jmath=54/25$ and $\pazocal{I}=1/2$.}
\label{Plot-DRNS}
\end{figure}
\begin{figure}[H]
\captionsetup[subfigure]{labelformat=empty}
\centering
\subfloat[\hspace{-0.25cm} \mbox{Event Horizon}]{{\includegraphics[height=0.4\textwidth]{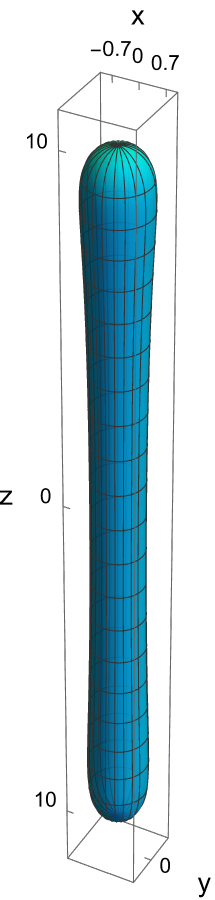}}}
\subfloat[\hspace{0.4cm} Ergoregions]{{\hspace{0.5cm}\includegraphics[width=0.3\textwidth]{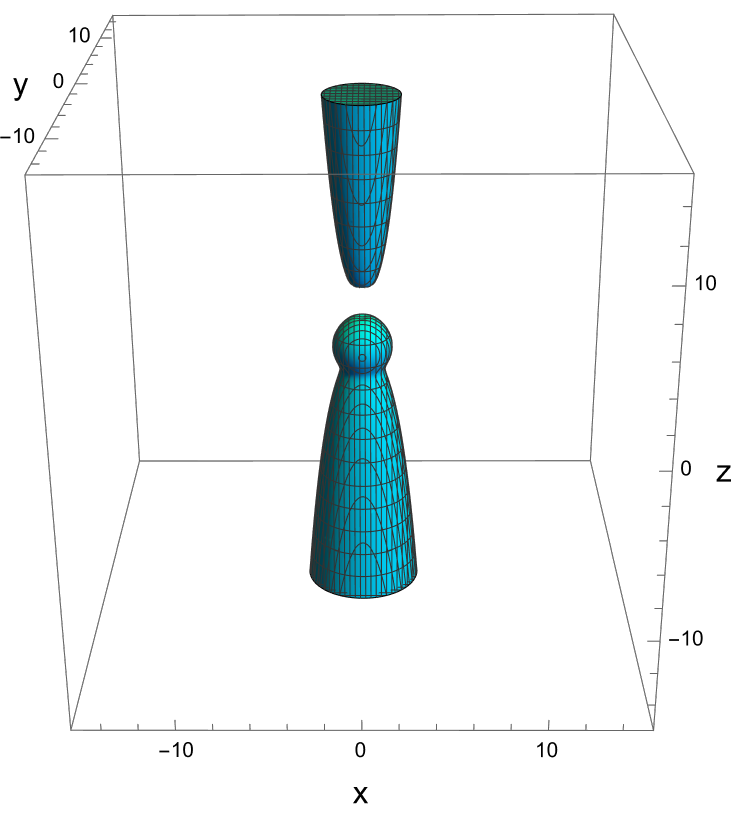}}}
\subfloat[\hspace{0.6cm} Ergoregions Cross-section $y=0$]{{\hspace{0.5cm}\includegraphics[width=0.3\textwidth]{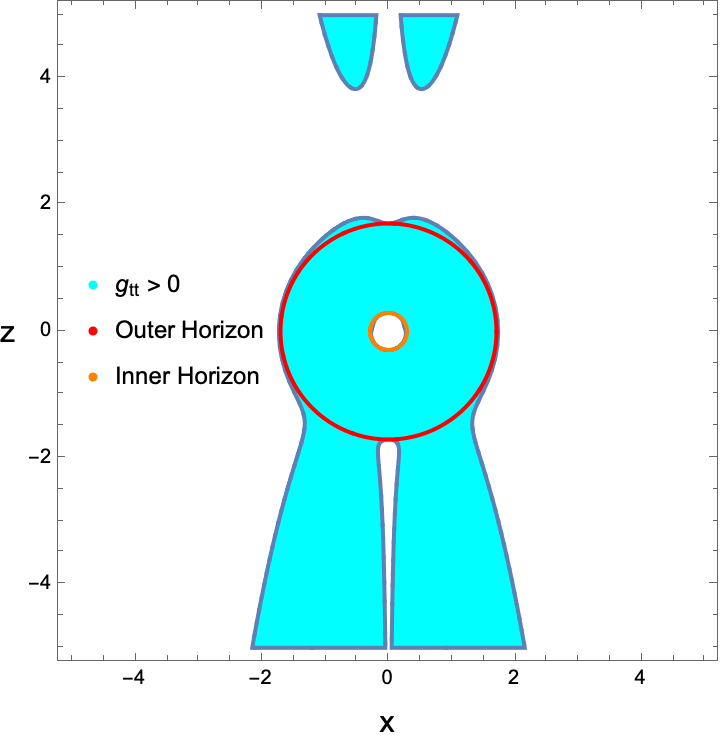}}}
\caption{\small Ergoregions and the event horizon of the dyonic Reissner-Nordstr\"om hole in a Melvin-swirling universe with parameters:
$B=4/5$, $M=1$, $Q=1/2$, $H=1/2$, $\jmath=54/25$ and $\pazocal{I}=1/2$.}
\label{Plot-DRNMS}
\end{figure}
\clearpage
\myparagraph{\small Extremal dyonic Reissner-Nordstr\"om in a (Melvin-)swirling universe ($a=0,\,M^2=Z^2$):}
\begin{figure}[H]
\captionsetup[subfigure]{labelformat=empty}
\centering
\subfloat[Event Horizon]{{\includegraphics[width=0.3\textwidth]{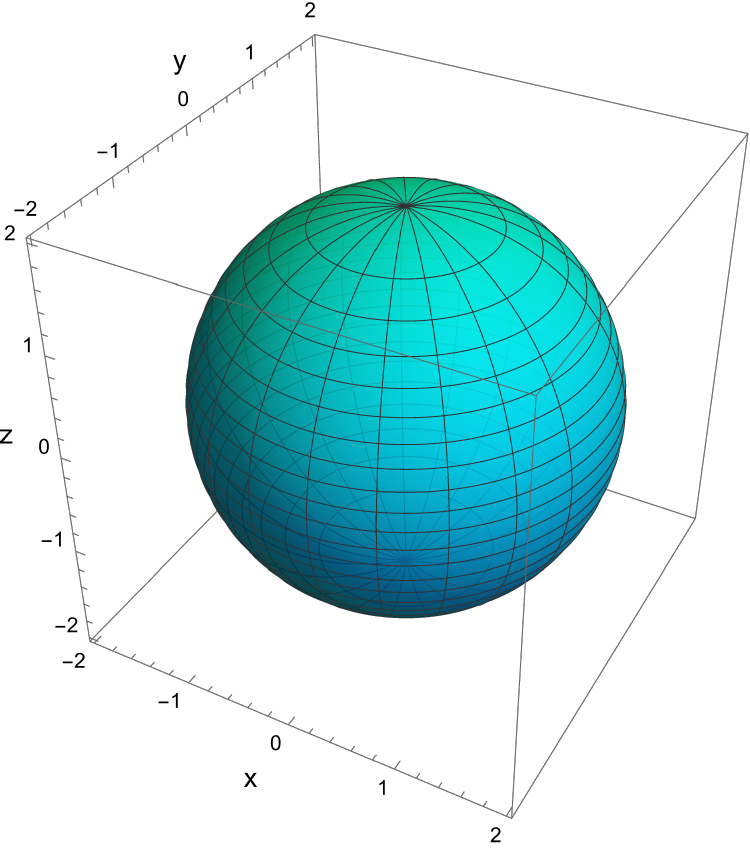}}}
\subfloat[\hspace{0.4cm} Ergoregions]{{\hspace{0.5cm}\includegraphics[width=0.3\textwidth]{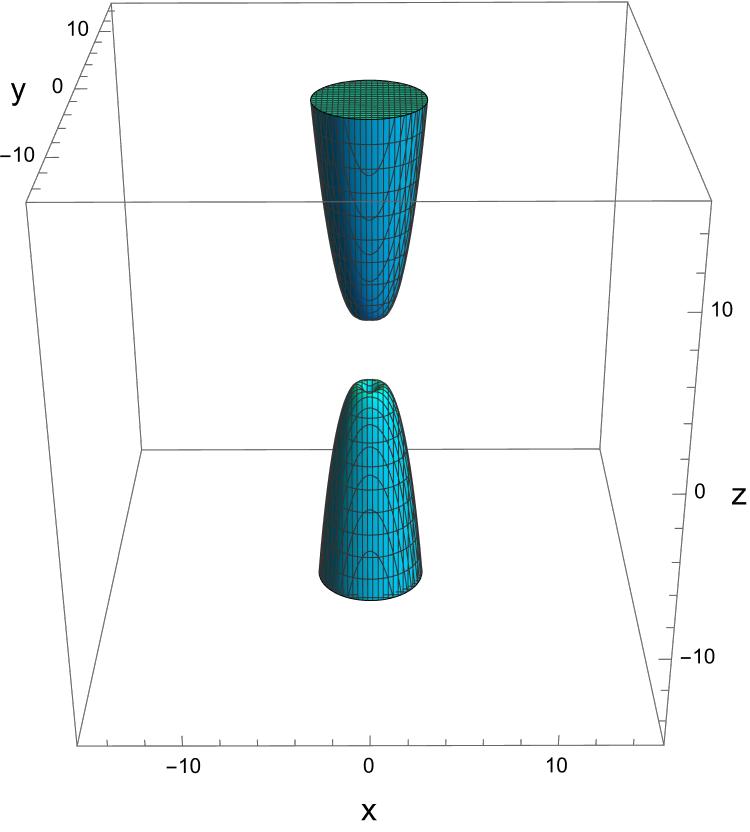}}}
\subfloat[\hspace{0.6cm} Ergoregions Cross-section $y=0$]{{\hspace{0.5cm}\includegraphics[width=0.3\textwidth]{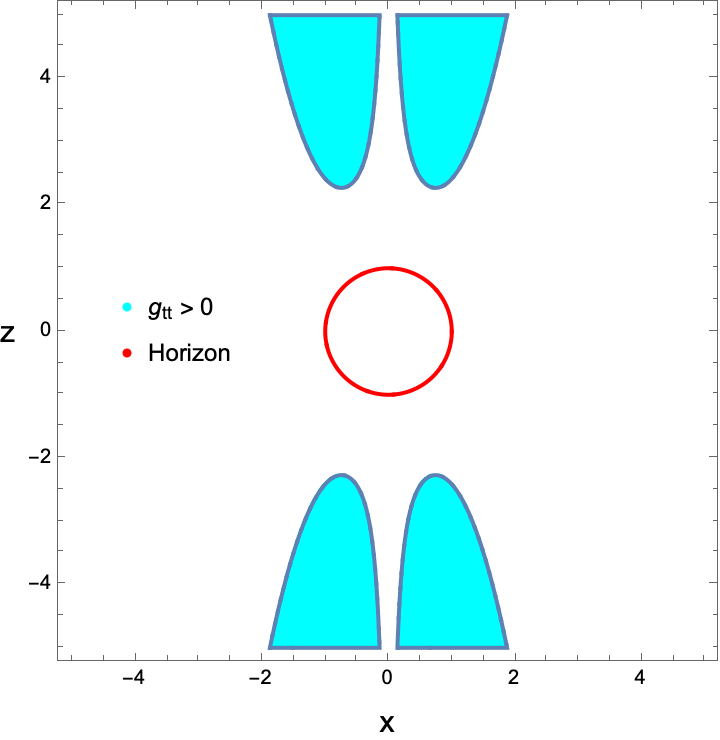}}}
\caption{\small Ergoregions and the event horizon of the extremal dyonic Reissner-Nordstr\"om hole in a swirling universe with parameters:
$B=0$, $M=1$, $Q=3/5$, $H=4/5$, $\jmath=116/75$ and $\pazocal{I}=1$.}
\label{Plot-DRNS-EXT-1}
\end{figure}
\begin{figure}[H]
\captionsetup[subfigure]{labelformat=empty}
\centering
\subfloat[Event Horizon]{{\includegraphics[width=0.3\textwidth]{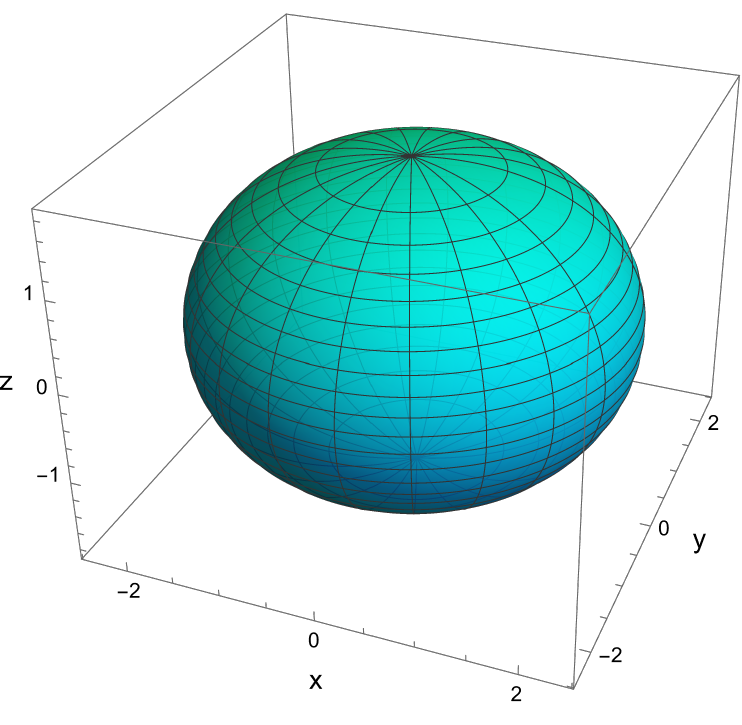}}}
\subfloat[\hspace{0.4cm} Ergoregions]{{\hspace{0.5cm}\includegraphics[width=0.3\textwidth]{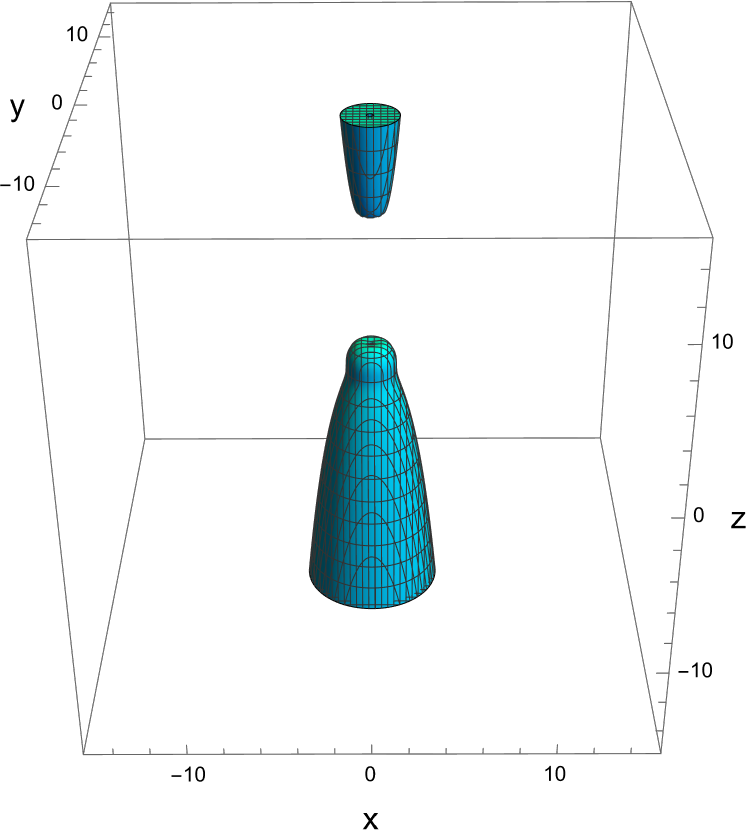}}}
\subfloat[\hspace{0.6cm} Ergoregions Cross-section $y=0$]{{\hspace{0.5cm}\includegraphics[width=0.3\textwidth]{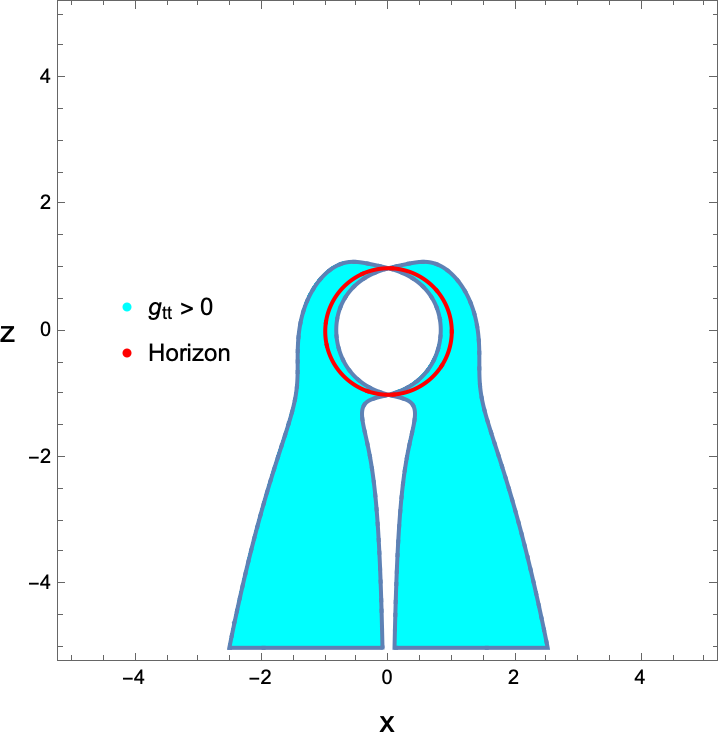}}}
\caption{\small Ergoregions and the event horizon of the extremal dyonic Reissner-Nordstr\"om hole in a Melvin-swirling universe with parameters:
$B=4/5$, $M=1$, $Q=3/5$, $H=4/5$, $\jmath=116/75$ and $\pazocal{I}=1$.}
\label{Plot-DRNMS-EXT-1}
\end{figure}
\begin{figure}[H]
\captionsetup[subfigure]{labelformat=empty}
\centering
\subfloat[Event Horizon]{{\includegraphics[width=0.3\textwidth]{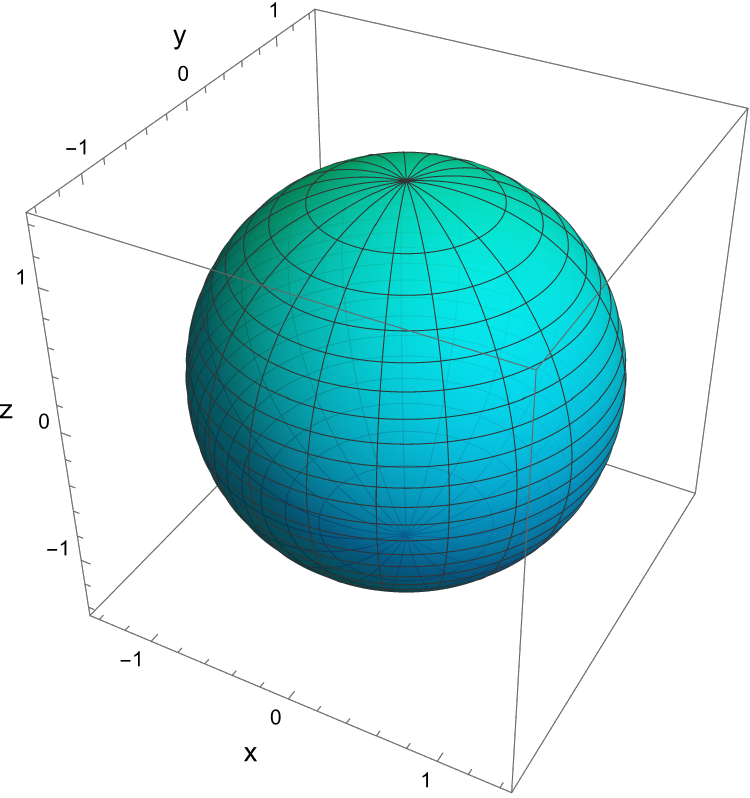}}}
\subfloat[\hspace{0.4cm} Ergoregions]{{\hspace{0.5cm}\includegraphics[width=0.3\textwidth]{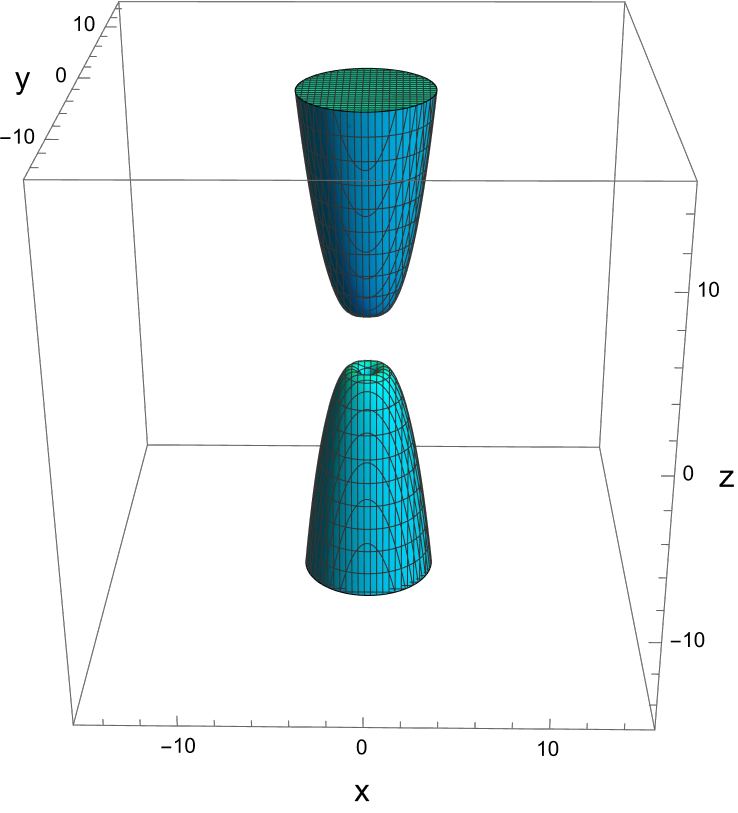}}}
\subfloat[\hspace{0.6cm} Ergoregions Cross-section $y=0$]{{\hspace{0.5cm}\includegraphics[width=0.3\textwidth]{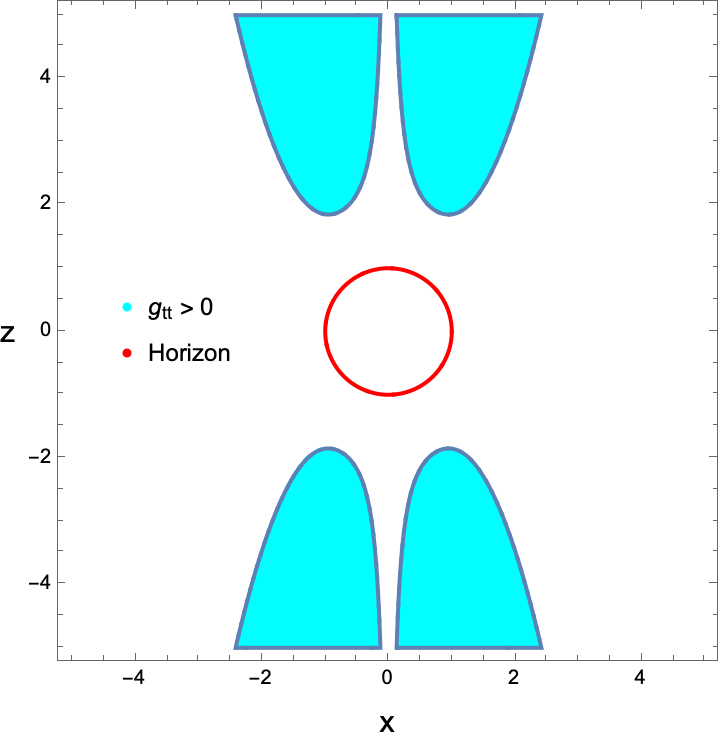}}}
\caption{\small Ergoregions and the event horizon of the extremal dyonic Reissner-Nordstr\"om hole in a swirling universe with parameters $B=0$, $M=1$, $Q=4/5$, $H=3/5$, $\jmath=87/100$ and $\pazocal{I}=1$.}
\label{Plot-DRNS-EXT-2}
\end{figure}
\begin{figure}[H]
\captionsetup[subfigure]{labelformat=empty}
\centering
\subfloat[Event Horizon]{{\includegraphics[width=0.3\textwidth]{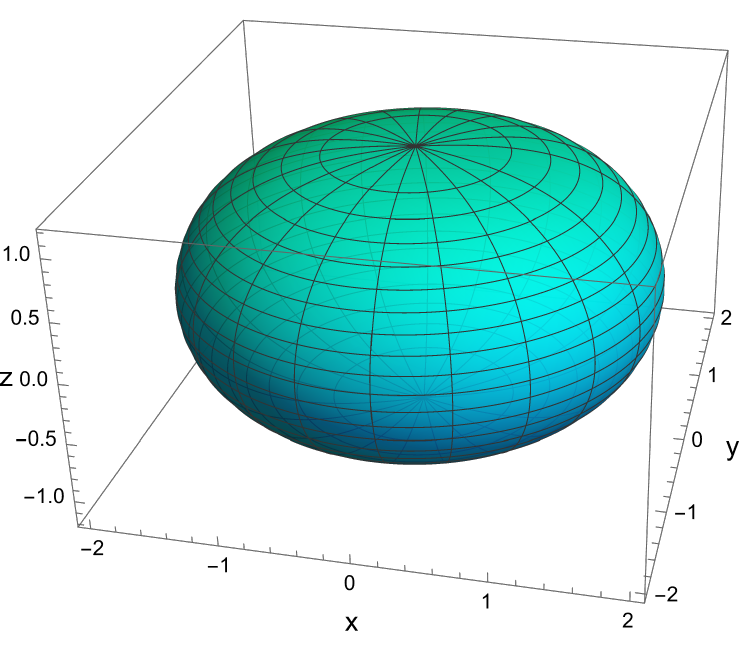}}}
\subfloat[\hspace{0.4cm} Ergoregions]{{\hspace{0.5cm}\includegraphics[width=0.3\textwidth]{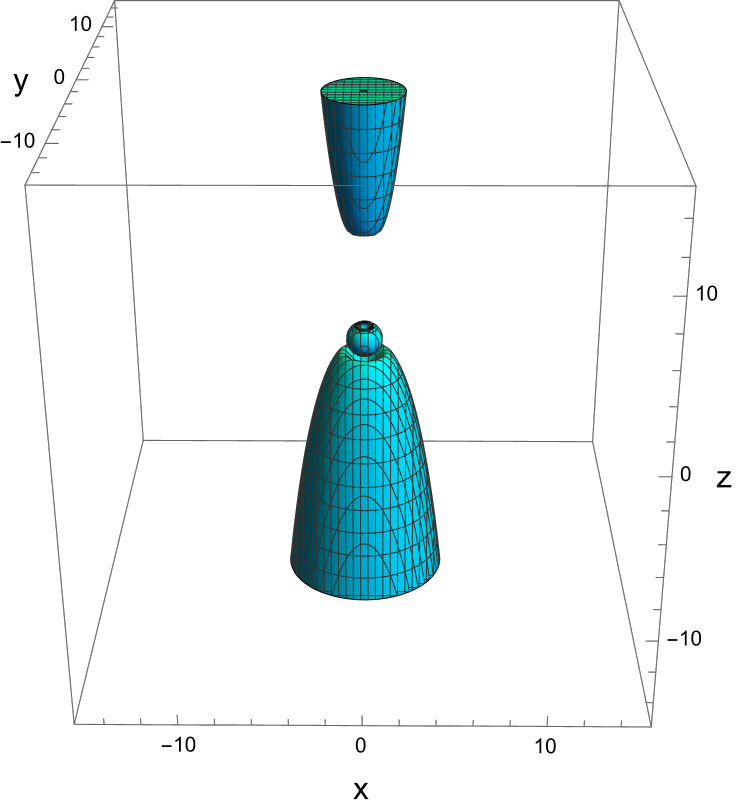}}}
\subfloat[\hspace{0.6cm} Ergoregions Cross-section $y=0$]{{\hspace{0.5cm}\includegraphics[width=0.3\textwidth]{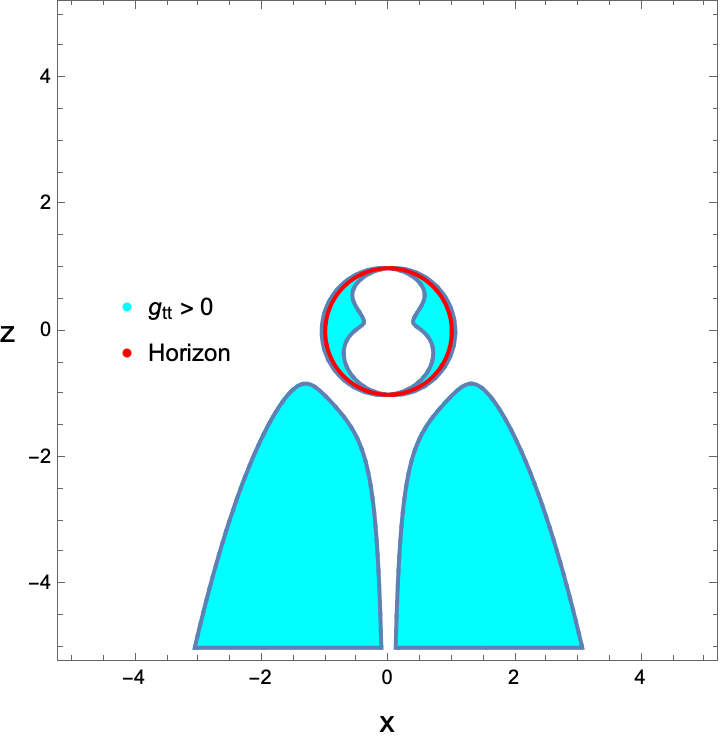}}}
\caption{\small Ergoregions and the event horizon of the extremal dyonic Reissner-Nordstr\"om hole in a Melvin-swirling universe with parameters $B=4/5$, $M=1$, $Q=4/5$, $H=3/5$, $\jmath87/100$ and $\pazocal{I}=1$.}
\label{Plot-DRNMS-EXT-2}
\end{figure}
Surprisingly, the plots in Fig.~\ref{Plot-DRNS-EXT-1} and Fig.~\ref{Plot-DRNS-EXT-2} seem to suggest that the extremal dyonic Reissner-Nordstr\"om black hole in a swirling universe resembles the classic
Reissner-Nordstr\"om black hole in the proximity of their event horizon.
Indeed, the horizon of this swirling black hole seems to be a perfect sphere, with the ergoregions not in the vicinity of the event horizon.
In contrast to what happens in the Melvin-swirling case as we can see in Fig.~\ref{Plot-DRNMS-EXT-1} and Fig.~\ref{Plot-DRNMS-EXT-2}.

We study here the geometric properties of the horizon.
\myparagraph{\small Extremal dyonic Reissner-Nordstr\"{o}m in a swirling universe ($B=a=0,\,M^2=Z^2$):}
For the extremal dyonic Reissner-Nordstr\"{o}m in a swirling universe, we obtain:
\begin{subequations}
\begin{align}
\pazocal{A} & = 4 \pi Z^2 \bigl(1 + \jmath^2 Z^4\bigr) \,,\\
\pazocal{C}_{equator} & = 2\,\pazocal{C}_{meridian} = 2 \pi |Z| \sqrt {1 + \jmath^2 Z^4 } \,,\\
\pazocal{R}_{1} & = |Z| \sqrt {1 + \jmath^2 Z^4} \,, \\
\pazocal{R}_{2} & = \frac{2\,\pazocal{C}_{meridian}}{\pazocal{C}_{equator}} = 1 \label{r1-DRNS} \,.
\end{align}
\end{subequations}
Hence, for this sub-case, the horizon is a perfect sphere ($\pazocal{R}_{2}=1$) as it is for the non-swirling black hole.
On the other hand, the physical radius $\pazocal{R}_{1}$ receives an additional contribution due to the swirling background.
Therefore, the effect of the swirling parameter in the extremal case is merely to change the volume, leaving the horizon perfectly spherical.
Furthermore, if we also remove the Dirac strings~\eqref{reg-DRNS}, we obtain

\begin{equation}
\pazocal{R}_{1} = \frac{Z^2}{|Q|} \label{R1-DRNS-reg}\,.
\end{equation}
\myparagraph{\small Extremal dyonic Reissner-Nordstr\"{o}m in a Melvin-swirling universe ($a=0,\,M^2=Z^2$):}
However, if the Melvin parameter $B$ is not zero, we have:
\begin{subequations}
\begin{align}
\pazocal{A} & = \frac{4 \pi Z^2 \Bigl[| - 4 H^2 + 4 \jmath H Q Z^2| + \jmath^2 Z^6 \Bigr]}{H^2} \,,\\
\pazocal{C}_{equator} & = \frac{2 \pi \Bigl[| - 4 H^2 + 4\jmath H Q Z^2| + \jmath^2 Z^6 \Bigr]}{|\jmath H| Z^2} \,,\\
\pazocal{C}_{meridian} &= \Bigg[\frac{2 Z \sqrt{| - 4 H^2 + 4 \jmath H Q Z^2| + \jmath^2 Z^6}}{|H|}\Biggr] \text{EllipticE}\bigg[\frac{ |-4 H^2 + 4 \jmath H Q Z^2|}{ |- 4 H^2 + 4 \jmath H Q Z^2| + \jmath^2 Z^6 }\bigg] \,, \\
\pazocal{R}_{1} & = \bigg|\frac{\jmath Z^4}{H}\bigg| \,, \label{R1-DRNMS}\\
\pazocal{R}_{2} & \neq 1 \,,
\end{align}
\end{subequations}
where 
\begin{equation}
\text{EllipticE}[x] = \int_{0}^{\frac{\pi}{2}}\sqrt{1 - x \sin^2 \theta}\, d\theta \,,
\end{equation}
is the complete elliptic integral of the second kind.

Thus, in contrast to the previous case, we now find that the horizon is \emph{not} a perfect sphere.
Nevertheless, it is interesting to notice that if we use the same value of the swirling parameter $\jmath$ that removes the Dirac strings in the non-Melvin case~\eqref{reg-DRNS}, we obtain that the ratio $\pazocal{R}_{1}$~\eqref{R1-DRNMS} for this black hole becomes exactly equal to that of the non-Melvin case as in Eq.~\eqref{R1-DRNS-reg}.
However, it should be noted that this choice of swirling parameter does not remove the Dirac strings for this charged Melvin-swirling black hole, as we proved in Eq~\eqref{not-rem-dirac-DRNMS}.

\section{Near-horizon extremal geometry for the sub-cases}
\label{app:nheg}

\myparagraph{NHEG dyonic Reissner-Nordstr\"{o}m in a swirling universe ($B=a=0$):}
For the dyonic Reissner-Nordstr\"{o}m in a swirling universe we have $\Omega_{H}^{(ext)} = A_{0 H}^{(ext)} = 0$, and the near-horizon extremal geometry is given by
\begin{subequations}
\begin{align}
\label{NHE-DRNS}
d\hat{s}{}^2 & = \Gamma \Bigl[ -\hat{r}^2\,d\hat{t}^2 + \frac{d\hat{r}^2}{\hat{r}^2} + d\theta^2 + \frac{Z^4 \sin^2\theta}{\Gamma^2}\,d\hat{\phi}^2 \Bigr] \,,\\
\hat{A} & = \Bigl[\jmath H Z^2 + Q \Bigr]  \hat{r} \,d\hat{t} + \frac{Z^2}{\Gamma} \Bigl[\jmath Q Z^2 - H \Bigr]  \cos\theta \, d\hat{\phi} \,,\\ %
\Gamma & = Z^2 \bigl( 1 + \jmath^2 Z^4 \bigr) \,.
\end{align}
\end{subequations}
Moreover, if we remove the conical singularities with the conditions given by Eq.~\eqref{conical-DRNS}, we obtain that the geometry becomes exactly $AdS_{2} \times S^{2}$:
\begin{subequations}
\label{NHEDRN}
\begin{align}
d\hat{s}{}^2 & = \Bigl[e^2 + p^2 \Bigr]\Bigl[ -\hat{r}^2\,d\hat{t}^2 + \frac{d\hat{r}^2}{\hat{r}^2} + d\theta^2 + \sin^2\theta d\hat{\phi}^2 \Bigr] \,,\\
\hat{A} & = e \, \hat{r} \, d\hat{t} + p \, \cos\theta \, d\hat{\phi} \,,\\
e & = \jmath H Z^2 + Q \,,\\
p & = \jmath Q Z^2 - H \,.
\end{align}
\end{subequations}
Therefore, the NHEG of the dyonic Reissner-Nordstr\"{o}m in a swirling universe is exactly the same as that of the dyonic Reissner-Nordstr\"{o}m without the swirling parameter.

Furthermore, this result also reveals another significant implication of embedding a charged black hole in the swirling universe.
Indeed, from the electromagnetic potential of this spacetime~\cite{Astorino:2022prj}, not necessarily in the extremal case, we have that a magnetic Ehlers transformation has the effect of mixing the charges, in the sense that if the seed solution is only electrically charged, $H=0$, we have that the resulting solution embedded in a swirling universe will also possess a magnetic charge proportional to the seed electric charge, and vice-versa if the seed solution is only magnetically charged, $Q=0$. 

However, from the NHE limit of this charged swirling black hole~\eqref{NHEDRN}, we can also conclude that, if both the magnetic and electric charge are present, the swirling parameter $\jmath$ also provides a way to remove the effect of one of the charges at the horizon, without actually setting the corresponding black hole charge to zero. 

Indeed, the condition to remove the Dirac strings~\eqref{reg-DRNS}, $ \jmath=\frac{H}{Q Z^2}$, exactly sets $p=0$ (i.e.~$\hat{A}_{\hat{\phi}}=0$) and $e=\frac{Z^2}{Q}$.
Thus, the removal of the Dirac string is equivalent to constraining the swirling parameter $\jmath$ in such a way that is null the effect of the magnetic charge $p$ in the extremal near-horizon limit.
Moreover, this value of the charge $e$ is also the same as the physical radius $\pazocal{R}_{1}$ we found for this black hole~\eqref{R1-DRNS-reg}.

Analogously, if we set $\jmath=-\frac{Q}{H Z^2}$ we obtain $e=0$ (i.e.~$\hat{A}_{\hat{t}} = 0$) and $p=-\frac{Z^2}{H}$, thus removing the effect of the electric charge at the horizon.
\myparagraph{NHEG magnetic Kerr-Newman in Melvin-swirling universe ($Q=0$):}
For the magnetic Kerr-Newman in Melvin-swirling universe sub-case, $Q=0$, after having removed the conical singularities with the constraints given by Eqs.~\eqref{conical-MKNMS}, it is possible to obtain $\frac{r_{0}^{4}}{\Gamma^2}\Bigr[\frac{4 \pi^2}{\delta \hat{\phi}^2}\Bigr]=1$ for 
\begin{equation} 
B = \pm \frac{2}{\sqrt{4 a^2 + 3 H^2}} \,.
\end{equation}
Moreover, this constraint of the parameters also sets $\kappa=0$, which implies that also for this black hole the NHE geometry becomes exactly $AdS_{2} \times S^{2}$.

Additionally, it is also possible to set $\hat{A}_{0}=0$, and therefore $\hat{A}_{\hat{t}}=0$ since $\kappa = 0$, if we also constrain the magnetic charge as
\begin{equation}
H = \pm a \sqrt{3 + \sqrt{17}} \,.
\end{equation}
\myparagraph{NHEG dyonic Kerr-Newman in a Melvin-swirling universe:}
Similarly, for the general rotating and dyonic case~\eqref{NHEDKNMS}, it is also possible to set $\Bigr[\frac{4 \pi^2}{\delta \hat{\phi}^2}\Bigr]=1$.
However, the possible values of $B$ for which this happens are \emph{not} compatible with the removal of the Dirac Strings~\eqref{reg-DKNMS}.
Indeed, for example, a possibility is given by
\begin{equation}
B = \pm \frac{2 a}{\sqrt{4 a^4 + a^2 \bigl ( 3 H^2 + 5 Q^2 \bigr) + 2 Q \biggl [ Q Z^2 + M \sqrt{ Q^2 Z^2 + a^2 \bigl( 3 H^2 + 4 Q^2 + 4 a^2 \bigr)}\, \biggr] }} \,,
\end{equation}
which is clearly not compatible with the constraints given by Eqs.~\eqref{reg-DKNMS}.
\myparagraph{NHEG of the others Melvin-swirling sub-cases without conical singularities:}
In all the other sub-cases where the Melvin parameter $B$ and the swirling parameter $\jmath$ are both present, and the conical singularities are removable, we have that the condition
\begin{equation}
\frac{r_{0}^{4}}{\Gamma^2}\biggl(\frac{2 \pi}{\delta \hat{\phi}}\biggr)^2 = 1 \,,
\end{equation}
can be satisfied only for $\theta = 0$ or  $\theta = \pi$, which means that the geometry can never be exactly $AdS_{2} \times S^{2}$.
Moreover, since $\sin\theta=0$ for these values of $\theta$, we have that there is not a single slice of the spacetime for which the geometry is a “warped” $AdS_{3}$.

Therefore, we can conclude that the near-horizon geometry of the extremal dyonic Reissner-Nordstr\"{o}m in a Melvin-swirling universe is not isometric to that of the ``classic'' dyonic Reissner-Nordstr\"{o}m, in contrast to what happens for the swirling $B=0$ sub-case~\eqref{NHEDRN}.

\section{Invariance under an Ehlers-Harrison transformation}
\label{app:ernst-property}
Using the definition of the Ernst potentials~\eqref{ernst-potentials}, we have that the function $f$ in the LWP ansatz is given by
\begin{equation}
\pm f = Re{(\E)} + |\Phi|^2 \,,
\end{equation}
where the positive sign is for the electric ansatz~\eqref{lwp-electric}, while the minus sign is for the magnetic ansatz~\eqref{lwp-magnetic}.

Consequently, after an Ehlers transformation~\eqref{ehlers}, the transformed function $f'$ is given by
\begin{equation}
f'_\text{Ehlers} = \frac{f}{|1 + i \,\jmath \,\E|^2} = \frac{f}{\bigl(1-\jmath\,h)^2+\jmath^2
\bigl(f^2+|\Phi|^2\bigr)} \,;
\end{equation} 
analogously, for a Harrison transformation~\eqref{harrison}, one has
\begin{equation}
f'_\text{Harrison} = \frac{f}{|1 - 2\alpha^*\Phi - |\alpha|^2\E|^2} \,.
\end{equation} 
Therefore, an Ehlers or Harrison transformation, whether magnetic or electric, never changes the sign of $f$ by construction.
Thus, after an electric transformation, i.e.~when one adds the NUT parameter or electromagnetic charges, the ergoregions are not modified, since $f=-g_{tt}$.
Similarly, a magnetic transformation, i.e.~when one embeds the seed in a swirling or Melvin background, does not alter the region in which closed timelike curves (CTCs) are allowed, since $f=g_{\phi \phi}$.

\end{document}